\documentclass[aps,10pt,twocolumn,groupedaddress,showpacs,reprintnumbers,amssymb,floatfix]{revtex4}
\usepackage{epsfig}
\usepackage{longtable}
\usepackage{graphicx}
\usepackage{lscape,ulem}
\usepackage{dcolumn}
\usepackage{amsmath,nccmath}
\begin{document}
\title{Incompressibility in finite nuclei and nuclear matter}
\author{J.~R.~Stone}
\author{N.~J.~Stone}
\affiliation{Department of Physics, University of Oxford, Oxford, OX1 3PU, UK}
\affiliation {Department of Physics and Astronomy, University of Tennessee, Knoxville, TN 37996, USA} 
\author{S.~A.~Moszkowski}
\affiliation{Department of Physics and Astronomy, UCLA, Los Angeles, CA 90095-1547, USA}
\date{\today}
 
\begin{abstract}
The incompressibility (compression modulus) $K_{\rm 0}$ of infinite symmetric nuclear matter at saturation density has become one of the major constraints on mean-field models of nuclear many-body systems as well as of models of high density matter in astrophysical objects and heavy-ion collisions. It is usually extracted from data on the Giant Monopole Resonance (GMR) or calculated using theoretical models. We present a comprehensive re-analysis of recent data on GMR energies in even-even $^{\rm 112-124}$Sn and $^{\rm 106,100-116}$Cd and earlier data on 58 $\le$ A $\le$ 208 nuclei. The incompressibility of finite nuclei $K_{\rm A}$ is calculated from experimental GMR energies and expressed in terms of $A^{\rm -1/3}$ and the asymmetry parameter $\beta$ = (N-Z)/A as a leptodermous expansion with volume, surface, isospin and Coulomb coefficients $K_{\rm vol}$, $K_{\rm surf}$, $K_\tau$ and $K_{\rm coul}$. Only data consistent with the scaling approximation, leading to a fast converging leptodermous expansion, with negligible higher-order-term contributions to $K_{\rm A}$,  were used in the present analysis. \textit{Assuming} that the volume coefficient  $K_{\rm vol}$ is identified with $K_{\rm 0}$,  the $K_{\rm coul}$ = -(5.2 $\pm$ 0.7) MeV and  the contribution from the curvature term K$_{\rm curv}$A$^{\rm -2/3}$ in the expansion is  neglected,  compelling evidence is found for $K_{\rm 0}$  to be in the range 250 $ < K_{\rm 0} < $ 315 MeV, the ratio of the surface and volume coefficients $c = K_{\rm surf}/K_{\rm vol}$ to be between -2.4 and -1.6 and $K_{\rm \tau}$  between  -840 and  -350 MeV.  In addition, estimation of the volume and surface parts of the isospin coefficient $K_\tau$, $K_{\rm \tau,v}$ and $K_{\rm \tau,s}$, is presented. 

We show that the generally accepted value of $K_{\rm 0}$ = (240 $\pm$ 20) MeV can be obtained from the fits provided $c \sim$ -1,  as predicted by the majority of mean-field models. However, the fits are significantly improved if $c$ is allowed to vary, leading to a range of  $K_{\rm 0}$, extended to higher values. The results demonstrate the importance of nuclear surface properties in determination of $K_{\rm 0}$ from fits to the leptodermous expansion of $K_{\rm A}$ . 

A self-consistent simple (toy) model has been developed, which shows that the density dependence of the surface diffuseness of a vibrating nucleus plays a major role in determination of the ratio K$_{\rm surf}/K_{\rm vol}$ and yields predictions consistent with our findings.

\end{abstract}
\pacs{21.60.Jz, 21.65.Cd, 21.65.Mn, 24.30.Cz}                           
\maketitle

\section{\label{intr}Introduction}
The incompressibility (compression modulus) K$_{\rm 0}$ of infinite symmetric nuclear matter (SNM) at saturation density has become one of the major constraints on mean-field models of nuclear many-body systems. Although infinite SNM does not exist in nature, its empirical properties, such as saturation density and saturation energy are rather well established (see e.g. \cite{dutra2012} and references. therein). Other quantities of interest,  such as the symmetry energy and its slope at saturation density \cite{tsang2012} and the compressibility modulus are much less constrained and are the subject of continued study. Traditionally, the experimental source of information on K$_{\rm 0}$ has been the Giant Monopole Resonance (GMR). A relatively large amount of data on GMR energies have been collected over the years with development in experimental technique followed by more complicated and accurate data analysis.

Alongside analysis and interpretation of GMR data which, admittedly, have some limitations, considerable effort has been put into theoretical calculation of $K_{\rm 0}$. The main model frameworks employed have been non-relativistic Hartree-Fock (HF) and relativistic mean-field (RMF) models with various effective interactions, extended beyond mean field by (Quasiparticle) Random-Phase approximation [(Q)RPA], and different variants of the liquid drop model. We summarize in Table~\ref{tab:surv} a representative selection of results of such calculations. Since the early 1960's, theoretical predictions of the compression modulus have fallen into three classes. The first comprises models based on so-called `realistic' potentials with parameters fitted to data on free nucleon-nucleon scattering (phase-shifts, effective ranges) and properties of the deuteron \cite{falk1961, bethe1971}, and the second models using effective density dependent nucleon-nucleon interactions, fitted to data on (doubly) closed shell nuclei and saturation properties of nuclear matter \cite{brink1967, vautherin1972, beiner1975, gogny1975, koehler1976}. The third class of models utilize the semi-empirical mass formula and its development to the liquid drop model and later the droplet model and its variants \cite{myers1966, myers1969, myers1974,myers1990}. `Realistic' models predicted systematically lower value of incompressibility (100 - 215 MeV) whereas models with effective interactions, mainly of the Skyrme type, predicted a wide range of higher values, up to 380 MeV. The empirical droplet-type models showed limited sensitivity to the value of $K_{\rm 0}$, which has been used as a chosen input parameter rather then a variable obtainable from the fit to atomic masses \cite{moller1995, moller2012}.  The preference of the early years was clearly for results of the `realistic' models which were seen as more fundamental. 

The first (to our knowledge) use of experimental data on GMR energies, in $^{\rm 40}$Ca, $^{\rm 90}$Zr and $^{\rm 208}$Pb, taken from an unpublished report by Marty et al. \cite{marty1975}, was performed by Blaizot et al. \cite{blaizot1976} who determined $K_{\rm 0}$ = (210 $\pm$ 30) MeV. 
\begingroup
\pagestyle{headings}
\squeezetable
\begin{longtable*}{llll}
\caption{\label{tab:surv}K$_{\rm 0}$ as calculated in selected representative theoretical approaches in chronological order. (S)HF stands for (Skyrme)Hartree-Fock, HFB for Hartree-Fock-Bogoliubov, (Q)RPA for (Quasiparticle) Random Phase Approximation, GCM Generator Coordinate Method, FRDM Finite Range Droplet Model, HB Hartree-Bogoliubov, PC Point coupling, EDF Energy Density Functional.  All entries are in MeV. For more detail see text and references therein.}\\
\hline\hline
\multicolumn{1}{l}{K$_{\rm 0}$} & \multicolumn{1}{c}{Method} & \multicolumn{1}{c}{Data} & \multicolumn{1}{c}{Reference}\\
\hline
\endfirsthead
\multicolumn{4}{c}%
{\tablename\ \thetable{} -- continued from previous page}\\
\hline\hline
\multicolumn{1}{l}{K$_{\rm 0}$} & \multicolumn{1}{c}{Method} & \multicolumn{1}{c}{Data} & \multicolumn{1}{c}{Reference}\\
\hline
\endhead
\hline
\endfoot
\hline\hline
\endlastfoot
214             & Puff-Martin model                            & Singlet and triplet scattering lengths and                                             &  Falk\&Wilets 1961 \cite{falk1961}       \\
                   & Yamaguchi potential                         & effective ranges; deuteron binding energy;                                           &                                                               \\
                   &                                                          & singlet phase shifts at 310 MeV.                                                            &                                                                \\
172 - 302   & Various early models                        &                                                                                                                & cited in \cite{falk1961}                 \\
150 - 380   & HF + simple 2-body potentials        & Properties of light nuclei;                                                                       & Brink\&Boecker 1967 \cite{brink1967}      \\
                   &                                                          & saturation properties of SNM.                                                                &                                           \\
295             & Thomas-Fermi and droplet models   &  Nuclear masses.                                                                                    &  Myers\&Swiatecki 1969 \cite{myers1969}   \\
100-200     & Various (realistic) models                  & Data prior to 1971 see \cite{bethe1971}.                                               & Bethe 1971 \cite{bethe1971}               \\
                   &HF: Skyrme                                        &  Binding energy and charge radius:                                                    & Vautherin\&Brink 1972 \cite{vautherin1972} \\
 370(a)        & SI (a)                                                  & $^{\rm 16}$O and $^{\rm 208}$Pb;                                                              &                                            \\ 
  342(b)       & SII (b)                                                 &  saturation properties of SNM;                                                                &                                            \\
                   &                                                           &  symmetry energy;                                                                                  &                                             \\
                   &                                                           &  spin-orbit splitting: 1p levels in $^{\rm 16}$O.                                    &                                            \\
240             & Droplet model for arbitrary shape     &Nuclear masses; fission barriers;                                                            & Myers\&Swiatecki 1974 \cite{myers1974}     \\
                   &                                                           & K$_{\rm 0}$ given in the list of preliminary                                                &                                          \\
                   &                                                           & input parameters.                                                                                   &                                            \\
306-364     & SHF:   SIII-SVI                                    & Binding energies and charge radii:                                                       & Beiner et al. 1975 \cite{beiner1975}       \\
                   &                                                           & $^{\rm 16}$O, $^{\rm 40}$Ca, $^{\rm 48}$Ca, $^{\rm 56}$Ni,                            &                                            \\
                   &                                                           & $^{\rm 90}$Zr, $^{\rm 140}$Ce, $^{\rm 208}$Pb.                                             &                                            \\       
 228             & HFB: Gogny D1                                  &  Properties of $^{\rm 16}$O and $^{\rm 90}$Zr;                                     & Gogny 1975 \cite{gogny1975}               \\
                   &                                                           &  spin-orbit splitting: 1d and 1p levels in $^{\rm 16}$O;                       &                                       \\
                   &                                                           &  2s neutron and  proton levels in $^{\rm 48}$Ca.                                 &                                             \\
                   &                                                            &  saturation properties of SNM; symmetry energy.                              &                                              \\  
 263             & SHF: Ska                                            &  Coefficients in the semi-empirical mass                                                & Koehler 1976 \cite{koehler1976}       \\
                   &                                                           &  formula \cite{myers1969}.                                                                       &                                                    \\
180 - 240   & HF + RPA: B1, D1, Ska, SIII, SIV          &  E$_{\rm GMR}$: $^{\rm 40}$Ca, $^{\rm 90}$Zr,$^{\rm 208}$Pb.                & Blaizot\&Grammaticos 1976 \cite{blaizot1976} \\ 
200 - 240   & Expansion of $K_{\rm A}$;                    &  $^{\rm 16}$O, $^{\rm 40}$Ca,$^{\rm 90}$Zr,$^{\rm 208}$Pb.                             &  Treiner et al. 1981 \cite{treiner1981} \\     
                   & asymptotic RPA sum rules                  &                                                                                                                 &                                          \\
275 - 325   & Expansion of $K_{\rm A}$                     &  E$_{\rm GMR}$:  $^{\rm 24}$Mg, $^{\rm {112-124}}$Sn, $^{\rm {144-152}}$Sm, $^{\rm 208}$Pb.  & Sharma et al. 1988 \cite{sharma1988}   \\
301             & Thomas-Fermi Model:                        & Ground state nuclear properties; neutron matter;                                    & Myers\&Swiatecki 1990 \cite{myers1990} \\
                   & Seyler-Blanchard effective                  &  diffuseness of nuclear density distributions;                                          &                                         \\
                   & interaction                                          & parameters of the optical potential.                                                         &                                    \\
200 - 350   & Expansion of $K_{\rm A}$                     & All E$_{\rm GMR}$ data available in 1993.                                          & Shlomo\&Youngblood 1993 \cite{shlomo1993} \\
280 - 310   & Constrained RMF+GCM:                      & E$_{\rm GMR}$: $^{\rm 40}$Ca,$^{\rm 90}$Zr, $^{\rm 208}$Pb.                           & Stoitsov et al. 1994, \cite{stoitsov1994} \\
                   & NL1, NL-SH, NL2, HS, L1                    &                                                                                                                 &                                           \\
240             & FRDM                                                  & Ground state atomic masses; fission barriers;                                         & Moller\&Nix 1995 \cite{moller1995} \\
                   &                                                            & low sensitivity to $K_{\rm 0}$ -  cannot rule out higher values.                 &                                                            \\
210 - 220   & HF(HFB)+RPA: Gogny D1S, D1,            & E$_{\rm GMR}$: $^{\rm 90}$Zr, $^{\rm 116}$Sn,$^{\rm 144}$Sm, $^{\rm 208}$Pb. & Blaizot et al. 1995 \cite{blaizot1995} \\
                   & D250, D260, D280, D300                     &                                                                                                                &                                      \\
200 - 230   & SHF + BCS; RPA:                                  &Masses and charge radii: $^{\rm 16}$O, $^{\rm 40,48}$Ca,$^{\rm 90}$Zr,         & Farine et al. 1997 \cite{farine1997} \\
                   & SkK180, SkK200, SkK220,                    & $^{\rm 112-124,132}$Sn, $^{\rm 144}$Sm,$^{\rm 208}$Pb.                                &                                      \\
                   & SkK240, SkKM                                      &                                                                                                                &                                     \\
250 - 270   & Time dependent RMF:                           & E$_{\rm GMR}$: $^{\rm 16}$Ca, $^{\rm 90}$Zr; $^{\rm 114}$Sn, $^{\rm 208}$Pb. & Vretenar et al. 1997 \cite{vretenar1997} \\
                   &  NL1, NL3;                                            &                                                                                                                &                                        \\
                   & Constrained RMF+GCM:                       &                                                                                                                &                                       \\
                   & NL1, NL3, NL-SH, NL2                         &                                                                                                                &                                         \\
225 - 236   & Comparison with \cite{blaizot1995};   & E$_{\rm GMR}$: $^{\rm 40}$Ca,$^{\rm 90}$Zr,$^{\rm 116}$Sn, $^{\rm 144}$Sm, $^{\rm 208}$Pb.    &  Youngblood et al. 1999 \cite{youngblood1999} \\
                   & E0 strength distribution                       &                                                                                                            &                                       \\ 
200 - 240   & EDF scaling approximation to GMR      & Nuclear masses and E$_{\rm GMR}$ data on                                          & Chung et al. 1999 \cite{chung1999} \\
                   &                                                              & 18 spherical nuclei with 89 $<$ A $<$ 209.                                         &                                    \\
268 - 308   & Expansion of  $K_{\rm A}$                      &  Nuclear masses.                                                                                    & Sapathy et al.   \cite{satpathy1999}  \\
240 - 275   & RMF: family of interactions;                 & E$_{\rm GMR}$: $^{\rm 208}$Pb.                                                                   & Piekarewicz  2002 \cite{piekarewicz2002} \\
                   &SHF + RPA:                                            & Binding energies and charge and neutron radii:                                       & Agrawal et al.2003\cite{agrawal2003}   \\
255(a)         & SK255(a)                                                & $^{\rm 16}$O,$^{\rm 40}$Ca,$^{\rm 48}$Ca,$^{\rm 90}$Zr,                             & \\
272(b)         & SK272(b)                                               & $^{\rm 116}$Sn,$^{\rm 132}$Sn,$^{\rm 208}$Pb;                                            &                                   \\
                   &                                                              & E$_{\rm GMR}$: $^{\rm 90}$Zr,$^{\rm 116}$Sn,$^{\rm 144}$Sm,$^{\rm 208}$Pb;  &        \\
                   &                                                              &  RMF NL3 used as `experimenatal data' .                                                     &                        \\
 230 - 250   &SHF + RPA:                                             &  Properties of infinite nuclear matter;                                                     & Colo et al. 2004 \cite{colo2004} \\
                    &                                                              & binding energies and charge radii:                                                          &                                                      \\
                   & over 40 parameter sets                         &  $^{\rm 40,48}$Ca, $^{\rm 56}$Ni, $^{\rm 208}$Pb;                                            &    \\
                   &                                                              &  binding energy of $^{\rm 132}$Sn;                                                               &     \\
                   &                                                              &  spin-orbit splitting of the neutron 3p shell in $^{\rm 208}$Pb;                &     \\
                   &                                                              & surface energy in the ETF approximation with SkM*.                              &             \\
                  & RMF (HB) with DD                                  & Properties of nuclear matter;                                                    & Lalazissis et al. 2005 \cite{lalazissis2005} \\
                   & meson-nucleon coupling:                     & nuclear binding energies; charge radii;                                                  &                                     \\
245 (a)        & DD-ME1(a)                                            & differences between  neutron and proton                                               &                                 \\
251 (b)        & DD-ME2  (b)                                          & density distributions for 18 nuclei.                                                       &     \\                                                              
220 - 260   & review                                                    & E$_{\rm GMR}$.                                                                                         & Shlomo et al. \cite{shlomo2006} \\
241             & RMF (OME) + PC                                    & OME potentials: radial dependence  of the                                             &  Hirose et al. 2007 \cite{ring2007} \\ 
                   &                                                               & non-relativistic G-matrix potentials;                                                    &                                 \\
                   &                                                               & PC: EOS of symmetric matter as calculated                                           &                                 \\
                   &                                                               & with  the Gogny force GT2.                                                                    &                                  \\
                   & SHFB+QRPA+DD pairing:                        & Volume, surface and mixed pairing;                                                     & Colo et al. 2008 \cite{colo2008} \\
230 - 240(a) &  SLy5  (a)                                                &  E$_{\rm GMR}$: $^{\rm 208}$Pb (a);                                                           &                                  \\ 
~220(b)        & SkM*   (b)                                                &  E$_{\rm GMR}$: $^{\rm 112-120}$Sn (b).                                                    &    \\                                                             
230 - 236    & RMF (BSP, IUFSU,IUFSU*)                        &  Binding energies and charge radii for nuclei                                         & Agrawal et al. 2012 \cite{agrawal2012}  \\
                    &                                                              & along several isotopic and isotonic chains;                                           &                                   \\
                    &                                                              & E$_{\rm GMR}$: $^{\rm 90}$Zr, $^{\rm 208}$Pb;                                              &                                   \\
                    &                                                              & properties of  dilute neutron matter;                                                      &                                   \\
                    &                                                              & bounds on the equations of state of the                                                &                                   \\
                    &                                                              & symmetric and asymmetric nuclear matter                                            &                                   \\
                    &                                                              & at supra-nuclear densities.                                                                    &                          \\
210 - 270    & SHF, RMF:                                              & E$_{\rm GMR}$.                                                                                          & Sagawa 2012 \cite{sagawa2012}      \\
                    & variety of interactions                           &                                                                                                                &                                                             \\
                    & SHFB+QRPA:                                           &                                                                                                               & Cao et al. 2012 \cite{cao2012}\\ 
217 (a)         &  SkM* (a)                                                &  E$_{\rm GMR}$: Sn isotopes (a);                                                                 &                                     \\
230 (b)         &  SLy5(b)                                                 &  E$_{\rm GMR}$: Cd, Pb isotopes (b).                                                         &                                    \\                
\hline 
\end{longtable*}
\pagebreak
\endgroup

They used theoretical values of $K_{\rm 0}$ calculated with B1 \cite{brink1967}, D1 \cite{gogny1975}, Ska \cite{koehler1976} and SIII and SIV \cite{beiner1975} effective forces in a Hartree-Fock + RPA model. This was welcomed as a step in the right direction, bringing a mean-field result in line with the `realistic' predictions. We will return to that analysis later in this paper (see Sec.~\ref{micro}) and show that modern calculation and current data move the limits on $K_{\rm 0}$ towards higher values. 

In later years theoretical calculations of $K_{\rm 0}$ developed in two basic directions. These were, first, microscopic calculations based on self-consistent methods with density dependent effective nucleon interactions, both non-relativistic and relativistic, and second, macroscopic models in which the incompressibility of a finite nucleus $K_{\rm A}$ is parameterized in the form of a leptodermous expansion in powers of A$^{\rm -1/3}$. The fundamental difference between the two approaches is that microscopic models yield variables describing vibrating nuclei, such as $K_{\rm 0}$, dependent on the parameters of the effective nucleon interaction. Description of the nuclear surface is not well developed in these models and volume and surface effects cannot be clearly separated. Macroscopic expansion contains individual contributions from the volume, surface, curvature, isospin and Coulomb terms which, in principle, can be obtained directly from a fit to values of $K_{\rm A}$, extracted  from experimental GMR energies. $K_{\rm 0}$ is then set equal to the leading term in the expansion, the volume term $K_{\rm vol}$.

The usual criticism of macroscopic models is that they do not describe vibrating nuclei adequately because they do not  include effects such as anharmonic vibrations, and that the values of the coefficients of the leptodermous expansion are dependent on the accuracy and methods of extraction of GMR energies, and thus $K_{\rm A}$, from raw experimental data \cite{blaizot1980}. The main objection is that the coefficients of the leptodermous expansion are correlated \cite{shlomo1993} and that all the terms in the expansion cannot be determined uniquely. More generally,
Satpathy et al. \cite{satpathy1999} pointed out that the semi-empirical mass formula, the basis for expansion of the incompressibility of a finite nucleus, has its problems and the form of leptodermous expansion of $K_{\rm A}$ is not uniquely determined. 

Since late 1970's, two ways of modeling nuclear matter density under compression have been singled out and extensively studied, the so-called scaling and constrained approximations \cite{blaizot1980, jennings1980, treiner1981}. The difference between the two concepts has a profound consequence on the behavior of the leptodermous expansion. In the constrained approximation the leptodermous expansion is converging slowly and higher order terms in A$^{\rm 1/3}$, in particular the curvature term depending on A$^{\rm 2/3}$ cannot be neglected. Unique determination of the coefficients in the expansion is indeed difficult and the extracted values may contain unwanted contributions from unresolved correlations. However, as was shown by Treiner et al. \cite{treiner1981}, in the scaling approximation the transition density clearly separates the volume from the surface region in a vibrating nucleus. The leptodermous expansion converges fast, higher order terms are negligible and the coefficients reflect properties of real nuclei. Thus the scaling model has been recommended for use in analysis of experimental GMR data as is done in the first part of this paper. 

Extensive discussion of the pros and cons of the macroscopic and microscopic methods has been given in several papers (see e.g. \cite{blaizot1980, blaizot1981, treiner1981, blaizot1989, blaizot1995}). Although the general tendency has been to prefer the microscopic approach, a fundamental problem emerged also there. The non-relativistic models, mainly using the Skyrme interaction, systematically predicted lower values of $K_{\rm 0}$, around 210 - 250 MeV, (see e.g. \cite{blaizot1995, farine1997, chung1999, colo2004}) but the relativistic models yielded higher values (see e.g. \cite{sharma1988, sharma1989, sharma1989a, stoitsov1994, vretenar1997, piekarewicz2002, sharma2009}). Re-analysis of experimental data available in 1989 using the leptodermous expansion was presented by Sharma et al. \cite{sharma1989, sharma1989a} showed that the best fit was achieved for $K_{\rm 0} \sim$ (300 $\pm$ 25) MeV, thus supporting predictions of relativistic models.

Currently a general consensus has developed to adopt a lower value of $K_{\rm 0}$, $K_{\rm 0}$ = (240 $\pm$ 20) MeV (e.g. \cite{shlomo2006}) which has been used as an initial condition/requirement in most models. Skyrme effective interactions were constructed to reproduce this 'canonical' value and attempts were made to reconcile \cite{agrawal2004} and modify effective Lagrangians \cite{agrawal2012} in relativistic models to comply with this adopted value.

These efforts however indicate the main weakness of the microscopic approaches. The effective interactions have a flexible form and too many variable parameters so that modifications can be introduced which yield a desired result but do not advance understanding of the underlying physics. The most recent illustration of the problem can be found in \cite{cao2012}, where even the state-of-the-art HFB+QRPA calculation did not succeed to reproduce GMR energies in Sn, Cd and Pb nuclei using the same Skyrme parameterization. The dependence of the calculated value of $K_{\rm 0}$ on the choice of the microscopic model is obvious from examination of Table~\ref{tab:surv}.

In parallel with $K_{\rm 0}$, investigation of the isospin incompressibility $K_\tau$, which quantifies the contribution from the neutron-proton difference to the incompressibility of a finite nucleus $K_{\rm A}$, has been performed.  We introduce here the term "isospin" incompressibility to avoid confusion with the "symmetry" incompressibility - the name sometimes used for the curvature of the symmetry energy at saturation density $K_{\rm sym}$.  This coefficient can be obtained in either the microscopic or the empirical approach \cite{blaizot1981, treiner1981, nayak1990, patra2002, sagawa2007, sharma2009}. Its recent extraction from empirical analysis of GMR data on Sn isotopes \cite{li2007, li2010} attracted a lot of attention as the value of $K_\tau$ was larger than predicted by most of the microscopic models. Determination of $K_\tau$ from experimental data on GMR is complicated by the fact that, as with the volume and surface contributions to $K_{\rm A}$, it also includes volume and surface terms and the latter cannot be easily evaluated in microscopic models \cite{blaizot1981, treiner1981, nayak1990, patra2002}.

In this paper we survey existing data on GMR energies in nuclei with A $\ge$ 56 and use them to set limits on $K_{\rm 0}$  and the isospin incompressibility coefficient $K_\tau$, using the macroscopic approach in the scaling approximation and employing a new method of analysis. In Sec.~\ref{basics} we present the basic expressions and the data selection for the analysis followed by Sec.~\ref{anal} containing the the main results. A schematic theoretical model of the ratio of the volume and surface contributions to $K_{\rm A}$ is presented in Sec.~\ref{toy}. Microscopic models are commented on in Sec.~\ref{micro}. Discussion of results and conclusions form Sec.~\ref{concl}.

\section{\label{basics} The basics}
The incompressibility $K_{\rm A}$ of a finite nucleus with mass A is related to the energy of the GMR resonance $E_{\rm GMR}$ of the nucleus \cite{blaizot1980}
\begin{equation}
K_{\rm A}=(M/\hbar^{\rm 2})<r^2>E^{\rm 2}_{\rm GMR},
\label{kexp}
\end{equation}
where $M$ is the nucleon mass and $r$ is rms \textit {matter} radius of the nucleus. $K_{\rm A}$ can be expanded in terms of $A^{\rm -1/3}$ and the asymmetry parameter $\beta=(N-Z)/A$ as \cite{blaizot1980}
\begin{eqnarray}
K_{\rm A} = K_{\rm vol}+K_{\rm surf}A^{-1/3}+K_{\rm curv}A^{-2/3}  \nonumber  \\
+K_{\rm coul}Z^2A^{\rm -4/3}+K_{\rm \tau} \beta^2.  
\label{ka}
\end{eqnarray}
 Higher order terms in $\beta$ can be safely neglected as their contribution to $K_{\rm A}$ is less then 1\% \cite{chen2009}. $K_{\rm vol}$, $K_{\rm surf}$, $K_{\rm curv}$, $K_{\rm \tau}$ and $K_{\rm coul}$ represent the volume, surface, curvature, isospin and Coulomb contributions to the incompressibility  $K_{\rm A}$. The coefficient $K_{\rm \tau}$ consists of two components, 
\begin{equation}
K_{\rm \tau} = K_{\rm \tau,v}+K_{\rm \tau,s}A^{\rm -1/3},
\label{ktau}
\end{equation}
where $K_{\rm \tau,v}$ ( $K_{\rm \tau,s}$ ) determine the volume (surface) isospin incompressibility. 

Assuming the expansion (\ref{ka}) theoretically justified, different coefficients can be extracted from comparison with experimental data. Care must be taken concerning the interpretation of $K_A$. The energy $E_{\rm GMR}$ is understood as a mean energy calculated from moments $m_{\rm k}$ of a strength function \cite{treiner1981}
\begin{equation}
m_{\rm k}=\int E^{\rm k}S(E)dE,
\label{mk}
\end{equation}
where the strength function $S(E)$ = $\sum_{n}|<n|\mathcal{\hat O}|0>|^2 \delta$(E - E$_{\rm n}$).
$|0>$ is the ground state of the nucleus and $E_{\rm n}$ is the energy of a state $n$. The monopole excitation operator $\mathcal{\hat O}$ is taken as $\sum_{i=1}^{A} r_{\rm i}^{\rm 2}$. Various mean energies $\tilde{E_{\rm k}}$ are calculated from moment ratios
\begin{equation}
\tilde{E_{\rm k}}=\sqrt{\frac{m_{\rm k}}{m_{\rm {k-2}}}}.
\label{moments}
\end{equation}
If the strength function is distributed in a narrow energy region, the mean energies $\tilde{E_{\rm k}}$ are close together and can be interpreted as  $E_{\rm GMR}$. In this case $K_{\rm A}$ is determined in principle unambiguously using (\ref{ka}) and $K_{\rm vol}$ in (\ref{ka}) is equal to the incompressibility of infinite symmetric nuclear matter $K_{\rm 0}$ at saturation density $\rho_{\rm 0}$ 
\begin{equation}       
K_{\rm 0} = 9 \rho_{\rm 0} \frac{d^2(\mathcal{E}/A)}{d\rho^{\rm 2}}|_{\rm \rho=\rho_{\rm 0}},
\end{equation}
where $\mathcal{E}/A$ the energy per particle. In a more realistic case when the strength function is somewhat spread out, (\ref{kexp}) must be written as
\begin{equation}
K_{\rm A}(k)=(M/\hbar^{\rm 2})<r^2>E^{\rm 2}_{\rm GMR}(k),
\label{kexpk}
\end{equation}
and the $K_{\rm A}$ can be determined only within a certain region of $k$. 

\subsection{Determination of $E_{\rm GMR}$}
Blaizot \cite{blaizot1980} and Treiner et al. \cite{treiner1981} studied two forms of the expansion (\ref{ka}): the scaling model, based on the cubic-energy-weighted sum rule (k=3), and the constrained model based on the linear-inverted-energy-weighted sum rule (k=-1). They showed that only in the scaling model does the series (\ref{ka}) converge rapidly and $K_{\rm vol}=K_{\rm 0}$. It follows that the contribution of the curvature term (which depends on $A^{\rm -2/3}$) can be neglected in the scaling model which simplifies the application of the model in  analysis of experimental data. We will adopt the scaling model throughout our analysis using $E_{\rm GMR}$ =  $\tilde{E_{\rm 3}}$ and interpret $K_{\rm vol} = K_{\rm 0}$ although we are aware of possible ambiguities in this approach \cite{blaizot1995}. 

There is another method of determination of $E_{\rm GMR}$, compatible with the scaling model. The GMR resonance in the strength function can be fitted, assuming Gaussian distribution, to obtain the peak energy $E_{\rm peak}$ and the full width at half maximum (FWHM) $\Gamma$. The GMR energy is then calculated as
\begin{equation}
\tilde{E_3}= (E_{\rm peak})^2 + 3\left(\frac{\Gamma}{2.35}\right)^2.
\label{ga}
\end{equation}
It can be shown that (\ref{ga}) is exact only for Gaussian distribution of the strength function, otherwise the relation between the energies obtained from (\ref{ga}) and (\ref{moments}) for k = 3 must be treated as an approximation. $\tilde{E_3}$ values obtained from (\ref{ga}) have larger uncertainties than values extracted from  moments, as both $E_{\rm peak}$ and $\Gamma$ have errors. However the expression (\ref{ga}) was regularly used in earlier, less accurate experiments, in which moment analysis was not possible, and results based on it are still often quoted for comparison with moment results (see e.g. \cite{lui2006}). 

Systematics of GMR energies for $A>56$, obtained from experiment using different methods of analysis's are shown in Figs.~\ref{fig:m3} ~-~\ref{fig:m0}: 
\begin{itemize}
\item{$\tilde{E_3}=\sqrt{\frac{m_{\rm 3}}{m_{\rm 1}}}$, energy in the scaling approximation (Fig.~\ref{fig:m3})}
\item{$\tilde{E_1}=\sqrt{\frac{m_{\rm 1}}{m_{\rm -1}}}$, energy in the constrained approximation (Fig.~\ref{fig:m1})}
\item{$\tilde{E_0}=\frac{m_{\rm 1}}{m_{\rm 0}}$, mean centroid energy (Fig.~\ref{fig:m0})}.
\end{itemize}
In addition, the values of $E_{\rm peak}$ and $\Gamma$ in (\ref{ga}), as extracted from different analyses, using Gaussian/Lorentzian/Breit-Wigner fit to the GMR strength distributions, are given in Fig.~\ref{fig:gmr_raw}. We note that Lorentzian and Breit-Wigner fit functions are quoted here in line with the original papers. Both are in the same form of a general Cauchy distribution
\begin{equation}
f(E,E_{\rm peak},\Gamma)=\frac{1}{\pi}\frac{\Gamma/2}{(E-E_{\rm peak})^{\rm 2}+(\Gamma/2)^{\rm 2}}.
\label{ca}
\end{equation}
This distribution differs from the Gaussian in a slower decrease in both tails away from the center \cite{kotz2006}. We note that moment $m_{\rm 3}$ cannot be calculated for a strength function in a Cauchy form (\ref{ca}) because the integral in (\ref{mk}) diverges. Thus $\tilde{E_3}$ cannot be evaluated accurately in this case using (\ref{moments}).
\begin{table}
\caption{\label{tab:data} List of all data groups selected for analysis in this work. The groups RCNP-M, TAMU0-M and GF-M, discussed in the text, contain the same data as RCNP-E, TAMU0-E and GF-E, respectively, but the entries for $^{56}$Fe and $^{58,60}$Ni are not included. $\tilde{E_{\rm 0}}$(average) labels a weighted average of GMR energies obtained from data given in \cite{youngblood2004, uchida2004}. (GF) indicates  that the $\tilde{E_{\rm 3}}$ energy was evaluated using expression (\ref{ga}). The number in the column `Data' indicates the total number of entries in each group. For more details see text.} 
\begin{ruledtabular}
\begin{tabular}{ccclr}
 Group &   Method               &        Data        & Isotope      &    Reference \\ \hline
RCNP   &  $\tilde{E_{\rm 3}}$  & 11     &    $^{112-124}$Sn                 &   \cite{li2007, li2010}  \\
       &  $\tilde{E_{\rm 3}}$  &        &    $^{106,110-116}$Cd             &   \cite{garg2011}    \\  \hline
RCNP-E &  $\tilde{E_{\rm 3}}$  & 16     &    $^{56}$Fe                      &   \cite{lui2006}\\                                                
       &  $\tilde{E_{\rm 3}}$  &        &    $^{58,60}$Ni                   &   \cite{lui2006}\\
       &  $\tilde{E_{\rm 3}}$  &        &    $^{112-124}$Sn                 &   \cite{li2007, li2010}  \\ 
       &  $\tilde{E_{\rm 3}}$  &        &    $^{106,110-116}$Cd             &   \cite{garg2011}    \\ 
       &  $\tilde{E_{\rm 0}}$(average)&  &    $^{208}$Pb                     &   \cite{youngblood2004, uchida2003} \\ \hline
TAMU3  &  $\tilde{E_{\rm 3}}$  & 4      &    $^{112,124}$Sn                 &   \cite{lui2004a}    \\
       &  $\tilde{E_{\rm 3}}$  &        &    $^{110,116}$Cd                 &   \cite{lui2004}    \\      \hline       
TAMU0  &  $\tilde{E_{\rm 0}}$  & 5      &    $^{112,124}$Sn                 &   \cite{lui2004a}   \\
       &  $\tilde{E_{\rm 0}}$  &        &    $^{116}$Sn                     &   \cite{youngblood1999} \\
       &  $\tilde{E_{\rm 0}}$  &        &    $^{110,116}$Cd                 &   \cite{lui2004}    \\     \hline
TAMU0-E & $\tilde{E_{\rm 0}}$  & 20     &    $^{56}$Fe                      &   \cite{lui2006}   \\
       &  $\tilde{E_{\rm 0}}$  &        &    $^{58,60}$Ni                   &   \cite{lui2006}  \\
       &  $\tilde{E_{\rm 0}}$  &        &    $^{106,110-116}$Cd             &   \cite{garg2011}  \\
       &  $\tilde{E_{\rm 0}}$  &        &    $^{110,116}$Cd                 &   \cite{lui2004}  \\
       &  $\tilde{E_{\rm 0}}$  &        &    $^{112-124}$Sn                 &   \cite{li2007, li2010}  \\ 
       &  $\tilde{E_{\rm 0}}$  &        &    $^{112,124}$Sn                 &   \cite{lui2004a}   \\
       &  $\tilde{E_{\rm 0}}$  &        &    $^{116}$Sn                     &   \cite{youngblood1999} \\
       &  $\tilde{E_{\rm 0}}$  &        &    $^{116}$Sn                     &   \cite{youngblood2004}  \\ \\
       &  $\tilde{E_{\rm 0}}$  &        &    $^{144}$Sm                     &   \cite{youngblood2004}\\
       &  $\tilde{E_{\rm 0}}$(average)&        &    $^{208}$Pb                     &   \cite{youngblood2004, uchida2003} \\ \hline
 GF    &    $\tilde{E_{\rm 3}}$(GF)  & 9      &    $^{110,116}$Cd                 &   \cite{lui2004}   \\
       &    $\tilde{E_{\rm 3}}$(GF)  &        &    $^{112,124}$Sn                 &   \cite{lui2004a}   \\ 
       &    $\tilde{E_{\rm 3}}$(GF)  &        &    $^{112-116,120,124}$Sn         &   \cite{sharma1988}  \\  \hline
GF-E   &    $\tilde{E_{\rm 3}}$(GF)  & 15     &    $^{56}$Fe                      &   \cite{lui2006}   \\        
       &    $\tilde{E_{\rm 3}}$(GF)  &        &    $^{58,60}$Ni                   &   \cite{lui2006}  \\
       &    $\tilde{E_{\rm 3}}$(GF)  &        &    $^{110,116}$Cd                 &   \cite{lui2004}   \\
       &    $\tilde{E_{\rm 3}}$(GF)  &        &    $^{112,124}$Sn                 &   \cite{lui2004a}   \\
       &    $\tilde{E_{\rm 3}}$(GF)  &        &    $^{112-116,120,124}$Sn         &   \cite{sharma1988}  \\
       &    $\tilde{E_{\rm 3}}$(GF)  &        &    $^{144}$Sm                     &   \cite{sharma1988} \\
       &    $\tilde{E_{\rm 3}}$(GF)  &        &    $^{148}$Sm                     &   \cite{sharma1988} \\
       &    $\tilde{E_{\rm 0}}$(average) &     &    $^{208}$Pb                     &   \cite{youngblood2004, uchida2003} \\
\end{tabular}
\end{ruledtabular} 
\end{table}

Examination of Figs.~\ref{fig:m3}~-~\ref{fig:gmr_raw} yields several interesting features. In general, the $\tilde{E_{\rm 3}}$, $\tilde{E_{\rm 1}}$ and $\tilde{E_{\rm 0}}$ show a non-negligible systematic difference for the same isotope. A possible explanation of these differences may be that higher moments are more sensitive to a spread of the GMR strength to higher excited states \cite{youngblood1996}. It follows that in evaluation of the incompressibility $K_{\rm A}$ of a finite nucleus, experimental GMR energies must be used consistently with the model adopted, e.g. $\tilde{E_{\rm 3}}$ in the scaling and $\tilde{E_{\rm 1}}$ in the constrained model. 

Furthermore, in particular for Sn and Cd isotopes, there is a systematic difference between results obtained by the Texas A\&M (TAMU) group (lower energies) and the Notre Dame/Japan/Groningen (RCNP)  group (higher energies). $\tilde{E_{\rm 3}}$ energies obtained from a Gaussian fit to the GMR strength distribution using (\ref{ga}) are between the two sets of results. However, the GMR peak energies obtained by the TAMU (RCNP) group from Gaussian (Lorentzian) fit to the strength distribution are remarkably close to each other and agree also with older data obtained by Sharma et al. \cite{sharma1988}. Differences occur in the strength distribution widths, shown in the right panel of Fig.~\ref{fig:gmr_raw}, pointing to a different philosophy in analysis of experimental data by different groups. Extraction of GMR energies with smaller errors from moments show these differences in, for example, background subtraction, more obviously.

\section{\label{anal}Analysis of experimental data}

In this section we review the method of analysis used by the RCNP group \cite{li2007, li2010, garg2011} to obtain $K_\tau$ from their $^{112 - 124}$Sn and $^{106,110-116}$Cd data. Next we describe the novel method of analysis of GMR data, introduced in this work, and apply it to  selected data sets, as detailed in Table~\ref{tab:data}, to extract both $K_{\rm 0}$ and $K_\tau$. In the two last sections we attempt to estimate limits on $K_{\rm \tau,v}$ and $K_{\rm \tau,s}$ and comment on the curvature term in the expansion (\ref{ka}).

\subsection{Method}
The previous analyses of the GMR data on $^{112 - 124}$Sn and $^{106,110-116}$Cd isotopes by Li et al. and Garg et al. \cite{li2007,li2010,garg2011} were based on a simplified formula
\begin{equation}
K_{\rm A} - K_{\rm coul}Z^2A^{-4/3} = K_{\rm vol}(1 + cA^{-1/3})+K_\tau\beta^2.
\label{kali}
\end{equation}
with $c$, the ratio of $K_{\rm surf}$ and $K_{\rm vol}$, set equal to -1 and $K_{\rm coul}$ taken from theory  to be -(5.2 $ \pm$ 0.7)  MeV \cite{sagawa2007}. This equation was approximated by a quadratic function of $x$, $y = a + bx^2$ with $b = K_\tau$ and $ a = K_{\rm vol}(1 + cA^{-1/3})$. The (weak) mass dependence of $a$ was neglected based on the argument that  $A^{-1/3}$ is changing only by $\sim$3.3\% over the range of Sn isotopes and just under 3\% for Cd nuclei \cite{garg2011}. Higher order terms, namely the curvature term and the surface contribution to $K_\tau$ were not included in the analysis. The experimental values of $K_{\rm A}$ were evaluated from (\ref{kexp}) using rms \textit{charge} radii taken from \cite{fricke1995}. Sensitivity of the data to the value of $K_{\rm 0}$ was not examined. $K_{\rm 0}$ was fixed to (240 $\pm $10) MeV and only the value of the isospin incompressibility $K_{\rm \tau}$ = -(550 $\pm$ 100) MeV was extracted from the data. Although the scaling approximation was used, experimental GMR energies were taken from $m_{\rm 1}/m_{\rm -1}$ ratios, which, as pointed out above, is internally inconsistent.

The approach of Li et al. was criticized by Pearson et al. \cite{pearson2010} who questioned the claims that the mass dependence of the first two terms in the leptodermous expansion for K$_{\rm A}$ (the volume and surface terms) is not significant and that the seven pieces of experimental data in $^{\rm 112-124}$Sn are enough to yield a unique value of K$_{\rm \tau}$. Pearson et al. did not make any distinction between the scaling and constrained approximations. In this case it is generally correct that if the higher order terms in the leptodermous expansion make a significant contribution to K$_{\rm A}$, and are not included in the fit, then the extracted value of $K_{\rm \tau}$ is only an 'effective' value, including implicitly the effects of the higher order terms. Although there is a possibility of a small contribution of higher order terms even in the scaling approximation \cite{treiner1981}, the coefficients of the leptodermous expansion are much less affected and are significantly closer to reality.

In the present work, the GMR data analysis, presented by Li et al., has been modified in several important ways.
\begin{table*} 
\caption{\label{tab:c1} Values of $K_{\rm 0}$ and $K_\tau$ obtained from the best fit to Sn, Cd and combined Sn+Cd data in RCNP, TAMU3, TAMU0 and GF data sets. Two entries for $K_\tau$ are given for each data set in rows (i) and (ii), with (i)  and without (ii)  a correlation with $K_{\rm 0}$. The value of $\sigma$ is very similar for these two entries and is given only for the first line. The range of  $K_{\rm 0}$, corresponding to 3$\sigma$ on both sides from the minimum (l=left, r=right) is shown in the last two columns of line one. In the third line are results from the MINUIT fit to $K_A$ calculated with charge radii. For more explanation, see text.}
 \begin{ruledtabular}
\begin{tabular}{lrccccc}
 Element &  & $K_{\rm 0}$ &    $K_\tau$   &  $\sigma$   & K$_{\rm 0}$(3$\sigma)_{\rm l}$ &  K$_{\rm 0}$(3$\sigma)_{\rm r}$  \\ \hline
         &  & [MeV]       &   [MeV]       &             &   [MeV]                         &             [MeV]          \\ \hline
Sn(RCNP) &(i)   &  209(6)     &  -595(177)    &    0.64     &                  202            &           215               \\
         &(ii)  &  209        &  -591(58)     &             &                                 &                             \\ 
         &(iii) &  216(6)     &  -537(177)    &    0.40     &                                 &                              \\  
Cd(RCNP) &(i)   &  211(11)    &  -463(405)    &    0.04     &                  207            &           215                 \\
         &(ii)  &  212        &  -460(120)    &             &                                 &                              \\
         &(iii) &  220(10)    &  -403(382)    &    0.04     &                                 &                               \\   
Sn+Cd(RCNP)&(i) &  211(5)     &  -633(157)    &    3.07     &                  199            &           222                 \\ 
         &(ii)  &  211        &  -598(52)     &             &                                 &                               \\
         &(iii) &  220(5)     &  -595(154)    &    3.82     &                                 &                                \\ 
Sn+Cd(TAMU3)&(i)&  193(7)     &  -652(193)    &    2.12     &                  179            &           207                  \\
         & (ii) &  193        &  -653(73)     &             &                                 &                                \\
         & (iii)&  200(7)     &  -594(194)    &    1.88     &                                 &                                \\
Sn+Cd(TAMU0)&(i)&  187(6)     &  -695(179)    &    8.03     &                  164            &           210                  \\
         & (ii) &  187        &  -690(72)     &             &                                 &                                \\
         & (iii)&  194 (6)    &  -641(177)    &    8.33     &                                 &                                \\
Sn+Cd(GF)&(i)   &  195(6)     &  -430(208)    &    0.76     &                  178            &           187                  \\ 
         &(ii)  &  195        &  -431(81)     &             &                                 &                                \\
         & (iii)&  202(6)     &  -355(213)    &    0.47     &                                 &                                \\ 
\end{tabular}
\end{ruledtabular}
\end{table*}

First, Eq.~(\ref{kali}) has been rewritten as
\begin{equation}
\frac{K_{\rm A}}{1 + cA^{-1/3}}- \frac{K_{\rm coul}Z^2A^{-4/3}}{1 + cA^{-1/3}} = K_{\rm vol}+K_\tau \frac{\beta^2}{1 + cA^{-1/3}}.
\label{kanjs}
\end{equation}
\begin{figure}
\centerline{\epsfig{file=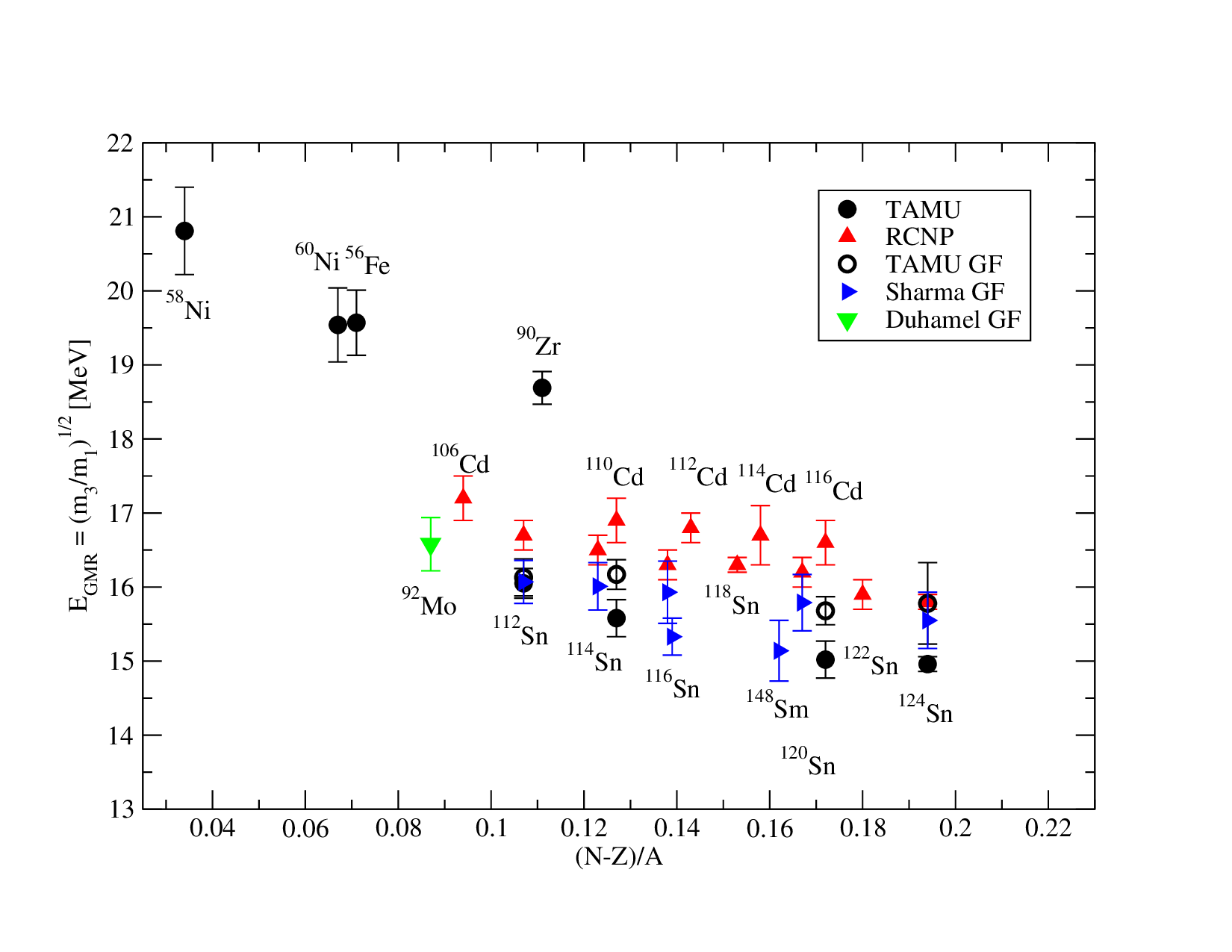,height=7cm,width=9cm}}
\caption{\label{fig:m3} (Color on-line) E$_{\rm GRM}$=$\tilde{E_3}=\sqrt{\frac{m_3}{m_{1}}}$ as a function of (N-Z)/A as reported in $^{58,60}$Ni, $^{56}$Fe \cite{lui2006} (TAMU), $^{90}$Zr \cite{youngblood2004} (TAMU), $^{106,110-116}$Cd \cite{garg2011} (RCNP), $^{110,116}$Cd \cite{lui2004} (TAMU), $^{112-124}$Sn \cite{li2007, li2010} (RCNP) and $^{112,124}$Sn \cite{lui2004a} (TAMU). We also display $E_{\rm GMR}$(GF) obtained from (\ref{ga}) for $^{92}$Mo \cite{duhamel1988} (Duhamel), $^{110,116}$Cd \cite{lui2004} (TAMU), $^{112,124}$Sn \cite{lui2004a} (TAMU), $^{112-116,120,124}$Sn, $^{144}$Sm and $^{148}$Sm \cite{sharma1988} (Sharma). For more details see text.}
\end{figure}

The equation can be symbolically written as a function $y=p + q x$ with $p = K_{\rm vol}$ and $q =  K_\tau$. The transformation has the advantage that both  $p$ and $q$  are independent of $A$ and $y$ is a linear function of $x$ with a slope determining $K_\tau$ and intercept equal to $K_{vol}$. As the scaling model is adopted in this work, we will assume that $K_{\rm vol}$ can be  taken equal to $K_{\rm 0}$ from now on and use $K_{\rm vol}$ and $K_{\rm 0}$ interchangeably according to the context. 

\begin{figure}
\centerline{\epsfig{file=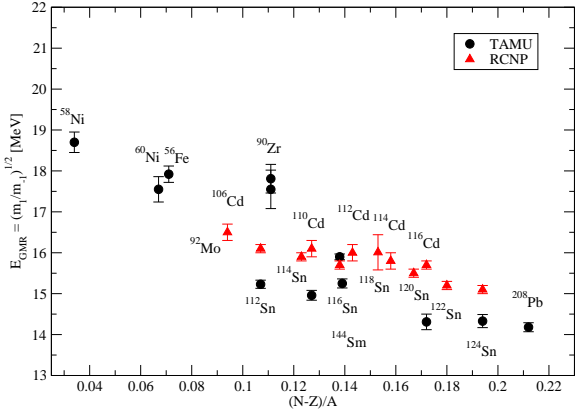,height=7cm,width=9cm}}
\caption{\label{fig:m1} (Color on-line) The same as Fig.~\ref{fig:m3} but for E$_{\rm GRM}$=$\tilde{E_1}=\sqrt{\frac{m_1}{m_{-1}}}$ as reported in $^{58,60}$Ni, $^{56}$Fe \cite{lui2006} (TAMU),  $^{90}$Zr \cite{youngblood2004, youngblood1999} (TAMU), $^{106,110-116}$Cd \cite{garg2011} (RCNP), $^{110,116}$Cd \cite{lui2004} (TAMU), $^{112-124}$Sn \cite{li2007, li2010} (RCNP), $^{112,124}$Sn \cite{lui2004a} (TAMU), $^{116}$Sn, $^{144}$Sm and $^{208}$Pb \cite{youngblood1999} (TAMU). Note the y-scale is the same as in Figs.~\ref{fig:m3} and ~\ref{fig:m0}.} 
\end{figure}
\begin{figure}
\centerline{\epsfig{file=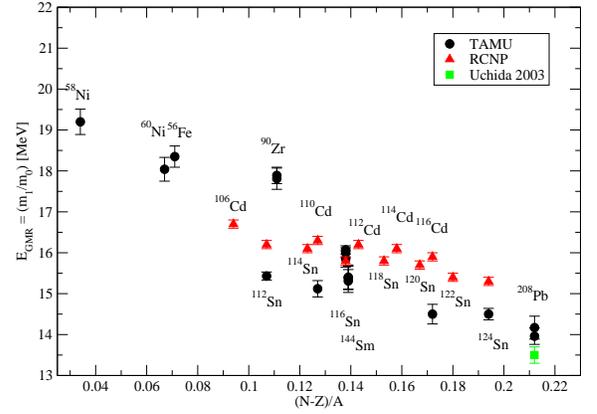,height=7cm,width=9cm}}
\caption{\label{fig:m0} (Color on-line) The same as Fig.~\ref{fig:m3} but for E$_{\rm GRM}$=$\tilde{E_0}=\frac{m_1}{m_{0}}$ as reported in $^{58,60}$Ni, $^{56}$Fe \cite{lui2006} (TAMU), $^{90}$Zr \cite{youngblood2004a, youngblood1999} (TAMU), $^{106,110-116}$Cd \cite{garg2011} (RCNP), $^{110,116}$Cd \cite{lui2004} (TAMU), $^{112-124}$Sn \cite{li2007, li2010} (RCNP), $^{112,124}$Sn \cite{lui2004a} (TAMU), $^{116}$Sn, $^{144}$Sm \cite{youngblood2004, youngblood1999} (TAMU) and $^{208}$Pb \cite{youngblood2004, youngblood1999} (TAMU) and \cite{uchida2003} (Uchida 2003). Note the y-scale is the same as in Figs.~\ref{fig:m3} and \ref{fig:m1}.}  
\end{figure}

The second significant difference is that we use $\tilde{E_3}$ GMR energies in calculation of $K_{\rm A}$ (\ref{kexp}), consistent with the scaling model, instead of $\tilde{E_1}$, used in \cite{li2007, li2010, garg2011}). As shown in Figs.~\ref{fig:m3} and \ref{fig:m1}, this makes a non-negligible difference in GMR energies and thus in $K_{\rm A}$.
\begin{figure}
\centerline{\epsfig{file=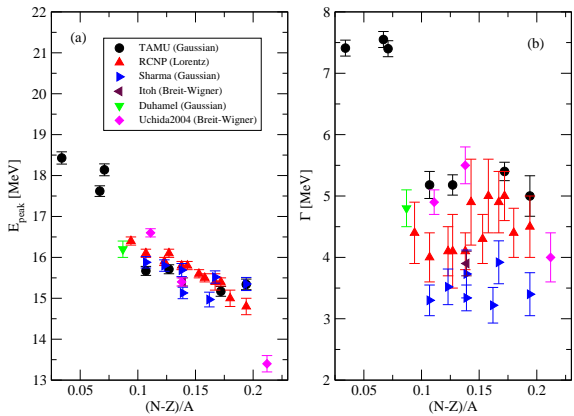,height=7cm,width=9cm}}
\caption{\label{fig:gmr_raw}  (Color on-line) (a) Peak energy $E_{\rm peak}$ and (b) corresponding width $\Gamma$ obtained from various fits to the experimental GMR strength function. Data are taken from $^{58,60}$Ni, $^{56}$Fe \cite{lui2006} (TAMU Gaussian), $^{92}$Mo \cite{duhamel1988} (Duhamel Gaussian), $^{90}$Zr \cite{uchida2004} (Uchida 2004 Breit-Wigner), $^{106,110-116}$Cd \cite{garg2011} (RCNP Lorentzian) , $^{110,116}$Cd \cite{lui2004} (TAMU Gaussian), $^{112-124}$Sn \cite{li2007,li2010} (RCNP Lorentzian), $^{112,124}$Sn \cite{lui2004a} (TAMU Gaussian), $^{112-116,120,124}$Sn \cite{sharma1988} (Sharma Gaussian), $^{116}$Sn \cite{uchida2004} (Uchida 2004 Breit-Wigner), $^{144}$Sm \cite{sharma1988} (Sharma Gaussian), $^{144}$Sm \cite{itoh2002} (Itoh Breit-Wigner), $^{148}$Sm \cite{sharma1988} (Sharma Gaussian), $^{208}$Pb (Breit-Wigner fit) \cite{uchida2004} (Uchida 2004 Breit-Wigner). The y-scale in the left panel is the same as in Figs.~\ref{fig:m3}, \ref{fig:m1} and \ref{fig:m0}.}
\end{figure}

In calculation of $K_{\rm A}$ the \textit{matter} radius is required by theory. However, Li et al. and Garg both used charge radii. As a third improvement (in principle) we examined two ways of estimation of matter radii (methods A and B), as detailed below. 

In Method A the rms radius of the matter distribution $<r^2_m>^{1/2}$ was evaluated using the expression in terms of the proton and neutron distribution radii 
\begin{equation}
<r^2_{\rm m}>^{1/2}=(Z <r^2_{\rm p}>^{1/2} - N <r^2_{\rm n}>^{1/2})/(Z+N),
\label{rm}
\end{equation}
where the rms neutron distribution radius $<r^2_{\rm n}>^{1/2} = <r^2_{\rm p}>^{1/2} + S$ is calculated from the proton distribution radius and the neutron skin $S$. 
The rms charge distribution radius $<r^2_{ch}>$ is obtained from a two-parameter Fermi distribution with half-density radius fitted to the experimental 2p-1s transition energy in muonic atoms and a width 2.30 fm \cite{fricke1995}. It can be converted into a rms proton distribution radius $<r^2_{\rm p}>^{1/2}$ using a simple Gaussian folding recipe \cite{reinhard1996}
\begin{equation}
<r^2_{\rm p}>=<r^2_{\rm ch}>-<r^2_{\rm ch,p}>+\frac{N}{Z}<r^2_{\rm ch,n}>,
\label{rch}
\end{equation}
where the intrinsic charge proton and neutron radii are (0.8768 $\pm $ 0.0069) fm$^2$ \cite{nakamura2010} and -(0.1161 $\pm $ 0.0022) fm$^2$ \cite{eidelman2004}, respectively. 

The neutron skin $S$ = $<r^2_{\rm n}>^{1/2}$ - $<r^2_{\rm p}>^{1/2}$ is determined from an  empirical relation between $S$ and  $\beta$: $S = (0.9 \pm 0.15)\beta -(0.03 \pm 0.02$) fm, obtained by interpolation of data from experiments with anti-protons \cite{centelles2009}. We adopted this empirical relation for isotopes for which the experimental value of the neutron skin is either not known or known with a large error. For $^{90}$Zr, $^{116}$Cd, $^{112,116,120,124}$Sn and $^{208}$Pb we took experimental values of the neutron skin \cite{trzcinska2001}. 

In Method B, radii of neutron matter distributions have been extracted from the angular distribution of 166 MeV alpha particle elastic scattering \cite{brissaud1972} and charge radii from an independent electron scattering experiment. 
Empirical dependence of the matter radii $r_{\rm mb}$ on $A^{1/3}$ has been approximated by
\begin{equation} 
<r^2_{\rm mb}>^{1/2}=(0.86 \pm 0.01)A^{\rm 1/3}+(0.47 \pm 0.05) {\rm fm}.
\label{rmb}
\end{equation}
obtained from a fit over a wide range of spherical nuclei. Treiner et al. \cite{treiner1981} used this relationship in their calculation of incompressibilities of finite nuclei but did not include the errors in the coefficients.   

\begin{figure}
\centerline{\epsfig{file=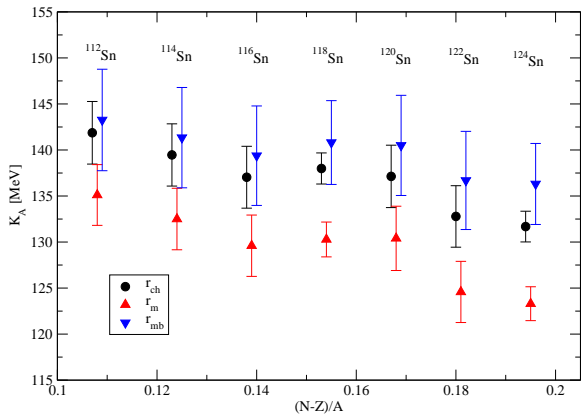,height=7cm,width=9cm}}
\caption{\label{fig:rad} (Color on-line) $K_{\rm A}$ calculated for Sn isotopes using  $\tilde{E_3}=\sqrt{\frac{m_3}{m_{1}}}$ and  the mean charge (circle), matter (triangle up), determined from nuclear skin, and matter (triangle down) radius obtained from elastic alpha-particle scattering. See text for more detail.} 
\end{figure}

The effect of the different way of evaluating $<r^2>$ on the calculated incompressibility of a finite nucleus is illustrated in Fig.~\ref{fig:rad} for Sn isotopes. It can be seen that values of $K_{\rm A}$ differ when charge radii and matter radii are used and the difference increases with $A$. The uncertainties on $K_{\rm A}$ calculated with $<r^2>$ obtained in method B \cite{brissaud1972} reflect all the constraints in the model used is their extraction and are considerably larger that the ones with matter radii from the neutron skin. Considering that formula (\ref{rmb}) arises from a global fit and is not recommended for use within isotopic sequences of a single element \cite{brissaud1972}, we choose $<r^2>$ obtained by method A in this work. The difference between $<r^2>$ from the two methods illustrates the known difficulty in determining matter radii. Therefore both matter and charge radii have been used in this work and consequences for the calculated values of $K_{\rm A}$ taken into account in the discussion. 
A fourth area of difference between this and previous work involves the adopted value of $K_{\rm coul}$. Because of the possible correlation between the volume and the Coulomb contribution to the expansion (\ref{kanjs}),  the value of $K_{\rm coul}$ is usually fixed in the fits. Sagawa et al. \cite{sagawa2007} obtained the value of $K_{\rm coul}$ = -(5.2 $\pm$ 0.7) MeV, in microscopic Skyrme-Hartree-Fock (SHF) and RMF calculations. They examined the correlation between $K_0$ and $K_{\rm coul}$ using 14(7) parameter sets in the SHF(RMF) models and found the variation of $K_{\rm coul}$ rather small, which is reflected in the quoted error. The caveat to this choice is that, although in principle the Coulomb contribution to the incompressibility of a finite nucleus is model independent, the value used here depends on the choice of the effective nuclear interaction through the expression \cite{sagawa2007}
\begin{equation}
K_{\rm coul}=\frac{3}{5}\frac{e^{\rm 2}}{r_{\rm 0}}\left(1-\frac{27\rho_{\rm nm}^{\rm 2}}
{K_{\rm 0}}\frac{d^{\rm 3}h}{d\rho^{\rm 3}}\big|_{\rho=\rho_{\rm nm}}\right),
\label{coulomb}
\end{equation}
where $h$ is the Hamiltonian density of symmetric nuclear matter and $\rho_{\rm nm}$ is the saturation density. The second term in the expression arises from expansion of the incompressibility of finite nuclei in terms of the difference between the equilibrium density $\rho_{\rm 0}$ and the saturation density of infinite nuclear matter. For the system to be stable, this difference must be positive \cite{blaizot1980}. The expansion introduces the dependence of the Coulomb incompressibility on the incompressibility of nuclear matter and thus the model dependence of $K_{\rm coul}$. However, this dependence is somewhat diluted by taking a wide spread of mean field models in \cite{sagawa2007}. This conclusion  is corroborated by a recent result by Vesely et al. \cite{vesely2012}, who used calculated values of K$_{\rm A}$ for $\sim$200 semi-magic nuclei across the nuclear chart in QRPA+ Hartree-Fock-Bogoliubov method with SLy4 and UNEDF0 forces and separable and zero-range pairing to determine K$_{\rm vol}$, K$_{\rm surf}$, K$_{\rm \tau,v}$, and K$_{\rm \tau,s}$ and  K$_{\rm coul}$ coefficients of the leptodermous expansion of K$_{\rm A}$. They obtained K$_{\rm coul}$=(-5.1$\pm$0.4) MeV in a very good agreement with Sagawa et al.  Shlomo and Youngblood \cite{shlomo1993} studied the  correlation $K_{\rm coul}$ - $K_{\rm 0}$ correlation, attempting to fit  the leptodermous expansion (\ref{ka}) to experimental data available in 1993. They found the $K_{\rm coul}$ -- $K_{\rm 0}$ correlation rather strong and were not able to constrain it in their fits. In order to investigate the effect of a possible stronger correlation than that found by Sagawa et al., our analysis was first carried out adopting the value  $K_{\rm coul}$ = -(5.2 $\pm$ 0.7) MeV. Next the analysis was repeated, increasing the error of $K_{\rm coul}$ to cover the range -7.3 $<$ $K_{\rm coul}$ $<$ -3.1 MeV, wide enough to account for possible effects of the $K_{\rm coul}$ -- $K_{\rm 0}$ correlation.
 
As a fifth extension of the procedure, the ratio of $K_{\rm surf}/K_{\rm vol}$, that was kept equal to -1 in the analysis by Li et al., was allowed to vary. Increasing the magnitude of $c$ above one had a significant effect, as described in Sec.~\ref{ext}. 

Finally, all fits to experimental data in this work were done in two stages: first, a `MESH' fit was performed when  variable parameters (e.g. $K_{\rm 0}$,  $K_\tau$ etc.) were changed in small steps in order to find the minimum of the function 
\begin{equation}
\sigma = \sum_{\rm i=1}^N \frac{(y_{\rm i}^{\rm exp} - (p+q x_{\rm_i}))^2}{(\Delta y_{\rm i}^{\rm exp})^{\rm 2}},
\label{sigma}
\end{equation}
where $N$ is the number of experimental points. The error $\Delta y_{\rm i}^{\rm exp}$ comes from two independent sources, the error in $K_{\rm A}$, determined by the uncertainty in GMR energy and the rms matter or charge radius, and the error in the Coulomb term. It is calculated as ($\Delta y_{\rm i}^{\rm exp})^{\rm 2} = \Delta (K_{\rm A})_{\rm i}^{\rm 2} + \Delta K_{\rm coul}^{\rm 2}$. $\sigma$ is not normalized to the number of experimental points and the number of variable parameters.

This procedure involves creating a multi-dimensional matrix with several million elements for each case. If the spacing between points is $\Delta$x, one of the points is sure to be within  $\Delta$x/2 of the true minimum, although in general it will not correspond to the lowest value. Each point in the matrix is then evaluated searching for a minimum taking small steps for each parameter.  It is essential that the range of each parameter is wide enough that the descent  to the minimum and ascent out it defines the minimum without doubt and the same minimum is found for all parameters. The lowest minimum common to all parameters is taken as the \textit{minimum minimorum} of the set. The MESH procedure ensures that other  local minima are not mistaken for the absolute minimum. If such a minimum cannot be found for a particular parameter (or a group of parameters) in a physically sensible parameter space, it might be an indication of a correlation between these parameters which is strong enough to prevent existence of a stationary point. The MESH method has been criticized for its inefficiency, especially for functions of many variables and a large demand on computer memory. On the other hand, this method is extremely simple and has absolute stability. It always converges within the desired tolerance in a known number of steps and is quite insensitive to the detailed behaviour of the function. 

The MESH method and the standard minimization methods using different algebraic procedures (single-parameter variations, simplex, gradient methods) should lead to exactly the same results. However, the MESH method maps all minima in the chosen parameter space and leads unambiguously to the absolute one. In contrast, the algebraic methods introduce the necessity to testing various various starting points to ensure that the minimum found is the \textit{minimum minimorum}.

When the minimum $\sigma$ is found in the MESH fit, the corresponding parameters are used as input to the CERN MINUIT package \cite{MINUIT} to obtain the final values of the fitted parameters (MINUIT fit), their errors and correlation coefficients, not available from the MESH fit. The main reason for breaking the minimization into two steps is that in some cases the parameter surface may have local minima or could be rather flat. An automatic routine, such as MINUIT, needs to be guided to the deepest minimum otherwise it may give misleading results. On the other hand, the MESH fit locates the \textit{minimum minimorum} rather accurately and the subsequent local improvement of the minimum using the MINUIT fit is reliable. Furthermore, in MINUIT the errors in fitted parameters are calculated from the error matrix \cite{MINUIT}. If there is more that one fitted parameter, the error includes non-diagonal elements of the error matrix which represent correlation between the parameters. 

We note that it has been reported in the past that attempts to fit all the parameters/coefficients in the expression for K$_{\rm A}$ to experimental data, taking them as independent variables, has not been productive. The parameters were said to correlated and the experimental data not accurate enough to constrain the correlations efficiently. However, our strategy of  multi-dimensional MESH fitting with all the parameters constrained within limits, expected from microscopic estimates, avoids most of the problem. Further examination of the minimum, already found in the MESH fit, using the MINUIT routine, produces precise values and errors of the parameters.
   
\subsection{\label{sncd}Sn and Cd data}
There is a considerable amount of data on GMR energies available on Sn and Cd isotopes. However, the data from different groups differ by several times their quoted errors and cannot be meaningfully averaged and treated simultaneously. We have divided them to three groups (see Table~\ref{tab:data}), each analyzed using our new analysis method. The objective of this section is to explore the degree to which the new method of analysis reproduces the results of \cite{li2007, li2010, garg2011} when the same constraints on $c$ and $K_{\rm coul}$ are retained. Relaxation of these constraints is studied in the next section.

\subsubsection{Data from the RCNP group}

\begin{figure}
\centerline{\epsfig{file=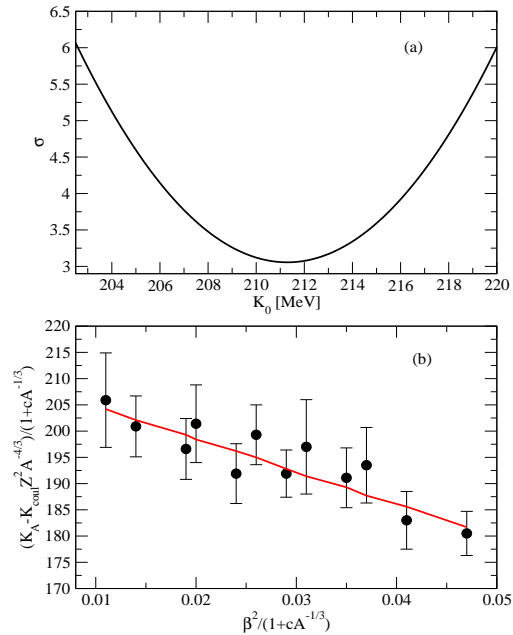,height=10cm,width=8cm}}
\caption{\label{fig:ndos_sncd} (Color on-line) (a) Minimization of $\sigma$ as a function of $K_{\rm 0}$ for combined data on Sn and Cd isotopes \cite{li2007,li2010, garg2011}. $K_{\rm 0}$ was varied with the step 0.1 MeV. (b) Fit to data, represented by the left-hand side of (\ref{kanjs}), as a function of $\frac{\beta^2}{1 + cA^{-1/3}}$ for the $K_{\rm 0}$, corresponding to minimum $\sigma$.}
\end{figure}

Sn ($^{\rm 112-124}$Sn) and Cd ($^{\rm 106,110-116}$Cd) data obtained by the RCNP group were analyzed separately [ RCNP(Sn) and RCNP(Cd) sets] and as  a combined Sn+Cd data set RCNP(Sn+Cd), not considered in \cite{li2010, garg2011}). $K_{\rm A}$ was fitted as a function of $K_{\rm 0}$ and $K_{\rm \tau}$ using (\ref{kanjs}). The MESH fit was performed for fixed values of $K_{\rm 0}$ in the region of 180 - 260 MeV with a step of 0.1 MeV. For each value of $K_{\rm 0}$,  $K_\tau$  was varied in the range -900$ \le K_\tau \le$ 0 MeV with a step of 2 MeV. In all cases values $c = -1$ and $K_{\rm c}$ = -(5.2$ \pm$ 0.7) MeV have been adopted. We found that $\sigma$ (\ref{sigma}) showed a well defined minimum in $K_{\rm 0}$ for each data set. This result is illustrated in Fig.~\ref{fig:ndos_sncd} for the combined Sn and Cd data set but the same behavior was observed for individual Sn and Cd sets as well as for all other data sets considered in this work for a
 fixed value of $c$ = -1. In the subsequent MINUIT fit, the values of $K_{\rm 0}$, $K_\tau$, obtained in the MESH fit were taken as input. 

Two ways of MINUIT fitting were adopted: (i) both $K_{\rm 0}$ and  $K_\tau$ were allowed to vary and (ii) $K_{\rm 0}$ was kept constant at the value corresponding to the minimal $\sigma$ in the MESH fit and only $K_{\rm \tau}$ varied. In case (i) the error in both $K_{\rm 0}$, $K_\tau$ included a correlation between them. In case (ii) there is only one variable parameter in the fit and therefore only diagonal elements of the error matrix enter the calculation of the error. In both cases the values of $K_{\rm A}$ were evaluated using matter radii obtained in method A. The results are summarized in Table~\ref{tab:c1}. Note that we also included results of the fit (i) to $K_{\rm A}$ obtained with charge radii (\ref{kexpk}) as line (iii). In the last two columns the range of $K_{\rm 0}$ corresponding to 3$\sigma$ above and below the minimum is given to indicate the quality of the fit to $K_{\rm 0}$ 

To examine how well results using the new method reproduce those of \cite{li2007, li2010, garg2011} we consider first $K_\tau$. The tabulated values are broadly in line with their -550 MeV for all data sets. However, the errors we find are, by a factor of two or more, larger than their $\pm$100 MeV when the correlation between $K_{\rm 0}$, $K_\tau$ is taken into account. 

Neither previous study examined the sensitivity of the data to the value of $K_{\rm 0}$. The new procedure showed clear sensitivity to $K_{\rm 0}$ returning best fit values in general consistent with the value assumed by Li et al. and Garg \cite{li2007, li2010, garg2011}. The best value from this analysis, including $K_{\rm 0}$ and $K_{\rm \tau}$  correlations and using matter radii [(i) in Table~\ref{tab:c1}] is $K_{\rm 0}$ = (210 $\pm$ 5) MeV. We stress that this result is based on the assumption that the ratio of volume to surface incompressibility, $c$, is equal to -1. We note that in these and all subsequent fits the use of charge radii systematically lowers the value of $K_{\rm \tau}$.
 
\subsubsection{Sn and Cd data by the TAMU group}
There are four pieces of data obtained by the TAMU group on $^{112,124}$Sn and $^{110,116}$Cd (set TAMU3) providing $\tilde{E_3}$ from the ratio of $m_3/m_1$ moments (\ref{moments}) which can be used to calculate $K_A$ compatible with the scaling model. 
Five pieces of data on $^{112,116,124}$Sn and $^{110,116}$Cd (set TAMU0) exist and can be used to calculate $\tilde{E_0}$ from the ratio of $m_1/m_0$ moments.

We analyzed the data in the same three ways as the RCNP data and present the results in Table \ref{tab:c1}. 
For both sets, taking c = -1, best fit values for K$_\tau$ consistent with the RCNP value but with larger errors and somewhat lower values of $K_{\rm 0}$  were returned by our preferred fit (i).

\subsubsection{Sn and Cd data from the Gaussian fit}

As stated in the Introduction, GMR energies obtained from a Gaussian fit to the strength functions using (\ref{ga}) also yield $K_{\rm A}$ compatible with the scaling model. $K_{\rm A}$ values for $^{\rm 112,114,116,120,124}$Sn and $^{\rm 110,116}$Cd isotopes \cite{lui2004, lui2004a, sharma1988} (see Fig.~\ref{fig:gmr_raw}) form a set of nine pieces of data, labeled GF. We report analysis of this set for completeness in Table~\ref{tab:c1}. Again, the same behavior is observed as for all the previous data sets. The value of $K_0$ is well determined. $K_\tau$ has a rather large error, as is expected because of a larger error in $\tilde{E_3}$, calculated using (\ref{ga}). 
   
\subsection{\label{ext}Extended data sets}

In this section we present results of investigation of two effects, outlined at the beginning of Sec.~\ref{anal}, the variation of $c$ and the correlation between $K_{\rm 0}$ and $K_{\rm coul}$. The former required detailed fitting of the data, the latter could be estimated by comparison of the fits to K$_{\rm A}$ calculated with  $\Delta K_{\rm coul}$=0.7 MeV or 2.1 MeV, as outlined above. In addition, all fits were performed using values of K$_{\rm A}$ obtained with both matter and charge radii.

To explore sensitivity to these effects, we constructed six representative data sets (see Table~\ref{tab:data}), each of which contained all available values of $K_{\rm A}$ calculated in the same way. The first three sets, RCNP-E (from $\tilde{E_3}$ (~\ref{moments})), TAMU0-E (from $\tilde{E_0}$) and GF-E (from $\tilde{E_3}$ (\ref{ga})), included combined data for Sn and Cd isotopes and, in addition, data on $^{58,60}$Ni and $^{56}$Fe \cite{lui2006}. The TAMU0-E set was further extended by data on $^{144}$Sm \cite{youngblood2004}. Furthermore, $K_{\rm A}$ values extracted from $\tilde{E_0}$ reported by the RCNP group on $^{106,110-116}$Cd \cite{garg2011} and $^{112-124}$Sn \cite{li2007, li2010} were added to the TAMU0-E group. The combination of the RCNP and TAMU data in this case was possible because they differed significantly less than the data obtained on Sn and Cd isotopes by the two groups from $\tilde{E_3}$ energies. The GF-E set also included $K_{\rm A}$ values obtained for $^{112-116,120,124}$Sn and $^{144,148}$Sm by Sharma et al. \cite{sharma1988} which were in very good agreement with the values from data by the RCNP and TAMU groups. Finally, all three data sets included the same $K_{\rm A}$ value for $^{\rm 208}$Pb. It was obtained by taking a weighted average of 
of values obtained from $\tilde{E_0}$ energies \cite{uchida2004, youngblood2004}, as neither $\tilde{E_3}$ energies 
nor data from a Gaussian fit are available.

The next three data sets, RCNP-M, TAMU0-M and GF-M were exactly the same as the first three, but the $K_{\rm A}$ values for the light $^{58,60}$Ni and $^{56}$Fe isotopes were excluded. The motivation for this modification has been that it is not yet quite clear whether the collective modes in light and heavy nuclei can be described by the same physics. It is often argued that data on lighter nuclei, with A less then about 100, do not provide reliable GMR energy as the GMR strength is fragmented (see e.g. \cite{hamamoto1997, paar2006}). Also, the validity of (\ref{ka}) may be questionable for lighter nuclei as they are less likely to behave as a liquid drop; shell and surface effects become increasingly important with decreasing $A$. Thus both options, taking all the data together, and considering only the heavier isotopes, were explored. 

Starting with the ratio $c$, we recall that the expansion (\ref{ka}) is, strictly speaking, valid only in the scaling approximation \cite{blaizot1981, treiner1981}, based on a simple scaling of the ground state density $\rho(r) \rightarrow \lambda^{\rm 3}\rho(\lambda r) $ following the transformation of coordinates $r \rightarrow r/\lambda$. In this approximation the curvature term is small and K$_{\rm surf}$ and  K$_{\rm vol}$ are proportional. Their ratio $c$ of has been estimated in different macroscopic and microscopic models. For example, Blaizot et al. \cite{blaizot1976, blaizot1980, blaizot1981} found $c$ to be between -1.4 and -1.6 and  Treiner et al. \cite{treiner1981} estimated -1.4 $\le$ c $\le$ -0.65. Myers and Swiatecki \cite{myers1990, myers1995} predicted $c$ $\sim$ -1.35 on the basis of a simple formula without adjustable parameters. Patra et al. \cite{patra2002} calculated $c$ in RMF Hartree and RETF (Relativistic Extended Thomas-Fermi) models and found $\sim$ -1.5 $< c <$ -0.5. Sagawa et al. \cite{sagawa2007} obtained c $\sim$ -1 for non-relativistic HF models within a few percent and c $\sim$ -1.16 for the NL3 RMF model. Sharma et al. \cite{sharma1989} performed a theoretical calculation of GMR energies and K$_{\rm A}$ using the Skyrme and hydrodynamic models and carried out various fits to these quantities to obtain coefficients in the leptodermous expansion in the scaling approximation. They obtained $c$ close to -1 in all cases. However, application of the fitting procedures to experimental data, available in 1989, yielded a very different result, $c$ = -(2.5 $\pm$ 0.3). RMF study by Sharma \cite{sharma2009} of GMR energies and coefficients of the leptodermous expansion showed a distinct dependence of the ratio $c$ on the choice of Lagrangian and yielded values -1.98 (NL3), -1.67 (SVI-2) and -1.00 (SiGO-c). Vesely et al. \cite{vesely2012} obtained the ratio the $c$ $\sim$ -1.6  in their QRPA+HFB calculations with SLy4 and UNEDF0 forces.

\begin{figure}
\centerline{\epsfig{file=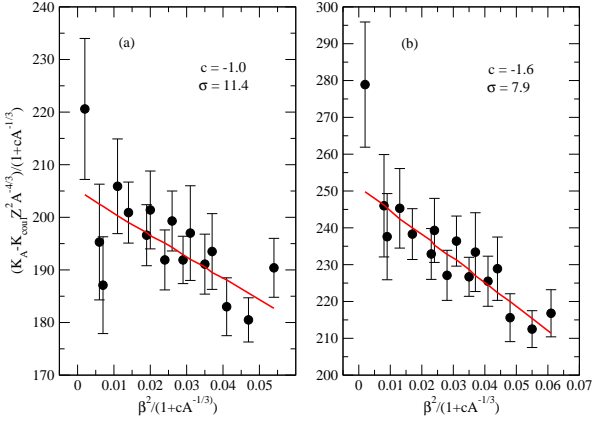,height=7cm,width=9cm}}
\caption{\label{fig:ndos_all}(Color on-line)  Fit to experimental data in set RCNP-E with $K_{\rm coul}$ = -(5.2 $\pm$ 0.7) MeV for (a) $c = -1.0$  and (b) $c = -1.6$ . Note that both x and y coordinates are $A$ and $c$ dependent. For more explanation see text.}
\end{figure}

\begin{figure}
\centerline{\epsfig{file=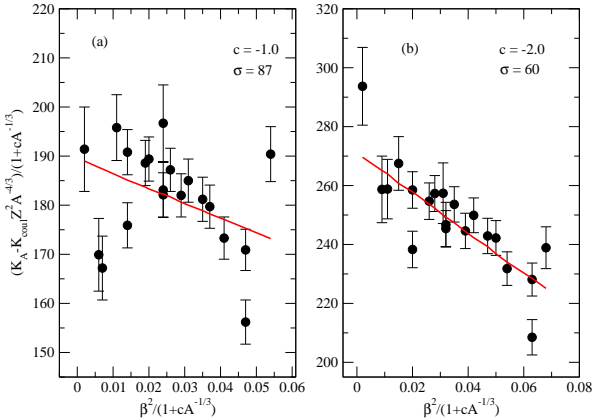,height=7cm,width=9cm}}
\caption{\label{fig:m1m0} (Color on-line) The same as Fig.~\ref{fig:ndos_all}, but for set TAMU-E and (b) $c$ = -2.0.}
\end{figure}

We sought the best MESH fit to RCNP-E, TAMU0-E and GF-E data sets for -2.4 $\ge$ c $\le$ -0.6 with a step -0.2. For each $c$ value, $K_{\rm 0}$ was varied in the range 150 $\le K_{\rm 0} \le$ 450 MeV with a step of 0.1 MeV and $K_\tau$ was varied within -1000 $\le K_\tau \le$ 200 MeV with a step of 0.5 MeV. A stable minimum for each $c$ as a function of $K_{\rm 0}$ was found, for both $K_{\rm coul}$ = -(5.2 $\pm$ 0.7) MeV and $K_{\rm coul}$ = -(5.2 $\pm$ 2.1) MeV, as documented in detail in Tables~\ref{tab:RCNP-E} - \ref{tab:GF-M}. 
Examination of the tables shows that the fit quality ($\sigma$) considerably improves for $c$ differing from -1. As a consequence, the best-fit value of $K_0$ is found at 
\begin{table}
\caption{\label{tab:RCNP-E} RCNP-E data set: Variation of $K_{\rm 0}$  and $K_\tau$ with fixed values of $c$. Typical errors can be found in Table~\ref{tab:c1} and, for the value of $c$ corresponding to the minimum $\sigma$, in Table~\ref{tab:summary}. Results have been obtained in the MESH fit with matter radii for both values of the error in $K_{\rm coul}$.}   
 \begin{ruledtabular}
\begin{tabular}{ccrcccrc}
\multicolumn{4}{c}{$\Delta$K$_{\rm coul}$=0.7 MeV}&\multicolumn{4}{c}{$\Delta$K$_{\rm coul}$=2.1 MeV} \\ \hline
 c & $K_{\rm 0}$ & $K_{\tau}$ & $\sigma$ &  $K_{\rm 0}$  & $K_{\tau}$ &  $\sigma$ \\
   &    [MeV]     &   [MeV]    &    &  [MeV]  &  [MeV]  \\          \hline
  -0.6  &   182.6  &  -297  & 15.33  &  180.9  &  -240   &  5.39 \\
  -0.8  &   193.1  &  -352  & 13.28  &  191.9  &  -310   &  4.75 \\   
  -1.0  &   205.0  &  -414  & 11.37  &  204.4  &  -390   &  4.16 \\   
  -1.2  &   218.4  &  -483  &  9.71  &  218.6  &  -480   &  3.66 \\     
  -1.4  &   233.7  &  -562  &  8.46  &  234.8  &  -580  &  3.30  \\     
  -1.6  &   251.1  &  -648  &  7.87  &  253.7  &  -699   &  3.19 \\   
  -1.8  &   271.3  &  -748  &  8.32  &  275.9  &  -840   &  3.64 \\   
  -2.0  &   294.8  &  -860  & 10.40  &  301.6  &  -993   &  4.37 \\    
  -2.2  &   322.6  &  -991  & 15.08  &  332.5 &  -1170  &  6.29 \\  
  -2.4  &   355.7  & -1140  & 23.90  &  370.3  & -1390  &  9.90 \\ 
  -2.6  &   395.6  & -1310  & 39.52  &  416.5  & -1390   & 16.39 \\
  -2.8  &   444.6  & -1510  & 66.56  &  474.3  & -1640   & 27.95 \\
  -3.0  &   505.0  & -1730  & 113.5  &  540.0  & -2180   & 49.00 \\
\end{tabular}
\end{ruledtabular}
\end{table}
the higher limit of the current estimates and beyond it. We illustrate the effect, similar in all three data sets, in Figs.~\ref{fig:ndos_all} and ~\ref{fig:m1m0} calculated with matter radii and $K_{\rm coul}$ = -(5.2 $\pm$ 0.7) MeV. Very similar behaviour is observed for $K_{\rm coul}$ = -(5.2 $\pm$ 2.1) MeV, although the uncertainties increase as expected.

\begin{figure}
\centerline{\epsfig{file=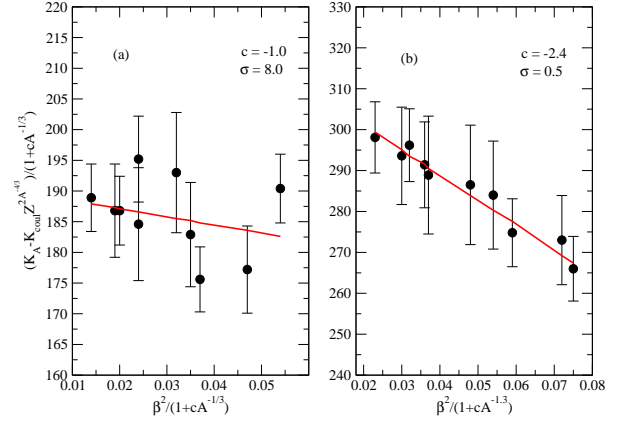,height=7cm,width=9cm}}
\caption{\label{fig:gf_m} (Color on-line) The same as Fig.~\ref{fig:ndos_all}, but for set GF-M and (b) $c = -2.4$.}   
\end{figure}

The above procedure was repeated using data sets RCNP-M, TAMU0-M and GF-M and full results are presented in Table~\ref{tab:summary}. All sets yielded the same results as for RCNP-E, TAMU0-E and GF-E sets in that the fits significantly improved with $c$ lower than -1. The values $K_0$ are higher that those returned from the fits to RCNP-E, TAMU0-E and GF-E sets, (see e.g. Fig.~\ref{fig:gf_m}) and are less sensitive to the difference between the matter and charge radius then values of $K_{\rm \tau}$. We stress that the scatter of entries in Table~\ref{tab:summary} is caused solely
\begin{table}
\caption{\label{tab:TAMU0-E}The same as Table ~\ref{tab:RCNP-E} but for TAMU0-E data set.}   
 \begin{ruledtabular}
\begin{tabular}{ccrcccrc}
\multicolumn{4}{c}{$\Delta$K$_{\rm coul}$=0.7 MeV}&\multicolumn{4}{c}{$\Delta$K$_{\rm coul}$=2.1 MeV} \\ \hline
 c & $K_{\rm 0}$ & $K_{\tau}$ & $\sigma$ & $K_{\rm 0}$  & $K_{\tau}$ &  $\sigma$ \\
   &     [MeV]   &   [MeV]    &          &    [MeV]     &  [MeV]  \\ \hline          
  -0.6  &   168.0  &  -220  & 101.3  & 164.1  &  -116   &  22.28 \\
  -0.8  &   177.7  &  -268  & 94.44  &  174.1  &  -180   &  20.57 \\   
  -1.0  &   188.5  &  -320  & 87.47  &  185.3  &  -248   &  18.82 \\   
  -1.2  &   200.6  &  -376  & 80.53  &  198.1  &  -328   &  17.08 \\     
  -1.4  &   214.5  &  -444  & 73.84  &  212.7  &  -416   &  15.39  \\     
  -1.6  &   230.1  &  -512  & 67.75  &  229.6  &  -520   &  13.86 \\   
  -1.8  &   248.3  &  -596  & 62.78  &  249.1  &  -632   &  12.64 \\   
  -2.0  &   269.4  &  -688  & 59.83  &  272.1  &  -764   &  11.99 \\    
  -2.2  &   294.2  &  -792  & 60.31  &  299.6  &  -920   &  12.36 \\  
  -2.4  &   323.7  &  -912  & 66.60  &  332.8  & -1100   &  14.51 \\ 
  -2.6  &   359.2  &  -1048 & 82.76  &  373.2  & -1304   &  19.82 \\
  -2.8  &   402.3  &  -1200 & 115.9  &  423.2  & -1544   &  30.86 \\
  -3.0  &   450.0  &  -1292 & 179.8  &  485.8  & -1812   &  52.56 \\    
\end{tabular}
\end{ruledtabular}
\end{table}
\begin{table}
\caption{\label{tab:GF-M}The same as Table~\ref{tab:RCNP-E} but for GF-M data set.}   
\begin{ruledtabular}
\begin{tabular}{ccrcccrc}
\multicolumn{4}{c}{$\Delta$K$_{\rm coul}$=0.7 MeV}&\multicolumn{4}{c}{$\Delta$K$_{\rm coul}$=2.1 MeV} \\ \hline
 c & $K_{\rm 0}$ & $K_{\tau}$ & $\sigma$ & $K_{\rm 0}$  & $K_{\tau}$ &  $\sigma$ \\
   &     [MeV]   &   [MeV]    &          &    [MeV]     &  [MeV]  \\ \hline          
  -0.6  &   170.4  &   -53  & 10.67  & 172.1  &  -132   &  2.45 \\
  -0.8  &   179.6  &   -91  &  9.35 &  181.2  &  -160   &  2.15 \\   
  -1.0  &   189.8  &  -133  &  8.00  &  191.5  &  -200   &  1.84 \\   
  -1.2  &   201.2  &  -179 &   6.61  &  202.7 &  -236  &  1.52 \\     
  -1.4  &   214.1  &  -231  &  5.22  &  215.5  &  -280  &  1.20 \\     
  -1.6  &   228.7  &  -289  &  3.85  &  229.9  &  -328   &  0.89 \\   
  -1.8  &   245.4  &  -355  &  2.57 &  246.2  &  -380   &  0.60 \\   
  -2.0  &   264.6  &  -429  &  1.48  &  265.0  &  -440   &  0.35 \\    
  -2.2  &   286.9  &  -792  &  0.70  &  286.9  &  -508   &  0.18 \\  
  -2.4  &   313.3  &  -610 &   0.51  &  312.4 & -584   &  0.14 \\ 
  -2.6  &   344.6  &  -721 &   1.31  &  342.8 &  -672  &  0.32 \\
  -2.8  &   382.5  &  -850 &   3.81  &  379.0 &  -768  &  0.87  \\
  -3.0  &   428.9  & -1000  & 9.28   &  423.3 &  -880  &  2.07  \\ \hline   
\end{tabular}
\end{ruledtabular}
\end{table}
 by differences in experimental data sets which are mutually incompatible within quoted errors. All correlations between fitted parameters are reflected in their errors.

\subsection{\label{kvs}Estimation of $K_{\rm \tau,v}$ and $K_{\rm \tau,s}$}
We also explored the volume and surface contributions to $K_\tau$ from the combined Sn and Cd data. $K_{\rm \tau}$ calculated in most microscopic  models to date, such as SHF and RMF contains essentially only the volume part $K_{\tau,v}$ as it is unclear how to calculate $K_{\tau,s}$ microscopically. In other words, the range of $K_\tau$, extracted from experiments \cite{chen2005, sagawa2007, li2007, centelles2009, chen2009, chen2010} contains both the volume and surface contributions and, strictly speaking, should not be compared with $K_{\rm \tau}$ calculated in microscopically.

There have been several attempts to extract values of $K_{\tau,v}$ and $K_{\tau,s}$ from various combinations of theory and experiment. Blaizot and Grammaticos \cite{blaizot1981} estimated $K_{\tau,v}$ and $K_{\tau,s}$ in a 
rather complicated way using (\ref{kali}). Treiner et al. \cite{treiner1981} used SIII and SkM Skyrme forces in self-consistent Thomas - Fermi calculation of $K_{\rm A}$ considering both the constrained and scaling models  (for details see \cite{treiner1981}). Nayak et al. \cite{nayak1990} used the leptodermous expansion of $K_{\rm A}$. The expansion coefficients were expressed in the framework of the scaling model in terms of quantities that are defined in infinite and semi-infinite matter. The coefficients were calculated in Extended Thomas-Fermi (ETF) approximation using SkM*, RATP, Ska and S3 Skyrme forces. However, as pointed out later by Pearson \cite{pearson2011} the models used for calculation of $K_{\tau,v}$ and $K_{\tau,s}$ in \cite{nayak1990} did not predict correct values of GMR energies. Patra et al. \cite{patra2002} calculated  $K_{\tau,v}$ and $K_{\tau,s}$ using a semiclassical RMF method with interaction NL1, NL3 and NL-SH.  Vesely et al. \cite{vesely2012} calculated the coefficients using QRPA + HFB method
 with two different pairing models and SLy4 and UNEDF0 Skyrme interactions.  We summarize all the results in Table~\ref{tab:ktvs}.  Note that other suggested limits on $K_{\rm \tau,v}$  = -(370 $\pm$ 120) MeV (in notation of the original paper \cite{chen2009} $ K_{\tau,2}^{sat}$) exist in the literature but they are calculated, not directly extracted from experimental data.
 
It is clear that the sensitivity of current GMR data to separate volume and surface contributions to the isospin incompressibility  in (\ref{ktau}) is limited and thus, in order to enhance this sensitivity, some additional constraints will be needed on the fit of $K_{\rm A}$ to obtain limits on $K_{\rm \tau,v}$ and $K_{\rm \tau,s}$.
First, we assumed that (\ref{ktau}) holds and looked for all combinations of $K_{\rm \tau,v}$ and $K_{\rm \tau,s}$, compatible with values of $K_{\rm \tau}$, within its errors, already obtained for each data set (see Table~\ref{tab:summary}). A MESH fit was performed in the region of  -1200 $ < K_{\rm \tau,v} <$ 0 MeV
\begin{table*}
\caption{\label{tab:summary} Summary of the values of K$_{\rm 0}$, K$_\tau$ and ratio of the volume and surface incompressibility $c$, as obtained from the MINUIT fit to data sets RCNP-E, GF-E, TAMU0-E and the M variant of these data sets. Results for each case are given for both matter and charge radii and both values of the error in $K_{\rm coul}$. }  
\begin{ruledtabular}
\begin{tabular}{lcccccc}
\multicolumn{7}{l}{$\Delta K_{\rm coul}$ = 0.7} \\  \hline
       &      \multicolumn{3}{c}{matter radii} &     \multicolumn{3}{c}{charge radii} \\  \hline
       &    $K_0$ &  $K_\tau$  &   c    &      $K_0$ &    $K_\tau$  &    c   \\  \hline
RCNP-E &   254(5) &  -664(121) &  -1.63  &     261(5) &   -632(116)  &   -1.59 \\
RCNP-M &   276(6) &  -700(138) &  -1.88  &     274(6) &   -644(135)  &   -1.74  \\
GF-E   &   251(5) &  -476(123) &  -1.80  &     252(4) &   -392(107)  &   -1.71  \\
GF-M   &   306(9) &  -584(169) &  -2.35  &     303(8) &   -500(173)  &   -2.24  \\ 
TAMU0-E &  278(4) &  -728(90)  &  -2.08  &     288(4) &   -716(84)   &   -2.05  \\
TAMU0-M &  347(5) &  -835(101) &  -2.60  &     344(6) &   -800(104)  &   -2.49  \\  \hline
\multicolumn{7}{l}{$\Delta K_{\rm coul}$ = 2.1} \\   \hline
       &      \multicolumn{3}{c}{matter radii} &    \multicolumn{3}{c}{charge radii} \\  \hline
       &     $K_0$ &  $K_\tau$  &   c    &       $K_0$  &    $K_\tau$  &    c   \\  \hline
RCNP-E &    252(8) &  -688(235) &  -1.58  &      260(8)  &   -648(228)  &   -1.56 \\
RCNP-M &    264(13)&  -664(305) &  -1.75  &      260(12) &   -604(310)  &   -1.58  \\
GF-E   &    249(9) &  -504(240) &  -1.77  &      253(8)  &   -414(227)  &   -1.72  \\
GF-M   &    306(18)&  -563(365) &  -2.35  &      304(18) &   -488(365)  &   -2.25  \\ 
TAMU0-E &   279(8) &  -802(198) &  -2.05  &      287(9)  &   -760(223)  &   -2.03  \\
TAMU0-M &   360(14)&  -903(252) &  -2.67  &      360(15) &   -856(272)  &   -2.59   \\  \hline
\end{tabular}
\end{ruledtabular}
\end{table*}
\begin{table*}
 \caption{\label{tab:ktvs} $K_{\tau,v}$ and $K_{\tau,s}$ as determined in different model approaches. All entries are in MeV. For more detail see text and the references therein.}
 \begin{ruledtabular}
\begin{tabular}{lrlcl}
 $K_{\tau,v}$ &  $K_{\tau,s}$ &  Method                & Force      &  Ref.\\  \hline
-420   &   850   & fit to $K_{\rm A}$(RPA)      & SIII       &  \cite{blaizot1981} \\
-508   &   1390  &                             & SIV        &     \\
-444   &   630   &                             & Ska        &     \\
-420   &   230   & fit to $K_{\rm A}$ (scaling) & SIII       &     \\
-508   &   670   &                             & SIV        &     \\
-444   &   640   &                             & Ska        &     \\ \hline
-319   &  -3540  & fit to $K_{\rm A}$ (constrained)  & SIII  & \cite{treiner1981} \\
-251   &  -1340  &                                  & SkM   &      \\
-456   &   420   & fit to $K_{\rm A}$ (scaling)      & SIII  &      \\  
-359   &   435   &                                  & SkM   &      \\   \hline
-349   &   497   & Extended Thomas-Fermi            & SkM*  & \cite{nayak1990}   \\
-338   &   313   &                                  & RATP  &       \\
-441   &   875   &                                  & Ska   &       \\
-456   &   383   &                                  & S3    &       \\ \hline
-676   &  1951   & RMF                              & NL1   &  \cite{patra2002}   \\ 
-690   &  1754   & RMF                              & NL3   &        \\
-794   &  1716   & RMF                              & NL-SH &        \\ \hline
-460(30) & 410(110) & fit to $K_{\rm A}$ from QRPA+HFB+sep. pair.   & SLy4 & \cite{vesely2012} \\
-510(30) & 570(120) &                            &  UNEDF0 &                      \\
-500(30) & 560(100) & fit to $K_{\rm A}$ from QRPA+HFB+z.r. pair.    & SLy4 & \\
-550(30) & 740(100) &                            & UNEDF0 \\
\label{KvsKss}
\end{tabular}
\end{ruledtabular}
\end{table*}
 and -1600  $ < K_{\rm \tau,s} <$ 1600 MeV, taking into account that 
$K_{\rm \tau,v}$ is expected to be negative in line with microscopic calculations. 

The second constraint was constructed assuming that (\ref{ktau}) applies and the expansion in terms of $A^{-1/3}$ \textit{converges at a reasonable rate}, i.e. no higher order terms are significant. The question of what is reasonable can be only answered in a somewhat arbitrary way as there is a large spread in values of $K_{\rm \tau,v}$ and $K_{\rm \tau,s}$ calculated microscopically (see Table~\ref{tab:ktvs}).  We looked at two scenarios: (i) the magnitude of the two coefficients is almost the same \cite{treiner1981, nayak1990} and (ii) $K_{\rm \tau,s}$ is roughly three times larger than $K_{\rm \tau,v}$ \cite{patra2002}. Taking the average mass number A=100, we obtain for the ratio (\ref{limit}) 0.2 for the former and 0.5 for the latter. Taking the higher value of the ratio, we choose to allow for a slower convergence of the expansion (\ref{ktau})
\begin{equation}
\frac{K_{\tau,s}A^{-1/3}}{K_{\tau,v}} \le 0.5.
\label{limit}
\end{equation}
Simultaneous application of (\ref{ktau}) and (\ref{limit}) yielded results presented in Table~\ref{tab:ktvs-res}. We conclude that the most likely limits on $K_{\rm \tau,s}$ are  -810  $ < K_{\rm \tau,v} <$ -370 MeV. Limits on the surface contribution to isospin incompressibility are -1020 $\le K_{\tau,s} \le$ 160 MeV.

We note that another possibility to determine $K_{\rm \tau,v}$ and $K_{\rm \tau,s}$ would be to fix $K_{\rm \tau,v}$ to a theoretical value, for example $K_{\rm \tau,v}$  = -(370 $\pm$ 120) MeV \cite{chen2009}. However, these values are naturally model dependent - the heavy ion collision data are no exception. The main objective of our paper is to explore what can be learned from the experimental data (GMR energies) alone using only the assumption that the leptodermous expansion is valid and converges fast.   
      
\subsection{The curvature term}
\begin{figure}
\centerline{\epsfig{file=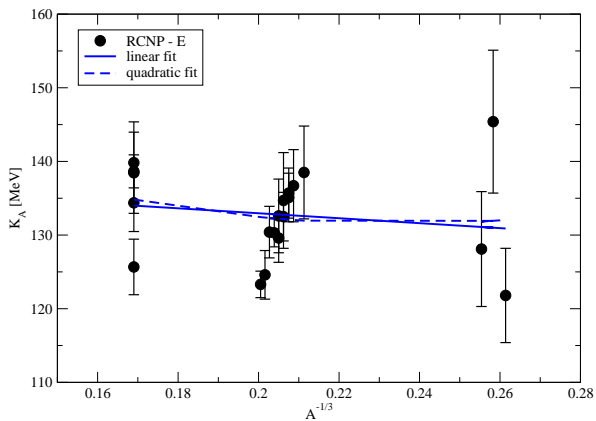,height=7cm,width=9cm}}
\caption{\label{fig:curv} (Color on-line) Linear (solid) and quadratic (dashed) fits to experimental $K_{\rm A}$ as a function of $A^{\rm -1/3}$ for data set RCNP-E.}
\end{figure}

We recall that (\ref{kali}) is an expansion in terms of powers of $A^{\rm -1/3}$. The second order term, which depends on $(A^{-1/3})^{\rm 2}$ is called the curvature term. The limited range  of A$^{\rm -1/3}$ considered in this work meant that no contribution of order higher than linear could be identified outside experimental error. As an example, linear and quadratic fits to the experimental $K_A$ as a function of A$^{\rm -1/3}$ are illustrated in Fig.~\ref{fig:curv} for the RCNP-E set.

A frequently raised objection to analysis of GMR data using the leptodermous formula (\ref{kali}) is that the omission of a very poorly known curvature term may lead to a substantial change in the surface term. Earlier work allows us to estimate this effect. Treiner et al. \cite{treiner1981} calculated the $K_{\rm curv}$ coefficient microscopically in the scaling model using the SIII and SkM Skyrme interactions. They found it to be positive and of the order of 300 MeV.  Sharma et al. \cite{sharma1989} also examined the
\begin{table}
 \caption{\label{tab:ktvs-res} K$_{\tau,v}$ and K$_{\tau,s}$ values (in MeV). Matter radii and $\Delta$K$_{\rm coul}$ = 0.7 MeV were used in the calculation. $\overline{A^{\rm -1/3}}$ was taken to be 0.2 in the mass region considered. For more detail see text and the references therein.}
\vspace*{5pt}
 \begin{ruledtabular}
\begin{tabular}{lcccccc}
       &  K$_{\tau,v}$  &  K$_{\tau,s}$ &  $\overline{A^{\rm -1/3}}$ $K_{\tau,s}$ &  K$_\tau$  &   ratio    &  $\sigma$ \\  \hline \hline
RCNP-E &    -500.0  &   -950.0 &   -190.0    &  -690.0    &  0.38  &   7.47  \\
RCNP-M &   -620.0  &  -410.0 &   -82.0    &  -702.0    &  0.13 &   4.02  \\
GF-E   &   -370.0  &  -700.0 &   -140.0    &  -510.0    &  0.38  &   5.17  \\
GF-M   &  -610.0  &    160.0 &    32.0    &  -578.0    & -0.053  &   0.49 \\
TAMU0-E  &  -550 &   -1020.0 &   -204.0    &  -754.0    &  0.37  &  58  \\
TAMU0-M   &  -810.0  &   -170 &   -34.0    &  -844.0    &  0.042  &  52.0  \\
\end{tabular}
\end{ruledtabular}
\end{table}
 consequence of including a curvature term and varied the coefficient between 350 and 400 MeV and found only a 1(4)\% change in $K_{\rm vol}$ ($K_{\rm surf}$) and $K_{\tau}$ almost unaffected. They adopted a value $K_{\rm curv}$ = 375 MeV which was kept constant during their final fits. If we accept as the best estimate of the $K_{\rm curv}$ the value +350 MeV the size of the curvature term is 24 MeV at A = 56 and 10 MeV at A = 208. At the same A values, with $K_{\rm surf}$ = 500 MeV, fits neglecting the curvature term give surface term  values 130 MeV and 85 MeV, respectively. The ratio of the curvature to the surface term is thus $\approx$ (15 $\pm$ 3) \% and inclusion of the curvature term would indeed increase the surface term but not to any great extent. The ratio $c$ would decrease also by a factor (1.15 $\pm$ 3)\% , shifting the range from -2.4 $<$ c $<$ -1.6 to -2.8 $<$ c $<$ -1.8 which is not a major change.

\begin{figure}
\centerline{\epsfig{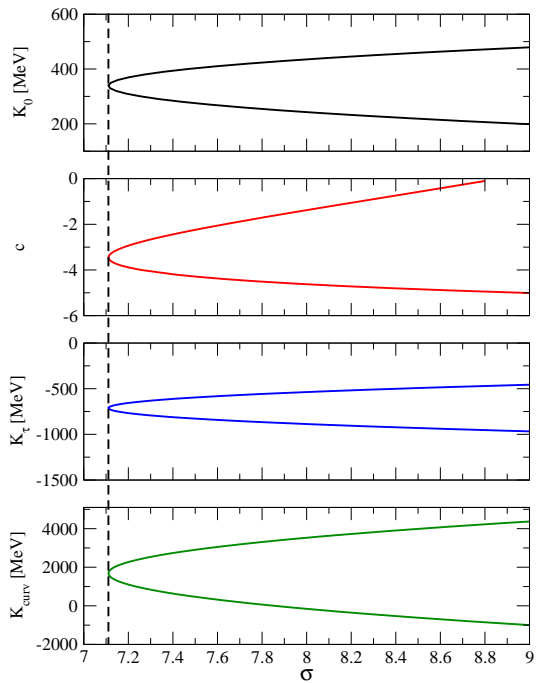}}
\caption{\label{fig:curv-slice} (Color on-line) MESH fit to RCNP-E data including the curvature term. Values of $K_{\rm 0}$, $c$, $K_{\tau}$  and $K_{\rm curv}$ as a function $\sigma$ are displayed, showing a well-defined unique minimum, indicated by the vertical dashed line.  For more detail see text.}
\end{figure}

To further explore the consequence of a range of  K$_{\rm curv}$ values, and to illustrate our fitting procedure in detail, we examined the extended equation
\begin{eqnarray}
\frac{K_{\rm A}}{1 + cA^{-1/3}}- \frac{K_{\rm coul}Z^{\rm 2}A^{\rm -4/3}}{1 + cA^{\rm -1/3}}-\frac{K_{\rm curv}A^{\rm -2/3}}{1 + cA^{-1/3}}& =&  \nonumber \\
 K_{\rm vol}+K_{\tau} \frac{\beta^{\rm 2}}{1 + cA^{\rm -1/3}}.
\label{kanjscurv}
\end{eqnarray}
and its fit to the RCNP-E data set. Keeping K$_{\rm coul}$ = -(5.2$\pm$0.7) MeV, we first performed the MESH fit in the four-parameter space, stepping  K$_{\rm 0}$ in the range 150 to 450 MeV (step 0.1 MeV),  K$_{\tau}$ in the range of -900 to -300 MeV (step 0.5 MeV),  c in the range of -4 to -0.1 (step 0.01) and  K$_{\rm curv}$ in the range of -1600 to 2000 MeV (step 100 MeV). Next we examined the stability of the minimum by making 'slices' across the four-parameter MESH along each parameter axis. The results are shown in Fig.~\ref{fig:curv-slice} demonstrating that exactly the same minimum is reached in each slice, i.e. the minimum is stable. The errors and the correlation coefficients were obtained in subsequent MINUIT fits in which one of the parameters was set at its minimum value in order to examine effects of various correlations.  Numerical results are given in Table~\ref{tab:curv}  (lines A-D). Lastly we performed a full four-parameter fit, varying  K$_{\rm 0}$, $c$,  K$_{\rm \tau}$ and  K$_{\rm curv}$ simultaneously in
 the MINUIT code (line E in Table~\ref{tab:curv}). The correlation coefficients obtained in all fits are shown in Table~\ref{tab:cor}.

The analysis has been repeated without the curvature term, performing parameter fits, with results in lines F-J in Table~\ref{tab:curv} and Table~\ref{tab:cor}. We observe that the least correlated parameter in both cases is K$_{\tau}$ and this level of correlation is somewhat smaller when the curvature term is included in the fit. This feature may
\begin{table}
\caption{\label{tab:curv} Results of fits to RCNP-E data: Results of fits A-E (including the curvature term) and  F-G  (without the curvature term). Entries without an error in bracket indicate which parameter was kept constant at their minimum value during the fits. For more explanation see text.} 
\vspace*{10pt}
 \begin{ruledtabular}
\begin{tabular}{cccllc}
& $\sigma$ &          K$_{\rm 0}$  &           $c$              &       $K_{\tau}$        &  K$_{\rm curv}$ \\  \hline
A  &  7.110586     &   339              &    -3.45(35)         &       -712(160)      &      1689(504)     \\
B   &  7.110592    &   339(23)         &   -3.45               &       -712(175)      &       1682(85)      \\ 
C  &  7.110587     &   339(93)        &     -3.45(1.6)       &       -712              &      1685(1958)   \\   
D   &  7.110586   &   339(26)        &   -3.46(67)         &       -712(176)       &         1686              \\
E   &  7.110588     &   339(106)      &     -3.45(1.66)     &     -712(187)        &      1685(2042)    \\  \hline
F   &  7.875914    &    253.8          &     -1.628(0.050)  &     -662(98)          &        $-$                \\
G   &   7.855915    &   253(5)         &     -1.628             &      -662 (121)      &        $-$               \\
H   & 7.855915     &    254(15)       &     -1.628(19)       &     -661.9             &         $-$          \\
J    &   7.855916    &   253(26)        &     -1.629(27)        &      -662(177)       &        $-$               \\  
\end{tabular}
\end{ruledtabular}
\end{table}
\begin{table}
\caption{\label{tab:cor} Correlation coefficients of parameters in fits shown in Table~\ref{tab:curv}.  K$_{\rm 0}$ (I),  $c$ (II),$K_{\tau}$  (III),   K$_{\rm curv}$ (IV).} 
\vspace*{5pt}
 \begin{ruledtabular}
\begin{tabular}{ccllccc}
     &        I-II       &       I-III      &    I - IV    &   II-III       &  II-IV        &    III-IV    \\ \hline 
A   &        $-$      &        $-$    &      $-$      &  0.842     &   0.995     &    0.790    \\
B   &        $-$      &      0.870     &    0.783    &    $-$       &   $-$     &     0.424    \\
C   &       0.993    &        $-$    &    0.988    &    $-$       &   0.999     &    $-$      \\ 
D   &       0.837    &     0.833     &    $-$      &   0.426     &   $-$         &     $-$       \\
E    &       0.977    &     0.516    &     0.969    &  0.350    &   0.999     &     0.334 \\    \hline
F  &          $-$     &       $-$     &     $-$      &     0.878   &    $-$       &     $-$      \\       
G  &         $-$       &       0.922  &     $-$      &       $-$    &    $-$       &     $-$       \\
H   &       0.992    &        $-$    &      $-$      &      $-$       &  $-$     &      $-$      \\
J    &       0.983       &     0.831   &    $-$        &   0.728  &    $-$        &     $-$       \\  
\end{tabular}
\end{ruledtabular}
\end{table}

 be associated with the fact that K$_{\tau}$ is not dependent on the mass number A in the first order.  More generaly, the inclusion of the curvature term in the fits does not dramatically influence the correlation between the rest of the parameters varied in a particular fit. 
 
Several conclusions can be drawn from Table~\ref{tab:curv} and Table~\ref{tab:cor}. First, the minima obtained in the both fits, with and without the curvature term, are each stable. The central values of K$_{\rm 0}$, $c$, K$_{\tau}$  and  K$_{\rm curv}$ remain almost constant and the extent to which they are correlated is only reflected in the errors. Second, the errors in the full fit including the curvature term, due to the insensitivity of the data to this term, are too large to allow practical determination of the four parameters in the expansion (\ref{kanjscurv}).  The fits with one parameter kept constant at its minimum value may indicates that the curvature term is likely to be positive but does not allow deduction of any useful value. For what it is worth, the effect including the curvature term on K$_{\tau}$ is small (of order 8\%), but the considerable changes of K$_{0}$ (33\%) and c (a factor of 2) take them ever further from the currently adopted values. In other words,  according to this analysis, K$_{\rm 0} \sim$ 220 - 240 MeV and $c \sim -1$ cannot be recovered by including the curvature term in the fit.

Finally, we note that our adopted method of fitting allows determination of only two-parameter correlation coefficients. It may be interesting to look for many-parameter correlations based on the two-parameter data. However, it is not clear whether any practically useful information would be obtained.  

We present this analysis as an example of the fitting routines and the trend of outcome of the fits when the curvature term is included.  We maintain as our main results Table~\ref{tab:summary}, obtained using (\ref{kanjs}), keeping in mind that the values of  K$_{\rm 0}$ and the magnitude of $c$ may be even higher.
\section{\label{toy} Incompressibility, surface energy and diffuseness with a "toy" model}.
In this section we explore a possible theoretical foundation for our empirical results suggesting that the magnitude of the ratio of the surface to volume incompressibility is different from one. The surface incompressibility has been investigated in the past by several authors (see e.g. \cite{blaizot1981, treiner1981, myers1995, nayak1990, sharma1989, patra2002}). The main purpose was to find the most realistic relation between the effective incompressibility of a finite nucleus and the GMR energy. It turns out that the changes of surface diffuseness of the nucleus under compression play an important role. Satchler \cite{satchler1973} and Blaizot and Grammaticos \cite{blaizot1981} discussed two modes of vibration, in which either the surface diffuseness remains constant and only the central density and radius are allowed to change or both central density and surface diffuseness vary. Here we explore the role of the surface diffuseness more generally in a simple model in a static (adiabatic) approximation. Clearly dynamical effects and a more comprehensive study of vibration modes in a compressed nucleus are important (see e.g. \cite{brack1983}) and will be a subject of future work.

\subsection{\label{1D} One-Dimensional model (1D)}
Let us assume the energy per particle in symmetric nuclear matter of density $\rho$ to have a simple form
\begin{equation}
W({\rho})=W_{\rm 0}(-2\hat{\rho}+\hat{\rho}^{\rm 2}),
\label{eq:eos}
\end{equation}
where $\hat{\rho}$=$\rho$/$\rho_{\rm NM}$ and W$_{\rm 0}$ and $\rho_{\rm NM}$ are the binding energy and density of symmetric nuclear matter at saturation
\begin{equation}
W(\hat{\rho}=1)=-W_{\rm 0}.
\end{equation}
Note that $\rho_{\rm NM}$ here is the 1D equivalent of the realistic value of saturation density in 3D.  For finite nuclei we have a constraint
\begin{equation}
\int{\rho(r)dr}=A.
\label{eq:intr}
\end{equation}
The energy density $\epsilon$ is written as 
\begin{equation}
\epsilon=\rho W(\rho) + c_{\rm s}\left(\frac{d\rho}{dr}\right)^{\rm 2}/\rho,
\label{eq:inh}
\end{equation}
where the last term is the inhomogeneity term designed to account for surface effects in finite nuclei \cite{weizsacker1935, berg1956}. For a system of non-interacting particles and neglecting any Fermi motion, c$_{\rm S}$ = $\hbar^{\rm 2}$/(8m) where $m$ is the nucleon mass. Berg and Wilets found, in order to obtain a good fit to nuclear properties, that c$_{\rm S}$ should be reduced by a factor between 1/2 and 1/8, dependent on the shape of the nuclear potential used \cite{berg1955}. The total energy is then given as
\begin{eqnarray} 
E(\rho)&=&\int{\epsilon dr}  \nonumber  \\
           &=& -2\frac{W_{\rm 0}}{\rho_{\rm NM}}\int{\rho^{\rm 2}dr} + \frac{W_{\rm 0}}{\rho_{\rm NM}^{\rm 2}}\int{\rho^{\rm 3}dr} \nonumber \\
           &+&c_{\rm s}\int{\rho^{\rm -1}\left(\frac{d\rho}{dr}\right)^{\rm 2}dr}.
\label{eq:etot}
\end{eqnarray}
We take the particle number density to have the Fermi distribution
\begin{equation}
\rho(r)=\frac{\rho_{\rm 0}}{\exp(\frac{r-R}{a})+1},
\end{equation}
where $\rho_{\rm 0}$ is the central density of the nucleus, $a$ is the diffuseness parameter and R = A/(k $\rho_{\rm 0}$) with k = 2 as the integral (\ref{eq:intr}) goes only over positive values of $r$ but the range of $r$ in the 1D model includes both positive and negative values (-$\infty$,+$\infty$). Note that in 1D model the densities $\rho_{\rm 0}$ and $\rho_{\rm NM}$ do not have their physical values but are defined as a number of particles per unit length. The diffuseness parameter $a$ is proportional to the surface thickness of the nucleus. For a Fermi distribution, the 90\% to 10\% thickness is 2 $\log(0.9/0.1) a$ = 4.4 $a$, much larger than $a$ itself. Thus we use the term diffuseness rather than thickness to discuss the surface properties.

It is easy to evaluate the integrals in (\ref{eq:etot}):
\begin{eqnarray}
\int{\rho^{\rm 2}dr} & = &\rho_{\rm 0}^{\rm 2}(R-a) =  \rho_{\rm 0} A (1-\frac{a}{R}) \nonumber \\
\int{\rho^{\rm 3}dr} & = &2\rho_{\rm 0}^{\rm 3}(R-\frac{3}{2}a) = \rho_{\rm 0} A (1-\frac{3a}{2R})
\label{eq:rhoint1}
\end{eqnarray}
up to first order in $a$. For the inhomogeneity term we have                
\begin{equation}
\int{\rho^{\rm -1}\left(\frac{d\rho}{dr}\right)^{\rm 2}dr}=\frac{\rho_{\rm 0}}{2a} = \frac{A}{2aR}.
\label{eq:rhoint2}
\end{equation}
The total energy (\ref{eq:etot}) for the Fermi density distribution takes the form
\begin{eqnarray}
\label{eq:etotfer}
E(\rho_{\rm 0})&=&-E_{\rm vol}(\rho_{\rm 0})+E_{\rm surf}(\rho_{\rm 0}) \\ 
               &=&(-2\hat{\rho}_{\rm 0}+\hat{\rho}_{\rm 0}^{\rm 2})W_{\rm 0}A + (4\hat{\rho}_{\rm 0}-3\hat{\rho}_{\rm 0}^{\rm 2})W_{\rm 0}\frac{aA}{2R} + c_{\rm s}\frac{A}{2aR}, \nonumber  
\end{eqnarray}
where $\hat{\rho_{\rm 0}}$ is in units of nuclear matter density at saturation, $\hat{\rho_{\rm 0}}$ = $\rho_{\rm 0}/\rho_{\rm NM}$. 
If $\rho_{\rm 0}$ = $\rho_{\rm NM}$, the nucleus saturates in the interior, but we lose some binding at the surface. In this case, the total energy $E(\rho_{\rm 0})$ simplifies to
\begin{equation}
E(\rho_{\rm NM})=-W_{\rm 0}A + \frac{W_{\rm 0}aA}{2R} + c_{\rm s}\frac{A}{2aR}.
\label{eq:etotsat}
\end{equation}
The first term in (\ref{eq:etotsat}) is the volume energy
\begin{equation}
E^{\rm 0}_{\rm vol}(\rho_{\rm NM})=W_{\rm 0}A
\label{eq:evolsat}
\end{equation}
and the second and third terms constitute the surface energy.  The second term accounts for deficiency of binding due to the sub-saturation density at the surface, and the last is the inhomogeneity term.  The energy is minimized for a diffuseness parameter $a$
\begin{equation}
a=a_{\rm 0}=\sqrt{c_{\rm s}/W_{\rm 0}}.
\label{eq:a0}
\end{equation}
Inserting (\ref{eq:a0}) into (\ref{eq:etotsat}) we see that in equilibrium the last two terms in (\ref{eq:etotsat}) contribute equally to the surface energy
\begin{equation}
E^{\rm 0}_{\rm surf}(\rho_{\rm NM})=\frac{\sqrt{c_{\rm S}W_{\rm 0}}A}{R_{\rm 0}}=\frac{W_{\rm 0}a_{\rm 0}A}{R_{\rm 0}}=W_{\rm 0}a_{\rm 0} k \rho_{\rm NM},
\label{eq:esursat}
\end{equation}
where $R = R_{\rm 0}$ = A/(k $\rho_{\rm NM}$).
For future use, it is convenient to introduce volume and surface energies as
\begin{equation}
W_{\rm vol}^{\rm 0}=E^{\rm 0}_{\rm vol}/A \qquad\mbox{and}\qquad W_{\rm surf}^{\rm 0}=E^{\rm 0}_{\rm surf}.
\label{eq:w}
\end{equation}

Let us now consider the volume and surface energy at arbitrary central density $\rho_{\rm 0} \neq \rho_{\rm NM}$. We re-write (\ref{eq:etotfer}) using (\ref{eq:esursat}) and $R$=$R_{\rm 0}/\hat{\rho_{\rm 0}}$ as

\begin{eqnarray}
E(\rho_{\rm 0})&=&(-2 \hat{\rho_{\rm 0}} + \hat{\rho_{\rm 0}}^{\rm 2})W_{\rm 0}A  \nonumber \\
&+& E^{\rm 0}_{\rm surf}\left[(4 \hat{\rho_{\rm 0}}^{\rm 2} - 3 \hat{\rho_{\rm 0}}^{\rm 3})\frac{a}{2a_{\rm 0}} + \hat{\rho_{\rm 0}}\frac{a_{\rm 0}}{2a}\right].
\label{eq:etotarb}
\end{eqnarray}
The energy is minimized for a diffuseness parameter $a$, which is now density dependent,
\begin{equation}  
a=a_{\rho}=a_{\rm 0}/\sqrt{4\hat{\rho_{\rm 0}}-3 \hat{\rho_{\rm 0}}^{\rm 2}}.
\label{eq:a}
\end{equation} 
$a$ can expanded about the saturation density $\hat\rho_{\rm 0}$ = 1 ($\hat{\rho_{\rm 0}}$ = 1 + $\delta\rho$) as 
\begin{equation}
a_{\rho}=a_{\rm 0}(1+\delta \rho + 3\delta \rho^{\rm 2} + 7\delta \rho^{\rm 3} + \ldots).
\label{eq:aexp}            
\end{equation}
We see that, to the first order, the toy model predicts the surface diffuseness directly proportional to density, i.e. it  increases with decreasing radius.

We first impose a slight deviation from the equilibrium condition (\ref{eq:etotarb}) and calculate the surface energy at arbitrary central density $\rho_{\rm 0}$ in (\ref{eq:etotarb}) taking $a = a_{\rm 0}$. We get
\begin{equation}
E_{\rm surf}(\rho_{\rm 0})=E^{\rm 0}_{\rm surf}\left(\frac{4\hat\rho_{\rm 0}^{\rm 2} - 3 \hat\rho_{\rm 0}^{\rm 3}}{2}+\frac{\hat\rho_{\rm 0}}{2}\right).
\label{eq:32a0}
\end{equation}  
Expansion in powers of $\delta\rho$ yields
\begin{equation}
E_{\rm surf}(\rho_{\rm 0})=E^{\rm 0}_{\rm surf}(1-\frac{5}{2}\delta \rho^{\rm 2} - \frac{3}{2}\delta \rho^{\rm 3} + \ldots)
\end{equation}
\begin{equation}
K_{\rm surf}=\hat{\rho}_{\rm 0}^{\rm 2}\frac{d^{\rm 2}E_{\rm surf}(\rho_{\rm 0})}{d\hat{\rho}_{\rm 0}^{\rm 2}}{\bigg |_{\rm \rho_{\rm 0}=\rho_{\rm NM},a=a_{\rm 0}}} = -5E^{\rm 0}_{\rm surf} =-5W_{\rm surf}^{\rm 0}.
\label{eq:ksurf}
\end{equation}
Blaizot and Grammaticos \cite{blaizot1981} (for notation see Sec.~5, Eqs.~5.1 and 5.13) use a simple, analytically soluble model based on a density dependent interaction, which for the parameter d = 1 is equivalent to the interaction employed here. For fixed surface diffuseness, K$_{\sigma}$ (equivalent to our K$_{\rm surf}$/W$_{\rm surf}^{\rm 0}$, when K$_{\rm NM}$ is chosen to be close to $18 B$), is the same as our result (\ref{eq:ksurf}).

When the density dependence of the diffuseness $a$ is included and $a = a_{\rm \rho}$ used,  expansion of the surface energy in powers $\delta\rho$ becomes
\begin{eqnarray}
E_{\rm surf}(\rho_{\rm 0})&=&E^{\rm 0}_{\rm surf}\sqrt{4 \hat{\rho_{\rm 0}}^{\rm 3}-3 \hat{\rho_{\rm 0}}^{\rm 4 }} \nonumber  \\
&=&(1-3\delta\rho^{\rm 2}-4\delta\rho^{\rm 3}-\ldots)E^{\rm 0}_{\rm surf}.
\label{eq:esurf}
\end{eqnarray}
Note that the surface energy vanishes for $\hat{\rho_{\rm 0}}$= 4/3. The surface incompressibility then becomes,
\begin{equation}
K_{\rm surf}=\hat{\rho}_{\rm 0}^{\rm 2}\frac{d^{\rm 2}E_{\rm surf}(\rho_{\rm 0})}{d\hat{\rho}_{\rm 0}^{\rm 2}}{\bigg |_{\rm \rho_{\rm 0}=\rho_{\rm NM},a=a_{\rm \rho}}} = -6 E^{\rm 0}_{\rm surf}=-6W_{\rm surf}^{\rm 0}.
\label{eq:ksurf1}
\end{equation}
This result, obtained in our self-consistent approach, is about 30\% higher then K$_{\sigma} \sim$ -4.2, calculated in the scaling approximation ( see Eq. 5.18 for d = 1 in \cite{blaizot1981}).

To determine the ratio of the surface and volume incompressibility, we can also expand the volume energy about the saturation value
\begin{eqnarray}
E_{\rm vol}(\rho_{\rm 0})&=&(-2\hat{\rho_{\rm 0}}+\hat{\rho_{\rm 0}}^{\rm 2})W_{\rm 0}=-(1-\delta\rho^{\rm 2})W_{\rm 0}A \nonumber \\
&=&-(1-\delta\rho^{\rm 2})E^{\rm 0}_{\rm vol}.
\label{eq:evol1}
\end{eqnarray}
and calculate the volume incompressibility
\begin{equation}
K_{\rm vol}=\hat{\rho}_{\rm 0}^{\rm 2}\frac{d^{\rm 2}(E_{\rm vol}(\rho_{\rm 0})/A)}{d\hat{\rho}_{\rm 0}^{\rm 2}}{\bigg |_{\rm \rho_{\rm 0}=\rho_{\rm NM}}} = 2 W_{\rm 0} = 2 W^{\rm 0}_{\rm vol}.
\label{eq:kvol}
\end{equation}
Combining (\ref{eq:ksurf}) and (\ref{eq:kvol}) we obtain the ratio $c$ of the surface to volume incompressibility at the saturation density in the case of diffuseness independent from density 
\begin{equation}
c = \frac{K_{\rm surf}}{K_{\rm vol}} = -\frac{5}{2} \frac{W_{\rm surf}^{\rm 0}}{W_{\rm vol}^{\rm 0}}.
\label{eq:c52}
\end{equation}
If the density dependence of $a$ is included and the equilibrium condition satisfied, the ratio increases to
\begin{equation}
c=\frac{K_{\rm surf}}{K_{\rm vol}}  = -3 \frac{W_{\rm surf}^{\rm 0}}{W_{\rm vol}^{\rm 0}}.
\label{eq:c3}
\end{equation}
                             
\subsection{\label{3D} D-Dimensional model} 
Having demonstrated the method of calculation of the surface to volume incompressibility ratio in the 1D model, it is straightforward to extend the model to any number of dimensions. In particular, the D = 3 model is of interest because it can be compared with actual data for finite nuclei. 
In the D$\neq$1 case special attention must be paid to the question of self-consistency of the model, requiring that the surface energy as a function of the bulk density $\rho_{\rm 0}$ is stationary at saturation \cite{myers1969, blaizot1981}. In other words, the term linear in $\delta \rho$ in the expansion of surface energy in terms of $\delta\rho$ must vanish. Fulfillment of this condition depends on the choice of the density dependence of the inhomogeneity term in (\ref{eq:inh}). If the term is inversely proportional to density (\ref{eq:inh}) the condition is automatically satisfied for the case D = 1 (\ref{eq:esurf}) but violated for D $\neq$ 1. We will examine this point in more detail.

For D $\neq$ 1 the total energy is given by a generalization of (\ref{eq:etot}). As for the inhomogeneity term, we are anticipating that self-consistency requires a different power of the density depending on the number of dimensions and that its strength c$_{\rm S}$ will be different than in the 1D case. Also, the condition (\ref{eq:intr}) has to be modified. We have $\int F(r)d^{\rm D}r = \int_{\rm 0}^\infty k D r^{\rm D-1} dr$. For F(r) = $\rho_{\rm 0}$ up to r = R (\ref{eq:intr}) it becomes $\int\rho d^{\rm D}r = k \rho_{\rm 0} R^{\rm D} = A$. It follows that in the D $\neq$ 1 case R = R$_{\rm 0}/\hat\rho_{\rm 0}^{\rm 1/D}$ and R$_{\rm 0}$ = (A/(k$\rho_{\rm NM}))^{\rm 1/D}$. We note that k = 4$\pi$/3 for D = 3.
A straightforward calculation shows that for D $\neq$ 1 and the inhomogeneity term inversely proportional to density in the form (\ref{eq:inh}), the integrals (\ref{eq:rhoint1}) become
\begin{eqnarray}
\int{\rho^{\rm 2}d^{\rm D}r} & = k\rho_{\rm 0}^{\rm 2}(R-a) =  \rho_{\rm 0} A (1-\frac{Da}{R}) \\  
\int{\rho^{\rm 3}d^{\rm D}r} & = k\rho_{\rm 0}^{\rm 3}(R-\frac{3}{2}a) = \rho_{\rm 0}A (1-\frac{3Da}{2R})
\end{eqnarray}
and
\begin{equation}
\int{\rho^{\rm -1}\left(\frac{d\rho}{dr}\right)^{\rm 2}d^{\rm D}r}=\frac{\rho_{\rm 0}k D R^{\rm D-1}}{2a} = \frac{D\rho_{\rm 0}^{\rm 1/D} A^{\rm 1-1/D} k^{\rm 1/D}}{2a}.
\label{eq:rhointD}
\end{equation}
The expression for the surface energy at arbitrary central density reads
\begin{fleqn}[0pt]
\begin{alignat}{2}
 &E^{\rm D}_{\rm surf}(\rho_0)  \nonumber \\
&=E^{\rm 0,D}_{\rm surf}(\rho_{\rm NM})\left[(4\hat\rho_{\rm 0}^{\rm 1+1/D}-3\hat\rho_{\rm 0}^{\rm 2+1/D})\frac{a}{2a^{\rm D}_{\rm 0}} + \hat\rho_{\rm 0}^{\rm 1/D}\frac{a^{\rm D}_{\rm 0}}{2a}\right].
\label{eq:esurfd}
\end{alignat}
\end{fleqn}
$E^{\rm 0,D}_{\rm surf}(\rho_{\rm NM})$ is the surface energy for the case $\rho_{\rm 0}$ = $\rho_{\rm NM}$ in equilibrium with $a = a^{\rm D}_{\rm 0}=\sqrt{c^{\rm D}_{\rm S}/W_{\rm 0}}$, 
\begin{equation}
E^{\rm 0,D}_{\rm surf}(\rho_{\rm NM})=\frac{DW_{\rm 0}a^{\rm D}_{\rm 0}A}{R_{\rm 0}}=DW_{\rm 0}a_{\rm 0}A^{\rm 1-1/D}k^{\rm 1/D}\rho_{\rm NM}^{\rm 1/D}.
\label{eq:eoD}
\end{equation}

Minimization of (\ref{eq:esurfd}) with respect to $a$ yields the equilibrium value of the diffuseness with the same density dependence as in the D = 1 case
\begin{equation}
a=a_{\rm \rho}^{\rm D}= a^{\rm D}_{\rm 0}/\sqrt{4\hat\rho_{\rm 0}-3\hat\rho_{\rm 0}^{\rm 2}}
\end{equation}
and the surface energy at equilibrium becomes
\begin{equation}
E^{\rm D}_{\rm surf}(\rho_0)=E^{\rm 0,D}_{\rm surf}\sqrt{4\hat\rho_{\rm 0}^{\rm 1+2/D}-3\hat\rho_{\rm 0}^{\rm 2+2/D}}.
\end{equation}
Expansion in powers of $\delta\rho$ leads, to the first order,
\begin{equation}
E^{\rm D}_{\rm surf}(\rho_0)=E^{\rm 0,D}_{\rm surf}(1+(-1+1/D)\delta\rho + \ldots),
\end{equation}
which violates the condition of self-consistency. 

Considering a general form of the density dependence of the inhomogeneity term $\rho^{\rm x}\left(\frac{d\rho}{dr}\right)^{\rm 2}$ and repeating the derivation above, it can be shown that the condition of self-consistency is satisfied for $x = 1 - 2/D$. In this case the integral over the inhomogeneity term takes the form
\begin{equation}
\int{\rho^{\rm 1-2/D}\left(\frac{d\rho}{dr}\right)^{\rm 2}d^{\rm D}r}=\frac{D^{\rm 3}\rho_{\rm 0}^{\rm (3-2/D)}k R^{\rm D-1}}{(3D-2)(4D-2)a}, 
\label{eq:rhointDs}
\end{equation}
and the expression for the surface energy in D dimensions becomes (compare (\ref{eq:etotfer}))
\begin{eqnarray}
\label{eq:esurfD}
E_{\rm surf}(\rho_{\rm 0})= (4\hat{\rho}_{\rm 0}-3\hat{\rho}_{\rm 0}^{\rm 2})W_{\rm 0}D\frac{Aa}{2R} \nonumber \\
 + c^{\rm D}_{\rm S}\frac{D^{\rm 3}\rho_{\rm 0}^{\rm 2-2/D}}{(3D-2)(2D-1)}\frac{A}{2aR}.
\end{eqnarray}
If $\rho_{\rm 0}$ = $\rho_{\rm NM}$, the surface energy is minimized for diffuseness parameter $a$
\begin{equation}
a^{\rm D}_{\rm 0}=\sqrt{c^{\rm D}_{\rm S}/W_{\rm 0}}\frac{D\rho_{\rm NM}^{\rm 1-1/D}}{\sqrt{(3D-2)(2D-1)}}
\end{equation}
and the surface energy in equilibrium is given as
\begin{equation}
E^{\rm 0,D}_{\rm surf}(\rho_{\rm NM})=\frac{DW_{\rm 0}a^{\rm D}_{\rm 0}A}{R_{\rm 0}}=DW_{\rm 0}a^{\rm D}_{\rm 0}A^{\rm 1-1/D}k^{\rm 1/D}\rho_{\rm NM}^{\rm 1/D}.
\end{equation}
Minimization of the surface energy at arbitrary density yields the diffuseness parameter $a$, which is now density dependent, equal to
\begin{equation}
a=a^{\rm D}_{\rm \rho}=a^{\rm D}_{\rm 0}\hat\rho^{\rm 1-1/D}_{\rm 0}/\sqrt{4\hat\rho_{\rm 0}-3\hat\rho_{\rm 0}^{\rm 2}}
\end{equation}
and the surface energy in equilibrium takes the form
\begin{equation}
E^{\rm D}_{\rm surf}=E^{\rm 0,D}_{\rm surf}\left[(4\hat{\rho}_{\rm 0}^{\rm 1+1/D}-3\hat{\rho}_{\rm 0}^{\rm 2+1/D})\frac{a^{\rm D}_{\rm \rho}}{2a^{\rm D}_{\rm 0}} + \hat{\rho}_{\rm 0}^{\rm 2-1/D}\frac{a^{\rm D}_{\rm 0}}{2a^{\rm D}_{\rm \rho}}\right].
\label{eq:esurfD2}
\end{equation}
If we neglect the density dependence of the diffuseness and calculate the surface energy (\ref{eq:esurfD2}) at $a^{\rm D}_{\rm \rho}=a^{\rm D}_{\rm 0}$ the surface energy becomes
\begin{equation}
E^{\rm D}_{\rm surf}=E^{\rm 0,D}_{\rm surf}\left[2\hat{\rho}_{\rm 0}^{\rm 1+1/D}-\frac{3}{2}\hat{\rho}_{\rm 0}^{\rm 2+1/D} + \frac{1}{2}\hat{\rho}_{\rm 0}^{\rm 2-1/D}\right].
\label{eq:esurfDa0}
\end{equation}
When the density dependence of the diffuseness is included $a^{\rm D}=a^{\rm D}_{\rm \rho}$, we obtain for the surface energy
\begin{equation}
E^{\rm D}_{\rm surf}=E^{\rm 0,D}_{\rm surf}\sqrt{4\hat{\rho_{\rm 0}}^{\rm 3}-3 \hat{\rho_{\rm 0}}^{\rm 4}}
\label{eq:edfinal}
\end{equation}
The expressions for the volume and surface incompressibility depend on the number of dimensions as
\begin{fleqn}[0pt]
\begin{alignat}{2}
  & K_{\rm vol}=D^{\rm 2}\hat{\rho}_{\rm 0}^{\rm 2}\frac{d^{\rm 2}(E_{\rm vol}/A)}{d\hat{\rho}_{\rm 0}^{\rm 2}} \nonumber  \\ 
  &K_{\rm surf}=D^{\rm 2}\hat\rho^{\rm 2}_{\rm 0}\frac{d^{\rm 2}(E^{\rm D}_{\rm surf}(\rho_{\rm 0})/A^{\rm 1-1/D})}{d\hat\rho^{\rm 2}_{\rm 0}}.
\label{eq:kvold}
\end{alignat}
\end{fleqn}
Taking into account that $E_{\rm vol}$ (\ref{eq:etotfer}) is the same in all dimensions we obtain from (\ref{eq:kvol})
\begin{equation}
K_{\rm vol}=2 D^{\rm 2} W_{\rm vol}^{\rm 0}.
\label{eq:kvold1}
\end{equation}
The surface incompressibility for constant and density dependent diffuseness can be written as
\begin{equation}
K_{\rm surf}=-5D^{\rm 2} E^{\rm 0,D}_{\rm surf}\quad\mbox{and}\quad K_{\rm surf}=-6 D^2 E^{\rm 0,D}_{\rm surf}.
\label{eq:comprfin}
\end{equation}
Using (\ref{eq:kvold1}), (\ref{eq:comprfin}) and 
\begin{equation}
W^{\rm 0,D}_{\rm surf} = E^{\rm 0,D}_{\rm surf}/{\rm A}^{\rm 1-1/D}
\label{eq:w0d}
\end{equation}
we finally obtain ratio $c$ for a constant and density dependent diffuseness
\begin{equation}
c=-\frac{5}{2}\frac{W_{\rm surf}^{\rm 0,D}}{W_{\rm vol}^{\rm 0}}\qquad\mbox{and}\qquad c=-3 \frac{W_{\rm surf}^{\rm 0,D}}{W_{\rm vol}^{\rm 0}}.
\label{eq:cd} 
\end{equation}
We can now evaluate the expressions for D = 3. The surface energy at saturation density can be estimated, taking $W_{\rm 0}$ = 16 MeV, $a^{\rm D}_{\rm 0}$ = 0.5 fm and $\rho_{\rm NM}$ = 0.16 fm$^{\rm -3}$, to be $\sim$ 21 A$^{\rm 2/3}$ MeV, which is in agreement to the surface energy coefficient in the FRDM obtained from fit to nuclear masses \cite{moller1995}. The volume energy does not depend on whether the diffuseness is constant or density dependent. It follows that the ratio of  $W_{\rm surf}^{\rm 0,D}$ and $W_{\rm vol}^{\rm 0}$ is about 21/16 $\sim$ 1.3. This leads to values of $c$ for diffuseness constant $c \sim -3$ and for density dependent $c \sim -4$. The latter value are somewhat more negative than the range of $c$ obtained from the analysis of experimental data in this work $-2.4 < c < -1.6$. However, as shown in the next section, incorporating a  second order correction to the finite radius of the nucleus brings the value of $c$ in line with the result of this work.

\subsection{Finite radius correction}
In Secs.~\ref{1D} and \ref{3D}, (Eqs.~\ref{eq:etotarb},~\ref{eq:esurf} and \ref{eq:edfinal}) we derived the following result for the toy model total energy including both volume and surface contributions:
\begin{equation}
E=W_0 A (-2\hat{\rho}_{\rm 0}+\hat{\rho}^{\rm 2}_{\rm 0})+E^{\rm 0,D}_{\rm surf}\sqrt{4\hat{\rho_{\rm 0}}^{\rm 3}-3 \hat{\rho_{\rm 0}}^{\rm 4}}.
\label{eq:rtotal}
\end{equation}
In this derivation, we considered only the effect of the density dependence of the surface diffuseness $a$ on the value of the surface incompressibility K$_{\rm surf}$. This did not affect the value of the volume incompressibility which depended only on the strength of the nucleon interaction W$_{\rm 0}$ at saturation. We will now consider a more general case in which, in addition, the total energy and, consequently, both the volume and surface incompressibilities, are dependent on the changing radius under compression in a complementary way to the derivation by Blaizot et al. \cite{blaizot1981}.

 Eq.~\ref{eq:rtotal} can be written in terms of the following dimensionless quantities:
\begin{eqnarray}
\hat{\rho_{\rm 0}}&=&(R/R_{\rm 0})^{\rm -D}=r^{\rm -D} \\
r&=&1+\delta{r}\\
\epsilon&=&E/(W_{\rm 0} A) \\
\alpha&=&E^{\rm 0,D}_{\rm surf}/(W_0 A) \\
\kappa_{\rm A}&=&K_{\rm A}/(W_{\rm 0} A),
\end{eqnarray}
where K$_{\rm A}$ is the total incompressibility of the finite nucleus. The energy $\epsilon$ and its expansion around r = 1 now reads
\begin{eqnarray}
\epsilon&=&(-2 r^{\rm -D} + r^{\rm -2D})+\alpha \sqrt{4r^{\rm -3D}-3r^{\rm -4D}} \nonumber  \\
&=&-1+\alpha+D^{\rm 2}(1-3\alpha)\delta r^{\rm 2}.
\label{repsilon}
\end{eqnarray}
The finite nucleus incompressibility is then given by:
\begin{equation}
\kappa_{\rm A}=r^{\rm 2}\frac{d^{\rm 2}\epsilon}{dr^{\rm 2}}(r=1)=2D^{\rm 2}-6D^{\rm 2}\alpha+\ldots=2D^{\rm 2}(1-3\alpha).
\end{equation}
We see that the ratio of the surface and volume contributions to the incompressibility $\kappa_{\rm A}$ is again equal to -3, independent of dimensions, as already shown in Secs.~\ref{1D} and \ref{3D}.

However, when the finite radius correction is included, we get instead of (\ref{repsilon}) the expression
\begin{eqnarray}
 &\epsilon=(-2 r^{\rm -D} + r^{\rm -2D})+\alpha \sqrt{4r^{\rm -3D}-3r^{\rm -4D}} r^{\rm D-1} \nonumber  \\
 &=-1+\alpha+(D-1)\alpha\delta r  \nonumber \\
 &+(D^{\rm 2}-(\frac{5}{2}D^{\rm 2}+\frac{3}{2}D-1)\alpha\delta r^{\rm 2}+\ldots.
\end{eqnarray}
For D$>$1, the radius correction reduces the equilibrium radius where the energy has a minimum to
\begin{equation}
r_{\rm eq}=1-\frac{(D-1)\alpha}{2D^{\rm 2}}+O(\alpha^{\rm 2})
\end{equation}
which leads to the expression of the finite nucleus incompressibility
\begin{eqnarray}
\kappa_{\rm A}&=&r^{\rm 2}\frac{d^{\rm 2}\epsilon}{dr^{\rm 2}}(r=r_{\rm eq})   \nonumber  \\
                      &=&2D^{\rm 2}-(2D^{\rm 2}+5D-1)\alpha + \ldots    \nonumber  \\
\kappa_{\rm A}&=&2-6\alpha\qquad\mbox{    (D=1)}  \nonumber \\
\kappa_{\rm A}&=&18-32\alpha\qquad\mbox{(D=3)}
\label{eq:kapa}
\end{eqnarray}
In comparison with (\ref{eq:cd}) we see that for D = 1 the coefficient in the ratio of surface to volume incompressibility is again -3, but for the realistic D = 3 case it is reduced to -16/9 = -1.78. Multiplication by the ratio of the surface and volume energy at saturation 21/16 finally yields c = -2.34 which is compatible with the empirical results obtained from the analysis of GMR energies presented in this work.
\section{\label{micro}Microscopic models}
As mentioned in the introduction and demonstrated in Table~\ref{tab:surv}, there has been considerable effort spent on developing microscopic models of the breathing mode of finite nuclei and its dependence on $K_{\rm 0}$. However, all microscopic calculations to date produce results dependent upon details of the model and the adopted effective nucleon-nucleon interaction. Recent investigation of 241 parameterizations of the Skyrme interaction by Dutra et al. \cite{dutra2012} and of 147 Lagrangians used in RMF models \cite{dutra2013} showed conclusively that there is a large variation in performance of these models in nuclear matter which has consequences for the breathing mode of finite nuclei since it depends on the incompressibility of symmetric nuclear matter.

\begin{figure}
\centerline{\epsfig{file=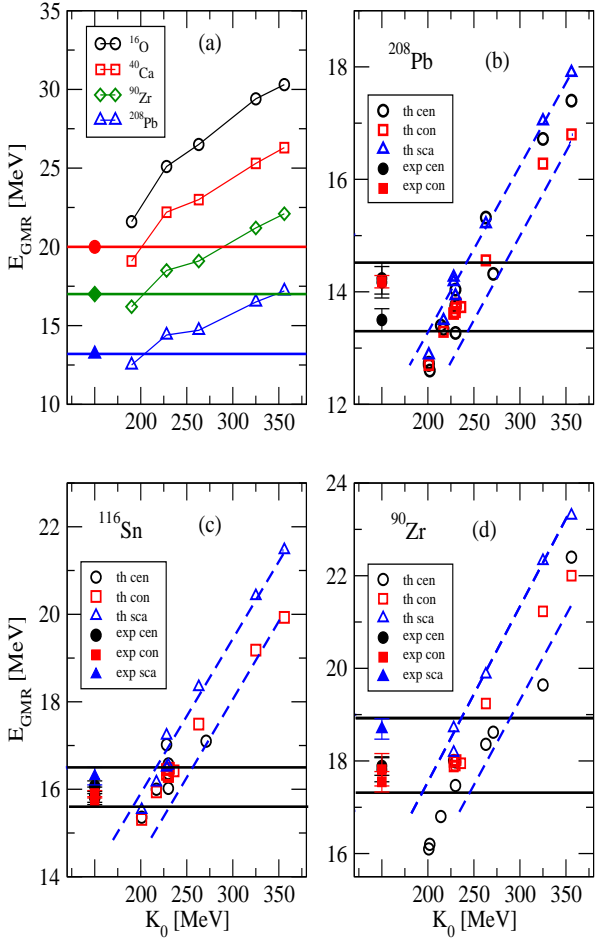,height=15cm,width=10cm}}
\caption{\label{fig:blaizot}(Color on-line) (a) Comparison of experimental and calculated GMR energies as a function of $K_{\rm 0}$  as presented in (\cite{blaizot1976}) and using current experimental and calculated values for (b) $^{\rm 208}$Pb, (c) $^{\rm 116}$Sn and (d) $^{\rm 90}$Zr. Experimental data are taken from \cite{youngblood1999, youngblood2004a} ($^{\rm 90}$Zr), \cite{youngblood1999, youngblood2004, li2010} ($^{\rm 116}$Sn) and 
\cite{youngblood1999, youngblood2004, uchida2003} ($^{\rm 208}$Pb). Horizontal lines (black) depict ranges of currently available GMR energies in $^{\rm 90}$Zr, $^{\rm 116}$Sn and  $^{\rm 208}$Pb, the dashed lines (blue) illustrate the range of theoretical predictions. Details of the calculation with references are given in Table~\ref{tab:blaizot}. For more information see text.}
\end{figure}

Based on the version of microscopic models and experimental GMR energies for $^{\rm 40}$Ca, $^{\rm 90}$Zr and $^{\rm 208}$Pb nuclei available in the 1970's,  Blaizot et al. \cite{blaizot1976} obtained a value of $K_{\rm 0}$ which has been accepted as standard for many years. Their approach was to use each of the five effective interactions B1, D1, Ska, SIV and SIII to calculate $K_{\rm 0}$ in nuclear matter and also, with the HF+RPA model, the GMR energies of these nuclei. The results plotted against each other gave a relatively smooth empirical relationship which bore out the expectation that lower $K_{\rm 0}$ was associated with lower E$_{\rm GMR}$. Intersection of the (single) experimental E$_{\rm GMR}$ with these empirical curves yielded the result $K_{\rm 0}$ = (210$ \pm$ 30) MeV. We reproduce this analysis in the upper left panel of Fig.~\ref{fig:blaizot}. Subsequent theoretical work has focused on attempts to obtain consistency with both this range for K$_{\rm 0}$ and the experimental E$_{\rm GMR}$ values. We note that the approximately linear relation between K$_{\rm 0}$ and E$_{\rm GMR}$ was also obtained using RMF + GCM (Generator Coordinate Method) \cite{stoitsov1994, vretenar1997}.

Using modern experimental data and microscopic theory (see Table~\ref{tab:blaizot} for details) we have repeated the analysis of Blaizot et al. \cite{blaizot1976}. We discarded $^{\rm 40}$Ca since it has become apparent that GMR strength is fragmented in lighter nuclei and the nature of collectivity may be different in light and heavy nuclei \cite{lui2006} and have added $^{\rm 116}$Sn. Of the five interactions used by Blaizot et al., we retained four SIII, SIV, Ska and D1 but have discarded B1, since is not density dependent and seriously under-binds $^{\rm 16}$O, $^{\rm 40}$Ca, $^{\rm 90}$Zr and  $^{\rm 208}$Pb (by 27-, 78-, 186- and 445 MeV, respectively). To these we have added SGII, SkP
and SKT5 which all have low incompressibility and also we used the KDE0v1 force \cite{agrawal2005}, one of the five Skyrme forces that passed all the constraints currently available on nuclear matter \cite{dutra2012}. Finally, to represent RMF models, FSUGold, NL3 and Hybrid \cite{piekarewicz2009} and BSP, IUFSU and IUFSU* Lagrangians \cite{agrawal2012} were added. For each nucleus and interaction the value of $K_{\rm 0}$ and up to three values of E$_{\rm GMR}$ (calculation the different centroid, constrained and scaling models) are plotted in three panels of Fig.~\ref{fig:blaizot}. In this sense, the spread of E$_{\rm GMR}$ values is a measure of a "theoretical error" in the best current model calculations. 
Also, at the left hand edge of each panel we show experimental values of E$_{\rm GMR}$ for the nucleus. All available experimental data evaluated with modern analysis methods are included. Several groups have obtained multiple experimental results and we find no reason to exclude any.

In each panel full horizontal lines show the spread of current experiment. As found by \cite{blaizot1976} the values of E$_{\rm GMR}$ and $K_{\rm 0}$ show a consistent variation for each isotope with the single line drawn
\begin{table}
\caption{\label{tab:blaizot} E$_{\rm GMR}$ of $^{\rm 90}$Zr and $^{\rm 208}$Pb, calculated in a HF+RPA model for Skyrme parameterizations SIII, SIV, Ska, and D1 Gogny force, used by Blaizot et al. \cite{blaizot1976}, in comparison with modern calculation using KDE0v1 \cite{anders2011, agrawal2011} and SGII, SkT5, SkP \cite{reinhard2012} forces. HF results for $^{\rm 90}$Zr, $^{\rm 116}$Sn and $^{\rm 208}$Pb for SIII, SIV and Ska \cite{agrawal2004, agrawal2012} are added for completeness. Results for $^{\rm 208}$Pb from HFB+QRPA with SLy4, SkM* and SkP Skyrme forces \cite{cao2012} are also given, as well as RMF+RPA values for GMR energies obtained with FSUGold, NL3 and Hybrid Lagrangians \cite{piekarewicz2009} and constrained RMF+GCM with NL2, NL-SH, NL-S1, NL3 and NL1 Lagrangians. In addition, calculations for $^{\rm 116}$Sn \cite{cao2012, anders2011, agrawal2011, piekarewicz2009}, are also shown. All entries are in MeV. For more explanation see text.}
\begin{ruledtabular}
\begin{tabular}{llllll}
  Skyrme    &   $K_{\rm 0}$  &   E$_{\rm GMR}$ & E$_{\rm GMR}$  &  E$_{\rm GMR}$  &  Method \\
  force       &                     &   ($^{\rm 208}$Pb)  &  ($^{\rm 116}$Sn)   &  ($^{\rm 90}$Zr)   &  \\   \hline
   NL2        &    399      &      16.6  &        &     21.9   &  centroid \cite{vretenar1997}   \\
   SIII       &    356      &      17.2  &        &     22.1   &  centroid \cite{blaizot1976}  \\ 
              &             &      17.90 & 21.47  &     23.30  &  (m$_{\rm 3}$/m$_{\rm 1}$)$^{\rm 1/2}$ \cite{agrawal2011}\\
              &             &      16.80 & 19.93  &     22.00  &   (m$_{\rm 1}$/m$_{\rm -1}$)$^{\rm 1/2}$ \cite{agrawal2011}\\ 
              &             &      17.40 &        &     22.4   &   centroid \cite{reinhard2012} \\ 
   NL-SH      &    355      &      15.0  &        &     19.5   &   centroid \cite{vretenar1997}   \\ 
              &             &      15.8  &        &     20.16  &   centroid \cite{stoitsov1994}   \\       
   SIV        &    325      &      16.5  &        &     21.2   &   centroid \cite{blaizot1976}      \\
              &             &      17.04 & 20.42  &     22.32  &  (m$_{\rm 3}$/m$_{\rm 1}$)$^{\rm 1/2}$ \cite{agrawal2011}\\
              &             &      16.28 & 19.18  &     21.23  &  (m$_{\rm 1}$/m$_{\rm -1}$)$^{\rm 1/2}$ \cite{agrawal2011} \\
              &             &      16.72 &        &     19.64  &  centroid \cite{reinhard2012} \\
   NL-S1      &    296      &      13.4  &        &     17.6   &  centroid \cite{stoitsov1994}  \\
   NL3        &    271      &      14.32 & 17.10  &     18.62  &  centroid  \cite{piekarewicz2009} \\
              &             &      13.0  &        &     16.9   &  centroid \cite{vretenar1997}   \\
   Ska        &    263      &      14.7  &        &     19.1   &  centroid \cite{blaizot1976}   \\
              &             &      15.21 & 18.34  &     19.87  &  (m$_{\rm 3}$/m$_{\rm 1}$)$^{\rm 1/2}$ \cite{agrawal2011}\\
              &             &      14.56 & 17.49  &     19.24  &  (m$_{\rm 1}$/m$_{\rm -1}$)$^{\rm 1/2}$ \cite{agrawal2011} \\
              &             &      15.32 &        &     18.36  &  centroid \cite{reinhard2012} \\
   IUFSU*     &    236      &      13.73 & 16.42  &     17.95  &  (m$_{\rm 1}$/m$_{\rm -1}$)$^{\rm 1/2}$  \cite{agrawal2012}\\ 
   IUFSU      &    231      &      13.79 & 16.48  &     18.02  &  (m$_{\rm 1}$/m$_{\rm -1}$)$^{\rm 1/2}$  \cite{agrawal2012}\\ 
   BSP        &    230      &      13.64 & 16.32  &     17.90  &   (m$_{\rm 1}$/m$_{\rm -1}$)$^{\rm 1/2}$  \cite{agrawal2012}\\ 
   SLy5       &    230      &      13.77 & 15.36  &            &   centroid  \cite{cao2012}\\
   SLy5       &             &      13.93 & 16.54  &            &   (m$_{\rm 3}$/m$_{\rm 1}$)$^{\rm 1/2}$ \cite{cao2012}\\
   SLy5       &             &      13.71 & 16.29  &            &   (m$_{\rm 1}$/m$_{\rm -1}$)$^{\rm 1/2}$ \cite{cao2012} \\
   FSUGold    &    230      &      14.04 & 16.58  &    17.98   &   centroid  \cite{piekarewicz2009} \\ 
   Hybrid     &    230      &      13.27 & 16.02  &    17.47   &   centroid \cite{piekarewicz2009} \\
   D1         &    228      &      14.4  &        &    18.5    &   centroid  \cite{blaizot1976}   \\
   KDE0v1     &    228      &     13.73  & 17.02  &     18.01  &   centroid \cite{anders2011} \\
   KDE0v1     &             &     14.18  & 16.50  &     18.18  &   (m$_{\rm 3}$/m$_{\rm  1}$)$^{\rm 1/2}$  \cite{anders2011} \\
   KDE0v1     &             &      13.61 & 16.34  &     17.88  &   (m$_{\rm 1}$/m$_{\rm -1}$)$^{\rm 1/2}$ \cite{anders2011}\\
   KDE0v1     &             &     14.26  & 17.23  &     18.71  &   (m$_{\rm 3}$/m$_{\rm  1}$)$^{\rm 1/2}$  \cite{agrawal2011} \\  
   KDE0v1     &             &     13.62  & 16.45  &     17.98  &   (m$_{\rm 1}$/m$_{\rm -1}$)$^{\rm 1/2}$ \cite{agrawal2011}\\ 
   SkM*       &    217      &     13.34  & 16.00  &            &    centroid  \cite{cao2012}\\
   SkM*       &             &     13.49  & 16.16  &            &    (m$_{\rm 3}$/m$_{\rm 1}$)$^{\rm 1/2}$ \cite{cao2012}  \\
   SkM*       &             &     13.29  & 15.94  &            &    (m$_{\rm 1}$/m$_{\rm -1}$)$^{\rm 1/2}$ \cite{cao2012} \\
   SGII       &    214      &     13.40  &        &     16.80  &    centroid \cite{reinhard2012} \\
   NL1        &    212      &     11.00  &        &     14.1   &    centroid \cite{vretenar1997}   \\
              &             &     11.7   &        &     14.7   &    centroid \cite{stoitsov1994}  \\
   SkT5       &    202      &     12.60  &        &     16.20  &    centroid \cite{reinhard2012}  \\
   SkP        &    201      &     12.80  &        &     16.10  &    centroid \cite{reinhard2012}  \\
   SkP        &             &     12.74  & 15.36  &            &    centroid  \cite{cao2012}\\
   SkP        &             &     12.88  & 15.53  &            &   (m$_{\rm 3}$/m$_{\rm 1}$)$^{\rm 1/2}$ \cite{cao2012} \\ 
   SkP        &             &     12.70  & 15.31  &            &   (m$_{\rm 1}$/m$_{\rm -1}$)$^{\rm 1/2}$ \cite{cao2012} \\ 
\end{tabular}
\end{ruledtabular}
\end{table}
 by Blaizot et al. replaced by bands limited by dashed lines. The intersection line of these bands with the range of experiment now replaces the simple crossings shown in the Blaizot et al. figure. Whereas the Blaizot et al. figure gave rise to $K_{\rm 0}$ all close to 200 MeV, up-to-date figures show $K_{\rm 0}$ ranges 180 - 270 MeV for $^{\rm 90}$Zr and 200 - 280 MeV for $^{\rm 208}$Pb. 
 
\begin{figure}
\centerline{\epsfig{file=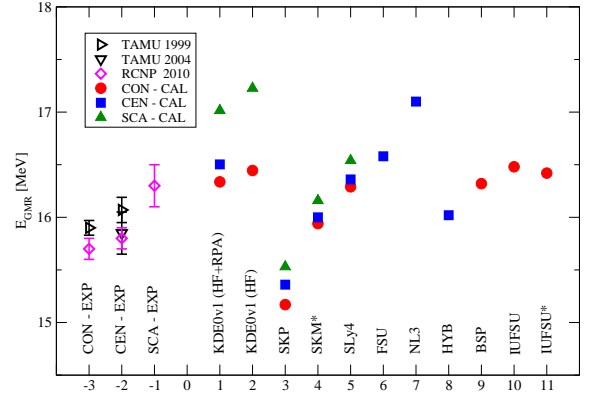,height=7cm,width=9cm}}
\caption{\label{fig:sn116} (Color on-line) Comparison of experimental and theoretical GMR energies in $^{\rm 116}$Sn for constrained (CON), centroid (CEN) and scaling (SCA) approximations. Experimental data are taken from \cite{youngblood1999} (TAMU1999), \cite{youngblood2004} (TAMU2004) and \cite{li2010} (RCNP2010). Hartree-Fock (HF)+RPA with KDE0v1 \cite{anders2011}, HF with KDE0v1, SIII, SIV, SkA Skyrme interactions \cite{agrawal2011, sil2006}  and the Hartree-Fock-Bogolyubov + QRPA \cite{cao2012} with SkP, SkM* and SLy4 Skyrme interactions and RMF with FSU, NL3 and Hybrid \cite{piekarewicz2009} and BSP, IUFSU and IUFSU* Lagrangians. Tick labels on x-axis indicate experimental data (-3, -2, -1) and various calculations (1 - 11).}
\end{figure}

$^{\rm 116}$Sn was added to our analysis because there has been some concern that microscopic models have difficulty in calculating 
E$_{\rm GMR}$ in agreement with experiment (\cite{cao2012, vesely2012, piekarewicz2010} and references therein). The $^{\rm 116}$Sn panel yields $K_{\rm 0}$ in the range 180 -260 MeV in good agreement with $^{\rm 90}$Zr and $^{\rm 208}$Pb. In Fig.~\ref{fig:sn116} the theoretical calculations of the GMR energies of $^{\rm 116}$Sn with a variety of models in detail. The selected models are those which give best agreement in E$_{\rm GMR}$ in  $^{\rm 90}$Zr and $^{\rm 208}$Pb. The model spread about 2 MeV spans the experimental range which does not suggest a peculiar character of Sn nuclei.
\section{\label{concl}Discussion and conclusions}
The main finding of this work is that the macroscopic model, using expansion of K$_{\rm A}$ in terms of A$^{\rm -1/3}$ and $\beta$, is sensitive to K$_{\rm 0}$ and the K$_{\rm surf}$/K$_{\rm vol}$ ratio, provided the expansion is written in such a way that K$_{\rm 0}$ is independent of $A$. This sensitivity is revealed by employing a MESH fit combined with the MINUIT fit. The fitting technique, used for the first time to extract coefficients of the leptodermous expansion of K$_{\rm A}$, has proven more efficient than the fitting procedures used before, especially in dealing with correlations between fitted parameters and including the effect of these correlations into calculation of errors. 

As we did not find a convincing reason for eliminating data differing by more than several standard deviations, available GMR energies were divided into groups, which were analyzed separately. The results within each group showed general consistency, however this procedure revealed some variations in extracted parameters. Other contributions to uncertainty were the question of adopting matter or charge radius in the calculation of $K_{\rm A}$ and the error in the theoretical value of $K_{\rm coul}$. Examining Table~\ref{tab:summary}, it is satisfying to see that neither of these uncertainties appreciably affects the values of $K_{\rm 0}$ and the ratio $c$, extracted from a fit to a particular data set. On the other hand, there is a systematic trend to higher values of (negative) $K_\tau$ when charge radii are used. The increased error in $K_{\rm coul}$ reflects in the increased error of $K_\tau$ but does not affect the range of best fit $K_\tau$ values.

The accuracy claimed for experimental GMR energies, extracted from moments of the strength function, improved considerably as compared to earlier results based on determination of the GMR peak position and width from fitting using a Gaussian or Lorentzian function. Ironically, the consistency between results obtained by different researchers, did not improve. On the contrary, the differences in the rather complex analysis of individual experiments, became more apparent.

Each entry in Table~\ref{tab:summary} represents an independent data set. However, the sets are not statistically distributed and thus the results cannot be averaged. Although \textit{a priori} all options for all groups should be taken into consideration, we choose, as representative, the results obtained for matter radii and $\Delta K_{\rm coul}$ = 0.7 MeV for five groups of data, not including the TAMU0-M which yields extreme values of K$_{\rm 0}$ and $K_\tau$.  

We deduce as our final results that $K_{\rm 0}$ lies in the range  250 -- 315 MeV and  the ratio of the surface and volume coefficients $c = K_{\rm surf}/K_{\rm vol}$ is between -1.6 and -2.4. Limits on the isospin coefficient $K_{\rm \tau}$ have been determined as  -840 $ < K_{\rm \tau} < $ -350 MeV. We wish to stress that the scatter of results in Table~\ref{tab:summary} is totally due to differences in experimetal data used in the fits and is \textit {not} because of correlations between the parameters in the fitting procedure. Correlations are reflected only in the quoted errors.

It is interesting to note that the  values of $K_{\rm 0}$ extracted from the M-variant of the data sets, which do not include A $\sim$ 60 nuclei, are systematically higher that those found using both light (Fe, Ni) and heavier (Cd, Sn, Sm and Pb) isotopes ranging from 270 to 315 MeV with ratio $c$ between -1.88 and -2.35. A similar trend has been observed, for example, by Paar et al. \cite{paar2006} who used a relativistic Hartree-Bogolyubov + QRPA model to calculate strength distribution and centroid and mean GMR energies and were unable to obtain agreement with experiment simultaneously for nuclei with A $\le$ 60 and A $\ge$ 90. The latter required an interaction with a higher value of $K_{\rm 0}$ than the former.  Repeat of the analysis \cite{blaizot1976}, which produced the often used range of $K_{\rm 0}$ between 180 -- 240 MeV, with modern input yields range 180 -- 280 MeV. 

The parameterized leptodermous expansion does not rely on any microscopic nuclear theory and offers in principle a direct connection with experimental data.  In this work it was used under several assumptions: (i) the liquid drop approach to description of the vibrating nucleus is valid and the relation between $K_{\rm A}$ and E$_{\rm GMR}$  (Eq.~\ref{kexp}) holds, (ii) the volume coefficient  $K_{\rm vol}$ can be identified with $K_{\rm 0}$, (iii)  the $K_{\rm coul}$ = -(5.2 $\pm$ 0.7) MeV and (iv) the leptodermous expansion for K$_{\rm A}$ converges fast enough that  contributions from the curvature  K$_{\rm curv}$A$^{\rm -2/3}$ and higher order terms in the expansion can  neglected. 

Our results depend strongly on a concept that under compression and decompression the surface and the bulk homogenous core of a nucleus can be treated separately and have, in principle, different properties. The scaling approximation allows such separation, but the assumption that  K$_{\rm vol}$ $\cong$ - K$_{\rm surf}$, i.e.  $c$ $\sim$ -1\textit{ is not} specified in this approximation which only predicts a linear dependence between the two coefficients. We show that the generally accepted value of $K_{\rm 0}$ = (240 $\pm$ 20) MeV can be obtained from the fits provided the ratio of $K_{\rm surf}/K_{\rm vol} \sim$ -1,  as predicted by a majority of mean-field models. However, the fits are significantly improved if $c$ is allowed to vary, leading to a range of  $K_{\rm 0}$, extended to significantly higher values. The results demonstrate the importance of nuclear surface properties in determination of $K_{\rm 0}$ from fits to the leptodermous expansion of $K_{\rm A}$ . 

It may strike the reader as strange that the we find that the surface incompressibility to be higher than the volume incompressibility. Intuitively one expects the surface of a nucleus, being less dense, to be more compressible. However, it is important to realize that in nuclear matter K$_{\rm 0}$, is inversely proportional not only to the compressibility of a uniform system  $\chi$ \cite{blaizot1976},
$\chi=-\frac{1}{\Omega}\frac{\partial \Omega}{\partial P}$, where $\Omega$ and $P$ are volume and pressure in the system, but also to the density, K$_{\rm 0} \sim 1/(\rho\chi)$. This means that for two systems with the same density, K$_{\rm 0}$ increases with decreasing $\chi$. However, for two systems with both $\chi$ and $\rho$ varying, the one with lower product $\rho\chi$ will have higher K$_{\rm 0}$. In finite nuclei, where the surface has lower density than the interior, the surface incompressibility will therefore be higher than the volume incompressibility provided $\chi$ increases more slowly than $\rho$ falls.

Results very similar to those obtained in this work, were reported in the early 1980's by Treiner et al. \cite{treiner1981}. In their three-parameter fit they extracted  K$_{\rm vol}$ = (300 $\pm$ 29) [(357 $\pm$ 35)] MeV, K$_{\rm surf}$ = -(608 $\pm$ 120) [(-833 $\pm$ 148)] MeV and K$_{\rm \tau}$ = -(475 $\pm$ 176) [(-833 $\pm$ 148)] MeV using Grenoble [Texas] data  (see Table 10 of \cite{treiner1981}). However, they also performed a one-parameter fit in which  the ratio K$_{\rm surf}$/ K$_{\rm vol}$ ranged only from  -1 to -1.2, and  K$_{\rm \tau}$  from -250 to -350 MeV. They then found  K$_{\rm vol}$= (220 $\pm$ 20) MeV ,  K$_{\rm surf}$ = -(240 $\pm$ 70) MeV and  K$_{\rm \tau}$ = -(300 $\pm$ 100) MeV. The limits in the one-parameter fit were motivated by the aim to reproduce predictions by the early Skyrme forces and the concern that the limited range of variation of A$^{\rm -1/3}$ and asymmetry (N-Z)/A for available data did not allow extraction of values of the different coefficients of the leptodermous expansion for  K$_{\rm A}$  with adequate accuracy \cite{treiner2013}. As can be seen in our Table~\ref{tab:c1}, results obtained for $c$ = -1,  compatible with the restrictions used in the one-parameter fit by Treiner et al. \cite{treiner1981}, are very similar to theirs.

Our results are also close to those obtained by  Sharma et al. \cite{sharma1989}  K$_{\rm surf}$ = -(750 $\pm$ 80) MeV, about 2.5 times larger than K$_{\rm vol}$ = (300 $\pm$ 25) MeV. It seems that experimental data favour the ratio $c$ different from -1 and K$_{\rm vol}$ above 250 MeV,  only weekly dependent of the data sets used and the groups who performed the analyses, in variance with theoretical predictions many mean-field models. The values obtained by Sharma et al. are slightly higher than our. This may be because Sharma et al. included a fixed curvature term in their calculations.  Our exploration of the effect of the curvature term showed that, although we were not able to determine   K$_{\rm curv}$ term in sufficient accuracy, its inclusion takes $c$ and  K$_{\rm vol}$ ever further from the currently adopted values.   

To search further for a physical origin of our results, we developed a simple self-consistent (toy) model, which revealed a connection between the density dependence of the surface diffuseness and the surface to volume incompressibility ratio. The model points to the important connection between the surface properties of a vibrating nucleus and its incompressibility as described by the leptodermous expansion, predicts surface diffuseness directly proportional to density and yields the surface to volume incompressibility ratio compatible with our results. Further development of the model, including dynamical (collective) degrees of freedom, goes beyond the scope of this work and will be published separately. 

A question may arise whether or not our results should be used as constraints on mean-field models. The leptodermous expansion is a parameterized description, which serves as a direct connection with experimental data. Microscopic models attempt to calculate the same parameters on the bases of a modeled nucleonic interaction. The success of any microscopic model will be judged by the extent to which the calculated parameters agree with experiment in this and other areas. However different mean-field models offer a wide range of results for each parameter (see e.g. Table~\ref{tab:surv}). A comparison of their predictions for GMR energies with experimental data (see Figs.~\ref{fig:blaizot} and \ref{fig:sn116}) indicates a certain spread of values. It is not obvious that any single model should be given preference in providing constraints on $K_{\rm vol}$ and $K_{\rm \tau}$. Rather, we believe results obtained by an experiment-based analysis, such as ours, are more logically useful to provide constraints.

In conclusion our work suggests that, based on the most precise and up-to-date data on GMR energies of Sn and Cd isotopes, together with a selected set of data from $^{\rm 56}$Ni to $^{\rm 208}$Pb,  the value of K$_{\rm 0}$ is higher than  generally accepted by a considerable margin. This result, 250 $<$ K$_{\rm 0} <$ 315 MeV has been obtained without any microscopic model assumptions, except (marginally) the Coulomb effect, and revealed the essential role of surface properties in vibrating nuclei. It is close to values calculated in most of the classical RMF models (before their modification to force a low value of K$_{\rm 0}$). It differs from the values given by conventional non-relativistic HF models with effective interactions such as the Skyrme or Gogny, although we should bear in mind that many of them have been constructed with the constraint of yielding a low value of K$_{\rm 0}$. The higher value of K$_{\rm 0}$ is also consistent with predictions of the Quark-Meson-Coupling model \cite{stone2007, whittenbury2014}. 

It would be highly desirable to revisit different microscopic models. It seems likely that their differences originate from the variety of ways in which surface properties are treated. Finally, a firmly established data set of GMR energies, confirmed in independent experiments and analyses by different groups, would be an invaluable contribution to understanding nuclear monopole vibration.
\section{Acknowledgement}
We are indebted to Bijay Agrawal, Mark Anders, Shalom Shlomo and P.-G. Reinhard for providing theoretical calculations of GMR energies used in this work prior to publication. Helpful discussions with Hiroyaki Sagawa, Shalom Shlomo, Jacques Treiner, Dario Vretenar,  Peter Moller, Bill Myers,  P.-G. Reinhard, Anthony Thomas,  Dave Youngblood, Y.-W.Lui and Umesh Garg are acknowledged with pleasure. Last but not least we wish to thank the anonymous referee for his/her very carefull reading of the manuscript and helpful comments, leading to its improvement. 

\bibliography{base}

\begin{thebibliography}{95}
\expandafter\ifx\csname natexlab\endcsname\relax\def\natexlab#1{#1}\fi
\expandafter\ifx\csname bibnamefont\endcsname\relax
  \def\bibnamefont#1{#1}\fi
\expandafter\ifx\csname bibfnamefont\endcsname\relax
  \def\bibfnamefont#1{#1}\fi
\expandafter\ifx\csname citenamefont\endcsname\relax
  \def\citenamefont#1{#1}\fi
\expandafter\ifx\csname url\endcsname\relax
  \def\url#1{\texttt{#1}}\fi
\expandafter\ifx\csname urlprefix\endcsname\relax\def\urlprefix{URL }\fi
\providecommand{\bibinfo}[2]{#2}
\providecommand{\eprint}[2][]{\url{#2}}

\bibitem[{\citenamefont{Dutra et~al.}(2012)\citenamefont{Dutra, Louren{\c c}o,
  {S\'a Martins}, Delfino, Stone, and Stevenson}}]{dutra2012}
\bibinfo{author}{\bibfnamefont{M.}~\bibnamefont{Dutra}},
  \bibinfo{author}{\bibfnamefont{O.}~\bibnamefont{Louren{\c c}o}},
  \bibinfo{author}{\bibfnamefont{J.~S.} \bibnamefont{{S\'a Martins}}},
  \bibinfo{author}{\bibfnamefont{A.}~\bibnamefont{Delfino}},
  \bibinfo{author}{\bibfnamefont{J.~R.} \bibnamefont{Stone}}, \bibnamefont{and}
  \bibinfo{author}{\bibfnamefont{P.~D.} \bibnamefont{Stevenson}},
  \bibinfo{journal}{Phys.\ Rev.\ C} \textbf{\bibinfo{volume}{85}},
  \bibinfo{pages}{035201} (\bibinfo{year}{2012}).

\bibitem[{\citenamefont{Tsang et~al.}(2012)\citenamefont{Tsang, Stone, Camera,
  Danielewicz, Gandolfi, Hebeler, Horowitz, Lee, Lynch, Kohley
  et~al.}}]{tsang2012}
\bibinfo{author}{\bibfnamefont{M.~B.} \bibnamefont{Tsang}},
  \bibinfo{author}{\bibfnamefont{J.~R.} \bibnamefont{Stone}},
  \bibinfo{author}{\bibfnamefont{F.}~\bibnamefont{Camera}},
  \bibinfo{author}{\bibfnamefont{P.}~\bibnamefont{Danielewicz}},
  \bibinfo{author}{\bibfnamefont{S.}~\bibnamefont{Gandolfi}},
  \bibinfo{author}{\bibfnamefont{K.}~\bibnamefont{Hebeler}},
  \bibinfo{author}{\bibfnamefont{C.~J.} \bibnamefont{Horowitz}},
  \bibinfo{author}{\bibfnamefont{J.}~\bibnamefont{Lee}},
  \bibinfo{author}{\bibfnamefont{W.~G.} \bibnamefont{Lynch}},
  \bibinfo{author}{\bibfnamefont{Z.}~\bibnamefont{Kohley}},
  \bibnamefont{et~al.}, \bibinfo{journal}{Phys.\ Rev.\ C}
  \textbf{\bibinfo{volume}{86}}, \bibinfo{pages}{015803}
  (\bibinfo{year}{2012}).

\bibitem[{\citenamefont{Falk and Wilets}(1961)}]{falk1961}
\bibinfo{author}{\bibfnamefont{D.~S.} \bibnamefont{Falk}} \bibnamefont{and}
  \bibinfo{author}{\bibfnamefont{L.}~\bibnamefont{Wilets}},
  \bibinfo{journal}{Phys.\ Rev.} \textbf{\bibinfo{volume}{124}},
  \bibinfo{pages}{1887} (\bibinfo{year}{1961}).

\bibitem[{\citenamefont{Bethe}(1971)}]{bethe1971}
\bibinfo{author}{\bibfnamefont{H.~A.} \bibnamefont{Bethe}},
  \bibinfo{journal}{Annu. \ Rev. \ Nucl.\ Sci.} \textbf{\bibinfo{volume}{21}},
  \bibinfo{pages}{93} (\bibinfo{year}{1971}).

\bibitem[{\citenamefont{Brink and Boeker}(1967)}]{brink1967}
\bibinfo{author}{\bibfnamefont{D.~M.} \bibnamefont{Brink}} \bibnamefont{and}
  \bibinfo{author}{\bibfnamefont{E.}~\bibnamefont{Boeker}},
  \bibinfo{journal}{Nucl.\ Phys.\ A} \textbf{\bibinfo{volume}{91}},
  \bibinfo{pages}{1} (\bibinfo{year}{1967}).

\bibitem[{\citenamefont{Vautherin and Brink}(1972)}]{vautherin1972}
\bibinfo{author}{\bibfnamefont{D.}~\bibnamefont{Vautherin}} \bibnamefont{and}
  \bibinfo{author}{\bibfnamefont{D.~M.} \bibnamefont{Brink}},
  \bibinfo{journal}{Phys.\ Rev.\ C} \textbf{\bibinfo{volume}{5}},
  \bibinfo{pages}{626} (\bibinfo{year}{1972}).

\bibitem[{\citenamefont{Beiner et~al.}(1975)\citenamefont{Beiner, Flocard,
  Giai, and Quentin}}]{beiner1975}
\bibinfo{author}{\bibfnamefont{M.}~\bibnamefont{Beiner}},
  \bibinfo{author}{\bibfnamefont{H.}~\bibnamefont{Flocard}},
  \bibinfo{author}{\bibfnamefont{N.~V.} \bibnamefont{Giai}}, \bibnamefont{and}
  \bibinfo{author}{\bibfnamefont{P.}~\bibnamefont{Quentin}},
  \bibinfo{journal}{Nucl.\ Phys.\ A} \textbf{\bibinfo{volume}{238}},
  \bibinfo{pages}{29} (\bibinfo{year}{1975}).

\bibitem[{\citenamefont{Gogny}(1975)}]{gogny1975}
\bibinfo{author}{\bibfnamefont{D.}~\bibnamefont{Gogny}}, in
  \emph{\bibinfo{booktitle}{Nuclear self-consistent fields}}, edited by
  \bibinfo{editor}{\bibfnamefont{G.}~\bibnamefont{Ripka}} \bibnamefont{and}
  \bibinfo{editor}{\bibfnamefont{M.}~\bibnamefont{Porneuf}}
  (\bibinfo{publisher}{North Holland}, \bibinfo{address}{Amsterdam},
  \bibinfo{year}{1975}), p. \bibinfo{pages}{333}.

\bibitem[{\citenamefont{Koehler}(1976)}]{koehler1976}
\bibinfo{author}{\bibfnamefont{H.~S.} \bibnamefont{Koehler}},
  \bibinfo{journal}{Nucl.\ Phys.\ A} \textbf{\bibinfo{volume}{258}},
  \bibinfo{pages}{301} (\bibinfo{year}{1976}).

\bibitem[{\citenamefont{Myers and Swiatecki}(1966)}]{myers1966}
\bibinfo{author}{\bibfnamefont{W.}~\bibnamefont{Myers}} \bibnamefont{and}
  \bibinfo{author}{\bibfnamefont{W.}~\bibnamefont{Swiatecki}},
  \bibinfo{journal}{Nucl.\ Phys.\ A} \textbf{\bibinfo{volume}{81}},
  \bibinfo{pages}{1} (\bibinfo{year}{1966}).

\bibitem[{\citenamefont{Myers and Swiatecki}(1969)}]{myers1969}
\bibinfo{author}{\bibfnamefont{W.}~\bibnamefont{Myers}} \bibnamefont{and}
  \bibinfo{author}{\bibfnamefont{W.}~\bibnamefont{Swiatecki}},
  \bibinfo{journal}{Ann.\ Phys.} \textbf{\bibinfo{volume}{55}},
  \bibinfo{pages}{395} (\bibinfo{year}{1969}).

\bibitem[{\citenamefont{Myers and Swiatecki}(1974)}]{myers1974}
\bibinfo{author}{\bibfnamefont{W.}~\bibnamefont{Myers}} \bibnamefont{and}
  \bibinfo{author}{\bibfnamefont{W.}~\bibnamefont{Swiatecki}},
  \bibinfo{journal}{Ann.\ Phys.} \textbf{\bibinfo{volume}{84}},
  \bibinfo{pages}{186} (\bibinfo{year}{1974}).

\bibitem[{\citenamefont{Myers and Swiatecki}(1990)}]{myers1990}
\bibinfo{author}{\bibfnamefont{W.}~\bibnamefont{Myers}} \bibnamefont{and}
  \bibinfo{author}{\bibfnamefont{W.}~\bibnamefont{Swiatecki}},
  \bibinfo{journal}{Ann.\ Phys.} \textbf{\bibinfo{volume}{204}},
  \bibinfo{pages}{401} (\bibinfo{year}{1990}).

\bibitem[{\citenamefont{Moller and Nix}(1995)}]{moller1995}
\bibinfo{author}{\bibfnamefont{P.}~\bibnamefont{Moller}} \bibnamefont{and}
  \bibinfo{author}{\bibfnamefont{J.~R.} \bibnamefont{Nix}},
  \bibinfo{journal}{At.\ Data \ Nucl.\ Data \ Tables}
  \textbf{\bibinfo{volume}{59}}, \bibinfo{pages}{185} (\bibinfo{year}{1995}).

\bibitem[{\citenamefont{Moller et~al.}(2012)\citenamefont{Moller, Myers,
  Sagawa, and Yoshida}}]{moller2012}
\bibinfo{author}{\bibfnamefont{P.}~\bibnamefont{Moller}},
  \bibinfo{author}{\bibfnamefont{W.~D.} \bibnamefont{Myers}},
  \bibinfo{author}{\bibfnamefont{H.}~\bibnamefont{Sagawa}}, \bibnamefont{and}
  \bibinfo{author}{\bibfnamefont{S.}~\bibnamefont{Yoshida}},
  \bibinfo{journal}{Phys.\ Rev.\ Lett.} \textbf{\bibinfo{volume}{108}},
  \bibinfo{pages}{052501} (\bibinfo{year}{2012}).

\bibitem[{\citenamefont{Marty et~al.}(1975)\citenamefont{Marty, Morlet, Willis,
  Comparat, Frascaria, and Kallne}}]{marty1975}
\bibinfo{author}{\bibfnamefont{N.}~\bibnamefont{Marty}},
  \bibinfo{author}{\bibfnamefont{M.}~\bibnamefont{Morlet}},
  \bibinfo{author}{\bibfnamefont{A.}~\bibnamefont{Willis}},
  \bibinfo{author}{\bibfnamefont{V.}~\bibnamefont{Comparat}},
  \bibinfo{author}{\bibfnamefont{R.}~\bibnamefont{Frascaria}},
  \bibnamefont{and} \bibinfo{author}{\bibfnamefont{J.}~\bibnamefont{Kallne}},
  \bibinfo{howpublished}{preprint IPNO-Ph No 75-11} (\bibinfo{year}{1975}).

\bibitem[{\citenamefont{Blaizot et~al.}(1976)\citenamefont{Blaizot, Gogny, and
  Grammaticos}}]{blaizot1976}
\bibinfo{author}{\bibfnamefont{J.~P.} \bibnamefont{Blaizot}},
  \bibinfo{author}{\bibfnamefont{D.}~\bibnamefont{Gogny}}, \bibnamefont{and}
  \bibinfo{author}{\bibfnamefont{B.}~\bibnamefont{Grammaticos}},
  \bibinfo{journal}{Nucl.\ Phys.\ A} \textbf{\bibinfo{volume}{265}},
  \bibinfo{pages}{315} (\bibinfo{year}{1976}).

\bibitem[{\citenamefont{Treiner et~al.}(1981)\citenamefont{Treiner, Krivine,
  Bohigas, and Martorell}}]{treiner1981}
\bibinfo{author}{\bibfnamefont{J.}~\bibnamefont{Treiner}},
  \bibinfo{author}{\bibfnamefont{H.}~\bibnamefont{Krivine}},
  \bibinfo{author}{\bibfnamefont{O.}~\bibnamefont{Bohigas}}, \bibnamefont{and}
  \bibinfo{author}{\bibfnamefont{J.}~\bibnamefont{Martorell}},
  \bibinfo{journal}{Nucl.\ Phys.\ A} \textbf{\bibinfo{volume}{371}},
  \bibinfo{pages}{253} (\bibinfo{year}{1981}).

\bibitem[{\citenamefont{Sharma et~al.}(1988)\citenamefont{Sharma, Borghols,
  Brandenburg, Crona, and van~der Woude}}]{sharma1988}
\bibinfo{author}{\bibfnamefont{M.~M.} \bibnamefont{Sharma}},
  \bibinfo{author}{\bibfnamefont{W.~T.~A.} \bibnamefont{Borghols}},
  \bibinfo{author}{\bibfnamefont{S.}~\bibnamefont{Brandenburg}},
  \bibinfo{author}{\bibfnamefont{S.}~\bibnamefont{Crona}}, \bibnamefont{and}
  \bibinfo{author}{\bibfnamefont{A.}~\bibnamefont{van~der Woude}},
  \bibinfo{journal}{Phys.\ Rev.\ C} \textbf{\bibinfo{volume}{38}},
  \bibinfo{pages}{2562} (\bibinfo{year}{1988}).

\bibitem[{\citenamefont{Shlomo and Youngblood}(1993)}]{shlomo1993}
\bibinfo{author}{\bibfnamefont{S.}~\bibnamefont{Shlomo}} \bibnamefont{and}
  \bibinfo{author}{\bibfnamefont{D.~H.} \bibnamefont{Youngblood}},
  \bibinfo{journal}{Phys.\ Rev.\ C} \textbf{\bibinfo{volume}{47}},
  \bibinfo{pages}{529} (\bibinfo{year}{1993}).

\bibitem[{\citenamefont{Stoitsov et~al.}(1994)\citenamefont{Stoitsov, Ring, and
  Sharma}}]{stoitsov1994}
\bibinfo{author}{\bibfnamefont{M.}~\bibnamefont{Stoitsov}},
  \bibinfo{author}{\bibfnamefont{P.}~\bibnamefont{Ring}}, \bibnamefont{and}
  \bibinfo{author}{\bibfnamefont{M.~M.} \bibnamefont{Sharma}},
  \bibinfo{journal}{Phys.\ Rev.\ C} \textbf{\bibinfo{volume}{50}},
  \bibinfo{pages}{1445} (\bibinfo{year}{1994}).

\bibitem[{\citenamefont{Blaizot et~al.}(1995)\citenamefont{Blaizot, Berger,
  Decharge, and Girod}}]{blaizot1995}
\bibinfo{author}{\bibfnamefont{J.~P.} \bibnamefont{Blaizot}},
  \bibinfo{author}{\bibfnamefont{J.~F.} \bibnamefont{Berger}},
  \bibinfo{author}{\bibfnamefont{J.}~\bibnamefont{Decharge}}, \bibnamefont{and}
  \bibinfo{author}{\bibfnamefont{M.}~\bibnamefont{Girod}},
  \bibinfo{journal}{Nucl.\ Phys.\ A} \textbf{\bibinfo{volume}{591}},
  \bibinfo{pages}{435} (\bibinfo{year}{1995}).

\bibitem[{\citenamefont{Farine et~al.}(1997)\citenamefont{Farine, Pearson, and
  Tondeur}}]{farine1997}
\bibinfo{author}{\bibfnamefont{M.}~\bibnamefont{Farine}},
  \bibinfo{author}{\bibfnamefont{J.~M.} \bibnamefont{Pearson}},
  \bibnamefont{and} \bibinfo{author}{\bibfnamefont{F.}~\bibnamefont{Tondeur}},
  \bibinfo{journal}{Nucl.\ Phys.\ A} \textbf{\bibinfo{volume}{615}},
  \bibinfo{pages}{135} (\bibinfo{year}{1997}).

\bibitem[{\citenamefont{Vretenar et~al.}(1997)\citenamefont{Vretenar,
  A.Lalazissis, Behnsch, Poeschl, and Ring}}]{vretenar1997}
\bibinfo{author}{\bibfnamefont{D.}~\bibnamefont{Vretenar}},
  \bibinfo{author}{\bibfnamefont{G.}~\bibnamefont{A.Lalazissis}},
  \bibinfo{author}{\bibfnamefont{R.}~\bibnamefont{Behnsch}},
  \bibinfo{author}{\bibfnamefont{W.}~\bibnamefont{Poeschl}}, \bibnamefont{and}
  \bibinfo{author}{\bibfnamefont{P.}~\bibnamefont{Ring}},
  \bibinfo{journal}{Nucl.\ Phys.\ A} \textbf{\bibinfo{volume}{621}},
  \bibinfo{pages}{853} (\bibinfo{year}{1997}).

\bibitem[{\citenamefont{Youngblood et~al.}(1999)\citenamefont{Youngblood,
  Clark, and Lui}}]{youngblood1999}
\bibinfo{author}{\bibfnamefont{D.~H.} \bibnamefont{Youngblood}},
  \bibinfo{author}{\bibfnamefont{H.~L.} \bibnamefont{Clark}}, \bibnamefont{and}
  \bibinfo{author}{\bibfnamefont{Y.-W.} \bibnamefont{Lui}},
  \bibinfo{journal}{Phys. \ Rev. \ Lett.} \textbf{\bibinfo{volume}{82}},
  \bibinfo{pages}{691} (\bibinfo{year}{1999}).

\bibitem[{\citenamefont{Chung et~al.}(1999)\citenamefont{Chung, Wang, and
  Santiago}}]{chung1999}
\bibinfo{author}{\bibfnamefont{K.~C.} \bibnamefont{Chung}},
  \bibinfo{author}{\bibfnamefont{C.~S.} \bibnamefont{Wang}}, \bibnamefont{and}
  \bibinfo{author}{\bibfnamefont{A.~J.} \bibnamefont{Santiago}},
  \bibinfo{journal}{Phys.\ Rev.\ C} \textbf{\bibinfo{volume}{59}},
  \bibinfo{pages}{714} (\bibinfo{year}{1999}).

\bibitem[{\citenamefont{Satpathy et~al.}(1999)\citenamefont{Satpathy,
  Maheswari, and Nayak}}]{satpathy1999}
\bibinfo{author}{\bibfnamefont{L.}~\bibnamefont{Satpathy}},
  \bibinfo{author}{\bibfnamefont{V.~S.~U.} \bibnamefont{Maheswari}},
  \bibnamefont{and} \bibinfo{author}{\bibfnamefont{R.~C.} \bibnamefont{Nayak}},
  \bibinfo{journal}{Physics Reports} \textbf{\bibinfo{volume}{319}},
  \bibinfo{pages}{85} (\bibinfo{year}{1999}).

\bibitem[{\citenamefont{Piekarewicz}(2002)}]{piekarewicz2002}
\bibinfo{author}{\bibfnamefont{J.}~\bibnamefont{Piekarewicz}},
  \bibinfo{journal}{Phys.\ Rev.\ C} \textbf{\bibinfo{volume}{66}},
  \bibinfo{pages}{034305} (\bibinfo{year}{2002}).

\bibitem[{\citenamefont{Agrawal et~al.}(2003)\citenamefont{Agrawal, Shlomo, and
  Au}}]{agrawal2003}
\bibinfo{author}{\bibfnamefont{B.~K.} \bibnamefont{Agrawal}},
  \bibinfo{author}{\bibfnamefont{S.}~\bibnamefont{Shlomo}}, \bibnamefont{and}
  \bibinfo{author}{\bibfnamefont{V.~K.} \bibnamefont{Au}},
  \bibinfo{journal}{Phys. \ Rev. \ C} \textbf{\bibinfo{volume}{68}},
  \bibinfo{pages}{031304} (\bibinfo{year}{2003}).

\bibitem[{\citenamefont{Colo et~al.}(2004)\citenamefont{Colo, Giai, Meyer,
  Bennaceur, and Bonche}}]{colo2004}
\bibinfo{author}{\bibfnamefont{G.}~\bibnamefont{Colo}},
  \bibinfo{author}{\bibfnamefont{N.~V.} \bibnamefont{Giai}},
  \bibinfo{author}{\bibfnamefont{J.}~\bibnamefont{Meyer}},
  \bibinfo{author}{\bibfnamefont{K.}~\bibnamefont{Bennaceur}},
  \bibnamefont{and} \bibinfo{author}{\bibfnamefont{P.}~\bibnamefont{Bonche}},
  \bibinfo{journal}{Phys.\ Rev.\ C} \textbf{\bibinfo{volume}{70}},
  \bibinfo{pages}{024307} (\bibinfo{year}{2004}).

\bibitem[{\citenamefont{Lalazissis et~al.}(2005)\citenamefont{Lalazissis,
  Niksic, Vretenar, and Ring}}]{lalazissis2005}
\bibinfo{author}{\bibfnamefont{G.~A.} \bibnamefont{Lalazissis}},
  \bibinfo{author}{\bibfnamefont{T.}~\bibnamefont{Niksic}},
  \bibinfo{author}{\bibfnamefont{D.}~\bibnamefont{Vretenar}}, \bibnamefont{and}
  \bibinfo{author}{\bibfnamefont{P.}~\bibnamefont{Ring}},
  \bibinfo{journal}{Phys. \ Rev. \ C} \textbf{\bibinfo{volume}{71}},
  \bibinfo{pages}{024312} (\bibinfo{year}{2005}).

\bibitem[{\citenamefont{Shlomo et~al.}(2006)\citenamefont{Shlomo, Kolomietz,
  and Colo}}]{shlomo2006}
\bibinfo{author}{\bibfnamefont{S.}~\bibnamefont{Shlomo}},
  \bibinfo{author}{\bibfnamefont{V.~M.} \bibnamefont{Kolomietz}},
  \bibnamefont{and} \bibinfo{author}{\bibfnamefont{G.}~\bibnamefont{Colo}},
  \bibinfo{journal}{Eur.\ Phys.\ J.\ A} \textbf{\bibinfo{volume}{30}},
  \bibinfo{pages}{23} (\bibinfo{year}{2006}).

\bibitem[{\citenamefont{Hirose et~al.}(2007)\citenamefont{Hirose, Serra, Ring,
  Otsuka, and Akaishi}}]{ring2007}
\bibinfo{author}{\bibfnamefont{S.}~\bibnamefont{Hirose}},
  \bibinfo{author}{\bibfnamefont{M.}~\bibnamefont{Serra}},
  \bibinfo{author}{\bibfnamefont{P.}~\bibnamefont{Ring}},
  \bibinfo{author}{\bibfnamefont{T.}~\bibnamefont{Otsuka}}, \bibnamefont{and}
  \bibinfo{author}{\bibfnamefont{Y.}~\bibnamefont{Akaishi}},
  \bibinfo{journal}{Phys. \ Rev. \ C} \textbf{\bibinfo{volume}{75}},
  \bibinfo{pages}{024301} (\bibinfo{year}{2007}).

\bibitem[{\citenamefont{Li et~al.}(2008)\citenamefont{Li, Colo, and
  Meng}}]{colo2008}
\bibinfo{author}{\bibfnamefont{J.}~\bibnamefont{Li}},
  \bibinfo{author}{\bibfnamefont{G.}~\bibnamefont{Colo}}, \bibnamefont{and}
  \bibinfo{author}{\bibfnamefont{J.}~\bibnamefont{Meng}},
  \bibinfo{journal}{Phys. \ Rev. \ C} \textbf{\bibinfo{volume}{78}},
  \bibinfo{pages}{064304} (\bibinfo{year}{2008}).

\bibitem[{\citenamefont{Agrawal et~al.}(2012)\citenamefont{Agrawal, Sulaksono,
  and Reinhard}}]{agrawal2012}
\bibinfo{author}{\bibfnamefont{B.~K.} \bibnamefont{Agrawal}},
  \bibinfo{author}{\bibfnamefont{A.}~\bibnamefont{Sulaksono}},
  \bibnamefont{and} \bibinfo{author}{\bibfnamefont{P.-G.}
  \bibnamefont{Reinhard}}, \bibinfo{journal}{Nucl. \ Phys.\ A}
  \textbf{\bibinfo{volume}{882}}, \bibinfo{pages}{1} (\bibinfo{year}{2012}).

\bibitem[{\citenamefont{Sagawa}(2012)}]{sagawa2012}
\bibinfo{author}{\bibfnamefont{H.}~\bibnamefont{Sagawa}},
  \bibinfo{howpublished}{invited talk at Compstar 2012, Tahiti}
  (\bibinfo{year}{2012}).

\bibitem[{\citenamefont{Cao et~al.}(2012)\citenamefont{Cao, Sagawa, and
  Colo}}]{cao2012}
\bibinfo{author}{\bibfnamefont{L.-G.} \bibnamefont{Cao}},
  \bibinfo{author}{\bibfnamefont{H.}~\bibnamefont{Sagawa}}, \bibnamefont{and}
  \bibinfo{author}{\bibfnamefont{G.}~\bibnamefont{Colo}},
  \bibinfo{journal}{Phys. \ Rev. \ C} \textbf{\bibinfo{volume}{86}},
  \bibinfo{pages}{054313} (\bibinfo{year}{2012}).

\bibitem[{\citenamefont{Blaizot}(1980)}]{blaizot1980}
\bibinfo{author}{\bibfnamefont{J.~P.} \bibnamefont{Blaizot}},
  \bibinfo{journal}{Phys.\ Rep.} \textbf{\bibinfo{volume}{64}},
  \bibinfo{pages}{171} (\bibinfo{year}{1980}).

\bibitem[{\citenamefont{Jennings and Jackson}(1980)}]{jennings1980}
\bibinfo{author}{\bibfnamefont{B.~K.} \bibnamefont{Jennings}} \bibnamefont{and}
  \bibinfo{author}{\bibfnamefont{A.~D.} \bibnamefont{Jackson}},
  \bibinfo{journal}{Phys.\ Rep.} \textbf{\bibinfo{volume}{66}},
  \bibinfo{pages}{141} (\bibinfo{year}{1980}).

\bibitem[{\citenamefont{Blaizot and Grammaticos}(1981)}]{blaizot1981}
\bibinfo{author}{\bibfnamefont{J.~P.} \bibnamefont{Blaizot}} \bibnamefont{and}
  \bibinfo{author}{\bibfnamefont{B.}~\bibnamefont{Grammaticos}},
  \bibinfo{journal}{Nucl.\ Phys.\ A} \textbf{\bibinfo{volume}{355}},
  \bibinfo{pages}{115} (\bibinfo{year}{1981}).

\bibitem[{\citenamefont{Blaizot}(1989)}]{blaizot1989}
\bibinfo{author}{\bibfnamefont{J.~P.} \bibnamefont{Blaizot}}, in
  \emph{\bibinfo{booktitle}{The Nuclear Equation of State Part A}}, edited by
  \bibinfo{editor}{\bibfnamefont{W.}~\bibnamefont{Greiner}} \bibnamefont{and}
  \bibinfo{editor}{\bibfnamefont{H.}~\bibnamefont{Stoecker}}
  (\bibinfo{publisher}{Plenum Press}, \bibinfo{address}{New York and London},
  \bibinfo{year}{1989}), vol. \bibinfo{volume}{216} of
  \emph{\bibinfo{series}{NATO ASI Series Part A}}, p. \bibinfo{pages}{679}.

\bibitem[{\citenamefont{Sharma et~al.}(1989)\citenamefont{Sharma, Stocker,
  Gleissl, and Brack}}]{sharma1989}
\bibinfo{author}{\bibfnamefont{M.~M.} \bibnamefont{Sharma}},
  \bibinfo{author}{\bibfnamefont{W.}~\bibnamefont{Stocker}},
  \bibinfo{author}{\bibfnamefont{P.}~\bibnamefont{Gleissl}}, \bibnamefont{and}
  \bibinfo{author}{\bibfnamefont{M.}~\bibnamefont{Brack}},
  \bibinfo{journal}{Nucl.\ Phys.\ A} \textbf{\bibinfo{volume}{504}},
  \bibinfo{pages}{337} (\bibinfo{year}{1989}).

\bibitem[{\citenamefont{Sharma}(1989)}]{sharma1989a}
\bibinfo{author}{\bibfnamefont{M.~M.} \bibnamefont{Sharma}}, in
  \emph{\bibinfo{booktitle}{The Nuclear Equation of State Part A}}, edited by
  \bibinfo{editor}{\bibfnamefont{W.}~\bibnamefont{Greiner}} \bibnamefont{and}
  \bibinfo{editor}{\bibfnamefont{H.}~\bibnamefont{Stoecker}}
  (\bibinfo{publisher}{Plenum Press}, \bibinfo{address}{New York and London},
  \bibinfo{year}{1989}), vol. \bibinfo{volume}{216} of
  \emph{\bibinfo{series}{NATO ASI Series Part A}}, p. \bibinfo{pages}{661}.

\bibitem[{\citenamefont{Sharma}(2009)}]{sharma2009}
\bibinfo{author}{\bibfnamefont{M.~M.} \bibnamefont{Sharma}},
  \bibinfo{journal}{Nucl.\ Phys.\ A} \textbf{\bibinfo{volume}{816}},
  \bibinfo{pages}{65} (\bibinfo{year}{2009}).

\bibitem[{\citenamefont{Agrawal and Shlomo}(2004)}]{agrawal2004}
\bibinfo{author}{\bibfnamefont{B.~K.} \bibnamefont{Agrawal}} \bibnamefont{and}
  \bibinfo{author}{\bibfnamefont{S.}~\bibnamefont{Shlomo}},
  \bibinfo{journal}{Phys.\ Rev. \ C} \textbf{\bibinfo{volume}{70}},
  \bibinfo{pages}{014308} (\bibinfo{year}{2004}).

\bibitem[{\citenamefont{Nayak et~al.}(1990)\citenamefont{Nayak, Pearson,
  Farine, Gleissl, and Brack}}]{nayak1990}
\bibinfo{author}{\bibfnamefont{R.~C.} \bibnamefont{Nayak}},
  \bibinfo{author}{\bibfnamefont{J.~M.} \bibnamefont{Pearson}},
  \bibinfo{author}{\bibfnamefont{M.}~\bibnamefont{Farine}},
  \bibinfo{author}{\bibfnamefont{P.}~\bibnamefont{Gleissl}}, \bibnamefont{and}
  \bibinfo{author}{\bibfnamefont{M.}~\bibnamefont{Brack}},
  \bibinfo{journal}{Nucl. \ Phys. \ A} \textbf{\bibinfo{volume}{516}},
  \bibinfo{pages}{62} (\bibinfo{year}{1990}).

\bibitem[{\citenamefont{Patra et~al.}(2002)\citenamefont{Patra, Centelles,
  Vi$\tilde{n}$as, and Estal}}]{patra2002}
\bibinfo{author}{\bibfnamefont{S.~K.} \bibnamefont{Patra}},
  \bibinfo{author}{\bibfnamefont{M.}~\bibnamefont{Centelles}},
  \bibinfo{author}{\bibfnamefont{X.}~\bibnamefont{Vi$\tilde{n}$as}},
  \bibnamefont{and} \bibinfo{author}{\bibfnamefont{M.~D.} \bibnamefont{Estal}},
  \bibinfo{journal}{Phys. \ Rev. \ C} \textbf{\bibinfo{volume}{65}},
  \bibinfo{pages}{044304} (\bibinfo{year}{2002}).

\bibitem[{\citenamefont{Sagawa et~al.}(2007)\citenamefont{Sagawa, Yoshida,
  Zeng, Gu, and Zhang}}]{sagawa2007}
\bibinfo{author}{\bibfnamefont{H.}~\bibnamefont{Sagawa}},
  \bibinfo{author}{\bibfnamefont{S.}~\bibnamefont{Yoshida}},
  \bibinfo{author}{\bibfnamefont{G.~M.} \bibnamefont{Zeng}},
  \bibinfo{author}{\bibfnamefont{J.~Z.} \bibnamefont{Gu}}, \bibnamefont{and}
  \bibinfo{author}{\bibfnamefont{X.-Z.} \bibnamefont{Zhang}},
  \bibinfo{journal}{Phys. \ Rev. \ C} \textbf{\bibinfo{volume}{76}},
  \bibinfo{pages}{034327} (\bibinfo{year}{2007}).

\bibitem[{\citenamefont{Li et~al.}(2007)\citenamefont{Li, Garg, Liu, Marks,
  Nayak, Rao, Fujiwara, Hashimoto, Kawase, Nakanishi et~al.}}]{li2007}
\bibinfo{author}{\bibfnamefont{T.}~\bibnamefont{Li}},
  \bibinfo{author}{\bibfnamefont{U.}~\bibnamefont{Garg}},
  \bibinfo{author}{\bibfnamefont{Y.}~\bibnamefont{Liu}},
  \bibinfo{author}{\bibfnamefont{R.}~\bibnamefont{Marks}},
  \bibinfo{author}{\bibfnamefont{B.~K.} \bibnamefont{Nayak}},
  \bibinfo{author}{\bibfnamefont{P.~V.~M.} \bibnamefont{Rao}},
  \bibinfo{author}{\bibfnamefont{M.}~\bibnamefont{Fujiwara}},
  \bibinfo{author}{\bibfnamefont{H.}~\bibnamefont{Hashimoto}},
  \bibinfo{author}{\bibfnamefont{K.}~\bibnamefont{Kawase}},
  \bibinfo{author}{\bibfnamefont{K.}~\bibnamefont{Nakanishi}},
  \bibnamefont{et~al.}, \bibinfo{journal}{Phys.\ Rev.\ Lett.}
  \textbf{\bibinfo{volume}{99}}, \bibinfo{pages}{162503}
  (\bibinfo{year}{2007}).

\bibitem[{\citenamefont{Li et~al.}(2010)\citenamefont{Li, Garg, Liu, Marks,
  Nayak, Rao, Fujiwara, Hashimoto, Nakanishi, Okumura et~al.}}]{li2010}
\bibinfo{author}{\bibfnamefont{T.}~\bibnamefont{Li}},
  \bibinfo{author}{\bibfnamefont{U.}~\bibnamefont{Garg}},
  \bibinfo{author}{\bibfnamefont{Y.}~\bibnamefont{Liu}},
  \bibinfo{author}{\bibfnamefont{R.}~\bibnamefont{Marks}},
  \bibinfo{author}{\bibfnamefont{B.~K.} \bibnamefont{Nayak}},
  \bibinfo{author}{\bibfnamefont{P.~V.~M.} \bibnamefont{Rao}},
  \bibinfo{author}{\bibfnamefont{M.}~\bibnamefont{Fujiwara}},
  \bibinfo{author}{\bibfnamefont{H.}~\bibnamefont{Hashimoto}},
  \bibinfo{author}{\bibfnamefont{K.}~\bibnamefont{Nakanishi}},
  \bibinfo{author}{\bibfnamefont{S.}~\bibnamefont{Okumura}},
  \bibnamefont{et~al.}, \bibinfo{journal}{Phys. \ Rev. \ C.}
  \textbf{\bibinfo{volume}{81}}, \bibinfo{pages}{034309}
  (\bibinfo{year}{2010}).

\bibitem[{\citenamefont{Chen et~al.}(2009)\citenamefont{Chen, Cai, Ko, Li,
  Shen, and Xu}}]{chen2009}
\bibinfo{author}{\bibfnamefont{L.~W.} \bibnamefont{Chen}},
  \bibinfo{author}{\bibfnamefont{B.~J.} \bibnamefont{Cai}},
  \bibinfo{author}{\bibfnamefont{C.~M.} \bibnamefont{Ko}},
  \bibinfo{author}{\bibfnamefont{B.~A.} \bibnamefont{Li}},
  \bibinfo{author}{\bibfnamefont{C.}~\bibnamefont{Shen}}, \bibnamefont{and}
  \bibinfo{author}{\bibfnamefont{J.}~\bibnamefont{Xu}}, \bibinfo{journal}{Phys.
  \ Rev. \ C} \textbf{\bibinfo{volume}{80}}, \bibinfo{pages}{014322}
  (\bibinfo{year}{2009}).

\bibitem[{\citenamefont{Lui et~al.}(2006)\citenamefont{Lui, Youngblood, Clark,
  Tokimoto, and John}}]{lui2006}
\bibinfo{author}{\bibfnamefont{Y.~W.} \bibnamefont{Lui}},
  \bibinfo{author}{\bibfnamefont{D.~H.} \bibnamefont{Youngblood}},
  \bibinfo{author}{\bibfnamefont{H.~L.} \bibnamefont{Clark}},
  \bibinfo{author}{\bibfnamefont{Y.}~\bibnamefont{Tokimoto}}, \bibnamefont{and}
  \bibinfo{author}{\bibfnamefont{B.}~\bibnamefont{John}},
  \bibinfo{journal}{Phys. \ Rev. \ C} \textbf{\bibinfo{volume}{73}},
  \bibinfo{pages}{014314} (\bibinfo{year}{2006}).

\bibitem[{\citenamefont{Kotz et~al.}(2006)}]{kotz2006}
\bibinfo{author}{\bibfnamefont{S.}~\bibnamefont{Kotz}} \bibnamefont{et~al.},
  \emph{\bibinfo{title}{Encyclopedia of Statistical Sciences (2nd ed.)}}
  (\bibinfo{publisher}{John Wiley and Sons}, \bibinfo{year}{2006}), ISBN
  \bibinfo{isbn}{978-0-471-15044-2}.

\bibitem[{\citenamefont{Youngblood
  et~al.}(2004{\natexlab{a}})\citenamefont{Youngblood, Lui, Clark, John,
  Tokimoto, and Chen}}]{youngblood2004}
\bibinfo{author}{\bibfnamefont{D.~H.} \bibnamefont{Youngblood}},
  \bibinfo{author}{\bibfnamefont{Y.~W.} \bibnamefont{Lui}},
  \bibinfo{author}{\bibfnamefont{H.~L.} \bibnamefont{Clark}},
  \bibinfo{author}{\bibfnamefont{B.}~\bibnamefont{John}},
  \bibinfo{author}{\bibfnamefont{Y.}~\bibnamefont{Tokimoto}}, \bibnamefont{and}
  \bibinfo{author}{\bibfnamefont{X.}~\bibnamefont{Chen}},
  \bibinfo{journal}{Phys. \ Rev. \ C} \textbf{\bibinfo{volume}{69}},
  \bibinfo{pages}{034315} (\bibinfo{year}{2004}{\natexlab{a}}).

\bibitem[{\citenamefont{Uchida et~al.}(2004)\citenamefont{Uchida, Sakaguchi,
  Itoh, Yosoi, Kawabata, Yasuda, Takeda, Murakami, Terashima, Kishi
  et~al.}}]{uchida2004}
\bibinfo{author}{\bibfnamefont{M.}~\bibnamefont{Uchida}},
  \bibinfo{author}{\bibfnamefont{H.}~\bibnamefont{Sakaguchi}},
  \bibinfo{author}{\bibfnamefont{M.}~\bibnamefont{Itoh}},
  \bibinfo{author}{\bibfnamefont{M.}~\bibnamefont{Yosoi}},
  \bibinfo{author}{\bibfnamefont{T.}~\bibnamefont{Kawabata}},
  \bibinfo{author}{\bibfnamefont{Y.}~\bibnamefont{Yasuda}},
  \bibinfo{author}{\bibfnamefont{H.}~\bibnamefont{Takeda}},
  \bibinfo{author}{\bibfnamefont{T.}~\bibnamefont{Murakami}},
  \bibinfo{author}{\bibfnamefont{S.}~\bibnamefont{Terashima}},
  \bibinfo{author}{\bibfnamefont{S.}~\bibnamefont{Kishi}},
  \bibnamefont{et~al.}, \bibinfo{journal}{Phys. \ Rev. \ C}
  \textbf{\bibinfo{volume}{69}}, \bibinfo{pages}{051301(R)}
  (\bibinfo{year}{2004}).

\bibitem[{\citenamefont{Garg}(2011)}]{garg2011}
\bibinfo{author}{\bibfnamefont{U.}~\bibnamefont{Garg}},
  \bibinfo{journal}{Acta.\ Phys.\ Polonica \ B} \textbf{\bibinfo{volume}{42}},
  \bibinfo{pages}{659} (\bibinfo{year}{2011}), \eprint{1101.3125}.

\bibitem[{\citenamefont{M.Uchida et~al.}(2003)\citenamefont{M.Uchida,
  H.Sakaguchi, M.Itoh, M.Yosoi, T.Kawabata, H.Takeda, Y.Yasuda, T.Murakami,
  T.Ishikawa, T.Taki et~al.}}]{uchida2003}
\bibinfo{author}{\bibnamefont{M.Uchida}},
  \bibinfo{author}{\bibnamefont{H.Sakaguchi}},
  \bibinfo{author}{\bibnamefont{M.Itoh}},
  \bibinfo{author}{\bibnamefont{M.Yosoi}},
  \bibinfo{author}{\bibnamefont{T.Kawabata}},
  \bibinfo{author}{\bibnamefont{H.Takeda}},
  \bibinfo{author}{\bibnamefont{Y.Yasuda}},
  \bibinfo{author}{\bibnamefont{T.Murakami}},
  \bibinfo{author}{\bibnamefont{T.Ishikawa}},
  \bibinfo{author}{\bibnamefont{T.Taki}}, \bibnamefont{et~al.},
  \bibinfo{journal}{Phys. \ Lett. \ B} \textbf{\bibinfo{volume}{557}},
  \bibinfo{pages}{12} (\bibinfo{year}{2003}).

\bibitem[{\citenamefont{Lui et~al.}(2004{\natexlab{a}})\citenamefont{Lui,
  Youngblood, Tokimoto, Clark, and John}}]{lui2004a}
\bibinfo{author}{\bibfnamefont{Y.~W.} \bibnamefont{Lui}},
  \bibinfo{author}{\bibfnamefont{D.~H.} \bibnamefont{Youngblood}},
  \bibinfo{author}{\bibfnamefont{Y.}~\bibnamefont{Tokimoto}},
  \bibinfo{author}{\bibfnamefont{H.~L.} \bibnamefont{Clark}}, \bibnamefont{and}
  \bibinfo{author}{\bibfnamefont{B.}~\bibnamefont{John}},
  \bibinfo{journal}{Phys. \ Rev. \ C} \textbf{\bibinfo{volume}{70}},
  \bibinfo{pages}{013407} (\bibinfo{year}{2004}{\natexlab{a}}).

\bibitem[{\citenamefont{Lui et~al.}(2004{\natexlab{b}})\citenamefont{Lui,
  Youngblood, Tokimoto, Clark, and John}}]{lui2004}
\bibinfo{author}{\bibfnamefont{Y.~W.} \bibnamefont{Lui}},
  \bibinfo{author}{\bibfnamefont{D.~H.} \bibnamefont{Youngblood}},
  \bibinfo{author}{\bibfnamefont{Y.}~\bibnamefont{Tokimoto}},
  \bibinfo{author}{\bibfnamefont{H.~L.} \bibnamefont{Clark}}, \bibnamefont{and}
  \bibinfo{author}{\bibfnamefont{B.}~\bibnamefont{John}},
  \bibinfo{journal}{Phys. \ Rev. \ C} \textbf{\bibinfo{volume}{69}},
  \bibinfo{pages}{034611} (\bibinfo{year}{2004}{\natexlab{b}}).

\bibitem[{\citenamefont{Youngblood et~al.}(1996)\citenamefont{Youngblood,
  Clark, and Lui}}]{youngblood1996}
\bibinfo{author}{\bibfnamefont{D.~H.} \bibnamefont{Youngblood}},
  \bibinfo{author}{\bibfnamefont{H.~L.} \bibnamefont{Clark}}, \bibnamefont{and}
  \bibinfo{author}{\bibfnamefont{Y.~W.} \bibnamefont{Lui}},
  \bibinfo{journal}{Phys. \ Rev. \ Lett.} \textbf{\bibinfo{volume}{76}},
  \bibinfo{pages}{1429} (\bibinfo{year}{1996}).

\bibitem[{\citenamefont{Fricke et~al.}(1995)\citenamefont{Fricke, Bernhardt,
  Heilig, Schaller, Schellenberg, and Shera}}]{fricke1995}
\bibinfo{author}{\bibfnamefont{G.}~\bibnamefont{Fricke}},
  \bibinfo{author}{\bibfnamefont{C.}~\bibnamefont{Bernhardt}},
  \bibinfo{author}{\bibfnamefont{K.}~\bibnamefont{Heilig}},
  \bibinfo{author}{\bibfnamefont{L.~A.} \bibnamefont{Schaller}},
  \bibinfo{author}{\bibfnamefont{L.}~\bibnamefont{Schellenberg}},
  \bibnamefont{and} \bibinfo{author}{\bibfnamefont{E.~B.} \bibnamefont{Shera}},
  \bibinfo{journal}{At. \ Data \ Nucl. \ Data \ Tables}
  \textbf{\bibinfo{volume}{60}}, \bibinfo{pages}{177} (\bibinfo{year}{1995}).

\bibitem[{\citenamefont{Pearson et~al.}(2010)\citenamefont{Pearson, Chamel, and
  Goriely}}]{pearson2010}
\bibinfo{author}{\bibfnamefont{J.~M.} \bibnamefont{Pearson}},
  \bibinfo{author}{\bibfnamefont{N.}~\bibnamefont{Chamel}}, \bibnamefont{and}
  \bibinfo{author}{\bibfnamefont{S.}~\bibnamefont{Goriely}},
  \bibinfo{journal}{Phys. \ Rev. \ C} \textbf{\bibinfo{volume}{82}},
  \bibinfo{pages}{037301} (\bibinfo{year}{2010}).

\bibitem[{\citenamefont{Duhamel et~al.}(1988)\citenamefont{Duhamel, Buenerd,
  de~Saintignon, Chauvin, Lebrun, Martin, and Perrin}}]{duhamel1988}
\bibinfo{author}{\bibfnamefont{G.}~\bibnamefont{Duhamel}},
  \bibinfo{author}{\bibfnamefont{M.}~\bibnamefont{Buenerd}},
  \bibinfo{author}{\bibfnamefont{P.}~\bibnamefont{de~Saintignon}},
  \bibinfo{author}{\bibfnamefont{J.}~\bibnamefont{Chauvin}},
  \bibinfo{author}{\bibfnamefont{D.}~\bibnamefont{Lebrun}},
  \bibinfo{author}{\bibfnamefont{P.}~\bibnamefont{Martin}}, \bibnamefont{and}
  \bibinfo{author}{\bibfnamefont{G.}~\bibnamefont{Perrin}},
  \bibinfo{journal}{Phys. \ Rev. \ C} \textbf{\bibinfo{volume}{38}},
  \bibinfo{pages}{2509} (\bibinfo{year}{1988}).

\bibitem[{\citenamefont{Youngblood
  et~al.}(2004{\natexlab{b}})\citenamefont{Youngblood, Lui, John, Tokimoto,
  Clark, and Chen}}]{youngblood2004a}
\bibinfo{author}{\bibfnamefont{D.~H.} \bibnamefont{Youngblood}},
  \bibinfo{author}{\bibfnamefont{Y.~W.} \bibnamefont{Lui}},
  \bibinfo{author}{\bibfnamefont{B.}~\bibnamefont{John}},
  \bibinfo{author}{\bibfnamefont{Y.}~\bibnamefont{Tokimoto}},
  \bibinfo{author}{\bibfnamefont{H.~L.} \bibnamefont{Clark}}, \bibnamefont{and}
  \bibinfo{author}{\bibfnamefont{X.}~\bibnamefont{Chen}},
  \bibinfo{journal}{Phys. \ Rev. \ C} \textbf{\bibinfo{volume}{69}},
  \bibinfo{pages}{054312} (\bibinfo{year}{2004}{\natexlab{b}}).

\bibitem[{\citenamefont{Itoh et~al.}(2002)\citenamefont{Itoh, Sakaguchi,
  Uchida, Ishikawa, Kawabata, Murakami, Takeda, Taki, Terashima, Tsukahara
  et~al.}}]{itoh2002}
\bibinfo{author}{\bibfnamefont{M.}~\bibnamefont{Itoh}},
  \bibinfo{author}{\bibfnamefont{H.}~\bibnamefont{Sakaguchi}},
  \bibinfo{author}{\bibfnamefont{M.}~\bibnamefont{Uchida}},
  \bibinfo{author}{\bibfnamefont{T.}~\bibnamefont{Ishikawa}},
  \bibinfo{author}{\bibfnamefont{T.}~\bibnamefont{Kawabata}},
  \bibinfo{author}{\bibfnamefont{T.}~\bibnamefont{Murakami}},
  \bibinfo{author}{\bibfnamefont{H.}~\bibnamefont{Takeda}},
  \bibinfo{author}{\bibfnamefont{T.}~\bibnamefont{Taki}},
  \bibinfo{author}{\bibfnamefont{S.}~\bibnamefont{Terashima}},
  \bibinfo{author}{\bibfnamefont{N.}~\bibnamefont{Tsukahara}},
  \bibnamefont{et~al.}, \bibinfo{journal}{Phys. \ Lett. \ B}
  \textbf{\bibinfo{volume}{549}}, \bibinfo{pages}{58} (\bibinfo{year}{2002}).

\bibitem[{\citenamefont{Reinhard}(1996)}]{reinhard1996}
\bibinfo{author}{\bibfnamefont{P.-G.} \bibnamefont{Reinhard}},
  \bibinfo{howpublished}{(unpublished)} (\bibinfo{year}{1996}).

\bibitem[{\citenamefont{Nakamura et~al.}(2010)}]{nakamura2010}
\bibinfo{author}{\bibfnamefont{K.}~\bibnamefont{Nakamura}}
  \bibnamefont{et~al.}, \bibinfo{journal}{J. \ Phys. \ G}
  \textbf{\bibinfo{volume}{37}}, \bibinfo{pages}{075021}
  (\bibinfo{year}{2010}).

\bibitem[{\citenamefont{Eidelman et~al.}(2004)}]{eidelman2004}
\bibinfo{author}{\bibfnamefont{S.}~\bibnamefont{Eidelman}}
  \bibnamefont{et~al.}, \bibinfo{journal}{Phys. \ Lett. \ B}
  \textbf{\bibinfo{volume}{592}}, \bibinfo{pages}{1} (\bibinfo{year}{2004}).

\bibitem[{\citenamefont{Centelles et~al.}(2009)\citenamefont{Centelles,
  Roca-Maza, Vi$\tilde{n}$as, and Warda}}]{centelles2009}
\bibinfo{author}{\bibfnamefont{M.}~\bibnamefont{Centelles}},
  \bibinfo{author}{\bibfnamefont{X.}~\bibnamefont{Roca-Maza}},
  \bibinfo{author}{\bibfnamefont{X.}~\bibnamefont{Vi$\tilde{n}$as}},
  \bibnamefont{and} \bibinfo{author}{\bibfnamefont{M.}~\bibnamefont{Warda}},
  \bibinfo{journal}{Phys. \ Rev. \ Lett.} \textbf{\bibinfo{volume}{102}},
  \bibinfo{pages}{122502} (\bibinfo{year}{2009}).

\bibitem[{\citenamefont{Trzcinska et~al.}(2001)\citenamefont{Trzcinska,
  Jastrzebski, Lubinski, Hartmann, Schmidt, von \~Egidy, and
  Klos}}]{trzcinska2001}
\bibinfo{author}{\bibfnamefont{A.}~\bibnamefont{Trzcinska}},
  \bibinfo{author}{\bibfnamefont{J.}~\bibnamefont{Jastrzebski}},
  \bibinfo{author}{\bibfnamefont{P.}~\bibnamefont{Lubinski}},
  \bibinfo{author}{\bibfnamefont{J.}~\bibnamefont{Hartmann}},
  \bibinfo{author}{\bibfnamefont{R.}~\bibnamefont{Schmidt}},
  \bibinfo{author}{\bibfnamefont{T.~.} \bibnamefont{von \~Egidy}},
  \bibnamefont{and} \bibinfo{author}{\bibfnamefont{B.}~\bibnamefont{Klos}},
  \bibinfo{journal}{Phys. \ Rev. \ Lett.} \textbf{\bibinfo{volume}{87}},
  \bibinfo{pages}{082501} (\bibinfo{year}{2001}).

\bibitem[{\citenamefont{Brissaud et~al.}(1972)\citenamefont{Brissaud,
  le~Bornec, Tatischeff, Bimbot, Brussel, and Duhamel}}]{brissaud1972}
\bibinfo{author}{\bibfnamefont{I.}~\bibnamefont{Brissaud}},
  \bibinfo{author}{\bibfnamefont{Y.}~\bibnamefont{le~Bornec}},
  \bibinfo{author}{\bibfnamefont{B.}~\bibnamefont{Tatischeff}},
  \bibinfo{author}{\bibfnamefont{L.}~\bibnamefont{Bimbot}},
  \bibinfo{author}{\bibfnamefont{M.~K.} \bibnamefont{Brussel}},
  \bibnamefont{and} \bibinfo{author}{\bibfnamefont{G.}~\bibnamefont{Duhamel}},
  \bibinfo{journal}{Nucl. \ Phys. \ A} \textbf{\bibinfo{volume}{191}},
  \bibinfo{pages}{145} (\bibinfo{year}{1972}).

\bibitem[{\citenamefont{Vesely et~al.}(2012)\citenamefont{Vesely, Toivanen,
  Carlsson, Dobaczewski, Michel, and Pastore}}]{vesely2012}
\bibinfo{author}{\bibfnamefont{P.}~\bibnamefont{Vesely}},
  \bibinfo{author}{\bibfnamefont{J.}~\bibnamefont{Toivanen}},
  \bibinfo{author}{\bibfnamefont{G.}~\bibnamefont{Carlsson}},
  \bibinfo{author}{\bibfnamefont{J.}~\bibnamefont{Dobaczewski}},
  \bibinfo{author}{\bibfnamefont{N.}~\bibnamefont{Michel}}, \bibnamefont{and}
  \bibinfo{author}{\bibfnamefont{A.}~\bibnamefont{Pastore}},
  \bibinfo{journal}{Phys. \ Rev. \ C} \textbf{\bibinfo{volume}{86}},
  \bibinfo{pages}{024303} (\bibinfo{year}{2012}).

\bibitem[{MIN()}]{MINUIT}
\bibinfo{howpublished}{Function Minimization and Error Analysis CERN Program
  Library entry D506 CERN Geneva}, \urlprefix\url{cernlib@cern.ch}.

\bibitem[{\citenamefont{Hamamoto et~al.}(1997)\citenamefont{Hamamoto, Sagawa,
  and Zhang}}]{hamamoto1997}
\bibinfo{author}{\bibfnamefont{I.}~\bibnamefont{Hamamoto}},
  \bibinfo{author}{\bibfnamefont{H.}~\bibnamefont{Sagawa}}, \bibnamefont{and}
  \bibinfo{author}{\bibfnamefont{X.~Z.} \bibnamefont{Zhang}},
  \bibinfo{journal}{Phys. \ Rev. \ C} \textbf{\bibinfo{volume}{56}},
  \bibinfo{pages}{3121} (\bibinfo{year}{1997}).

\bibitem[{\citenamefont{Paar et~al.}(2006)\citenamefont{Paar, Vretenar, Niksic,
  and Ring}}]{paar2006}
\bibinfo{author}{\bibfnamefont{N.}~\bibnamefont{Paar}},
  \bibinfo{author}{\bibfnamefont{D.}~\bibnamefont{Vretenar}},
  \bibinfo{author}{\bibfnamefont{T.}~\bibnamefont{Niksic}}, \bibnamefont{and}
  \bibinfo{author}{\bibfnamefont{P.}~\bibnamefont{Ring}},
  \bibinfo{journal}{Phys. \ Rev. \ C} \textbf{\bibinfo{volume}{74}},
  \bibinfo{pages}{037303} (\bibinfo{year}{2006}).

\bibitem[{\citenamefont{Myers and Swiatecki}(1995)}]{myers1995}
\bibinfo{author}{\bibfnamefont{W.}~\bibnamefont{Myers}} \bibnamefont{and}
  \bibinfo{author}{\bibfnamefont{W.}~\bibnamefont{Swiatecki}},
  \bibinfo{journal}{Nucl.\ Phys.\ A} \textbf{\bibinfo{volume}{587}},
  \bibinfo{pages}{92} (\bibinfo{year}{1995}).

\bibitem[{\citenamefont{Chen et~al.}(2005)\citenamefont{Chen, Ko, and
  Li}}]{chen2005}
\bibinfo{author}{\bibfnamefont{L.-W.} \bibnamefont{Chen}},
  \bibinfo{author}{\bibfnamefont{C.~M.} \bibnamefont{Ko}}, \bibnamefont{and}
  \bibinfo{author}{\bibfnamefont{B.-A.} \bibnamefont{Li}},
  \bibinfo{journal}{Phys. \ Rev. \ C} \textbf{\bibinfo{volume}{72}},
  \bibinfo{pages}{064309} (\bibinfo{year}{2005}).

\bibitem[{\citenamefont{Chen et~al.}(2010)\citenamefont{Chen, Ko, Li, and
  Xu}}]{chen2010}
\bibinfo{author}{\bibfnamefont{L.-W.} \bibnamefont{Chen}},
  \bibinfo{author}{\bibfnamefont{C.~M.} \bibnamefont{Ko}},
  \bibinfo{author}{\bibfnamefont{B.-A.} \bibnamefont{Li}}, \bibnamefont{and}
  \bibinfo{author}{\bibfnamefont{J.}~\bibnamefont{Xu}}, \bibinfo{journal}{Phys.
  \ Rev. \ C} \textbf{\bibinfo{volume}{82}}, \bibinfo{pages}{024321}
  (\bibinfo{year}{2010}).

\bibitem[{\citenamefont{Pearson}(2011)}]{pearson2011}
\bibinfo{author}{\bibfnamefont{J.~M.} \bibnamefont{Pearson}},
  \bibinfo{howpublished}{private communication} (\bibinfo{year}{2011}).

\bibitem[{\citenamefont{Satchler}(1973)}]{satchler1973}
\bibinfo{author}{\bibfnamefont{G.~R.} \bibnamefont{Satchler}},
  \bibinfo{journal}{Particles and Nuclei} \textbf{\bibinfo{volume}{5}},
  \bibinfo{pages}{105} (\bibinfo{year}{1973}).

\bibitem[{\citenamefont{Brack and Stocker}(1983)}]{brack1983}
\bibinfo{author}{\bibfnamefont{M.}~\bibnamefont{Brack}} \bibnamefont{and}
  \bibinfo{author}{\bibfnamefont{W.}~\bibnamefont{Stocker}},
  \bibinfo{journal}{Nucl. \ Phys. \ A} \textbf{\bibinfo{volume}{406}},
  \bibinfo{pages}{413} (\bibinfo{year}{1983}).

\bibitem[{\citenamefont{von Weizsacker}(1935)}]{weizsacker1935}
\bibinfo{author}{\bibfnamefont{C.~F.} \bibnamefont{von Weizsacker}},
  \bibinfo{journal}{Z. \ Phys.} \textbf{\bibinfo{volume}{96}},
  \bibinfo{pages}{431} (\bibinfo{year}{1935}).

\bibitem[{\citenamefont{Berg and Wilets}(1956)}]{berg1956}
\bibinfo{author}{\bibfnamefont{R.~A.} \bibnamefont{Berg}} \bibnamefont{and}
  \bibinfo{author}{\bibfnamefont{L.}~\bibnamefont{Wilets}},
  \bibinfo{journal}{Phys. \ Rev.} \textbf{\bibinfo{volume}{101}},
  \bibinfo{pages}{201} (\bibinfo{year}{1956}).

\bibitem[{\citenamefont{Berg and Wilets}(1955)}]{berg1955}
\bibinfo{author}{\bibfnamefont{R.~A.} \bibnamefont{Berg}} \bibnamefont{and}
  \bibinfo{author}{\bibfnamefont{L.}~\bibnamefont{Wilets}},
  \bibinfo{journal}{Proc. \ Phys. \ Soc. \ (London) \ A}
  \textbf{\bibinfo{volume}{68}}, \bibinfo{pages}{229} (\bibinfo{year}{1955}).

\bibitem[{\citenamefont{Dutra et~al.}(2013)\citenamefont{Dutra, Lourencco,
  Carlson, Delfino, Menezes, Avancini, Stone, Providência, and
  Typel}}]{dutra2013}
\bibinfo{author}{\bibfnamefont{M.}~\bibnamefont{Dutra}},
  \bibinfo{author}{\bibfnamefont{O.}~\bibnamefont{Lourencco}},
  \bibinfo{author}{\bibfnamefont{B.~V.} \bibnamefont{Carlson}},
  \bibinfo{author}{\bibfnamefont{A.}~\bibnamefont{Delfino}},
  \bibinfo{author}{\bibfnamefont{D.~P.} \bibnamefont{Menezes}},
  \bibinfo{author}{\bibfnamefont{S.~S.} \bibnamefont{Avancini}},
  \bibinfo{author}{\bibfnamefont{J.~R.} \bibnamefont{Stone}},
  \bibinfo{author}{\bibfnamefont{C.}~\bibnamefont{Providência}},
  \bibnamefont{and} \bibinfo{author}{\bibfnamefont{S.}~\bibnamefont{Typel}}
  (\bibinfo{year}{2013}), \eprint{arXiv/1303.2562}.

\bibitem[{\citenamefont{Agrawal et~al.}(2005)\citenamefont{Agrawal, Shlomo, and
  Au}}]{agrawal2005}
\bibinfo{author}{\bibfnamefont{B.~K.} \bibnamefont{Agrawal}},
  \bibinfo{author}{\bibfnamefont{S.}~\bibnamefont{Shlomo}}, \bibnamefont{and}
  \bibinfo{author}{\bibfnamefont{V.~K.} \bibnamefont{Au}},
  \bibinfo{journal}{Phys. \ Rev. \ C} \textbf{\bibinfo{volume}{72}},
  \bibinfo{pages}{014310} (\bibinfo{year}{2005}).

\bibitem[{\citenamefont{Piekarewicz and Centelles}(2009)}]{piekarewicz2009}
\bibinfo{author}{\bibfnamefont{J.}~\bibnamefont{Piekarewicz}} \bibnamefont{and}
  \bibinfo{author}{\bibfnamefont{M.}~\bibnamefont{Centelles}},
  \bibinfo{journal}{Phys. \ Rev. \ C} \textbf{\bibinfo{volume}{79}},
  \bibinfo{pages}{054311} (\bibinfo{year}{2009}).

\bibitem[{\citenamefont{Anders and Shlomo}(2011)}]{anders2011}
\bibinfo{author}{\bibfnamefont{M.}~\bibnamefont{Anders}} \bibnamefont{and}
  \bibinfo{author}{\bibfnamefont{S.}~\bibnamefont{Shlomo}},
  \bibinfo{howpublished}{private communication} (\bibinfo{year}{2011}).

\bibitem[{\citenamefont{Agrawal}(2011)}]{agrawal2011}
\bibinfo{author}{\bibfnamefont{B.~K.} \bibnamefont{Agrawal}},
  \bibinfo{howpublished}{private communication} (\bibinfo{year}{2011}).

\bibitem[{\citenamefont{Reinhard}(2012)}]{reinhard2012}
\bibinfo{author}{\bibfnamefont{P.-G.} \bibnamefont{Reinhard}},
  \bibinfo{howpublished}{private communication} (\bibinfo{year}{2012}).

\bibitem[{\citenamefont{Piekarewicz}(2010)}]{piekarewicz2010}
\bibinfo{author}{\bibfnamefont{J.}~\bibnamefont{Piekarewicz}},
  \bibinfo{journal}{J. \ Phys. \ G} \textbf{\bibinfo{volume}{37}},
  \bibinfo{pages}{064038} (\bibinfo{year}{2010}).

\bibitem[{\citenamefont{Sil et~al.}(2006)\citenamefont{Sil, Shlomo, Agrawal,
  and Reinhard}}]{sil2006}
\bibinfo{author}{\bibfnamefont{T.}~\bibnamefont{Sil}},
  \bibinfo{author}{\bibfnamefont{S.}~\bibnamefont{Shlomo}},
  \bibinfo{author}{\bibfnamefont{B.~K.} \bibnamefont{Agrawal}},
  \bibnamefont{and} \bibinfo{author}{\bibfnamefont{P.-G.}
  \bibnamefont{Reinhard}}, \bibinfo{journal}{Phys. \ Rev. \ C}
  \textbf{\bibinfo{volume}{73}}, \bibinfo{pages}{034316}
  (\bibinfo{year}{2006}).

\bibitem[{\citenamefont{Treiner}(2013)}]{treiner2013}
\bibinfo{author}{\bibfnamefont{J.}~\bibnamefont{Treiner}},
  \bibinfo{howpublished}{private communication} (\bibinfo{year}{2013}).

\bibitem[{\citenamefont{Stone et~al.}(2007)\citenamefont{Stone, Guichon,
  Matevosyan, and Thomas}}]{stone2007}
\bibinfo{author}{\bibfnamefont{J.~R.} \bibnamefont{Stone}},
  \bibinfo{author}{\bibfnamefont{P.~A.~M.} \bibnamefont{Guichon}},
  \bibinfo{author}{\bibfnamefont{H.~H.} \bibnamefont{Matevosyan}},
  \bibnamefont{and} \bibinfo{author}{\bibfnamefont{A.~W.}
  \bibnamefont{Thomas}}, \bibinfo{journal}{Nucl. \ Phys. \ A}
  \textbf{\bibinfo{volume}{792}}, \bibinfo{pages}{341} (\bibinfo{year}{2007}).

\bibitem[{\citenamefont{Whittenbury et~al.}(2014)\citenamefont{Whittenbury,
  Carroll, Thomas, Tsushima, and Stone}}]{whittenbury2014}
\bibinfo{author}{\bibfnamefont{D.~L.} \bibnamefont{Whittenbury}},
  \bibinfo{author}{\bibfnamefont{J.~D.} \bibnamefont{Carroll}},
  \bibinfo{author}{\bibfnamefont{A.~W.} \bibnamefont{Thomas}},
  \bibinfo{author}{\bibfnamefont{K.}~\bibnamefont{Tsushima}}, \bibnamefont{and}
  \bibinfo{author}{\bibfnamefont{J.~R.} \bibnamefont{Stone}},
  \bibinfo{howpublished}{submitted to Phys.Rev.C} (\bibinfo{year}{2014}).

\end{thebibliography}
\end{document}